\documentstyle[graphics]{mn}
\onecolumn
\input{psfig}

\def\aeta{A\&A }

\def\xp{\M{x}_\perp}
\def\los{LOS}
\def\loss{LOSs}

\def\iint{\int\!\!\!\!\!\int}	% double integral
\def\iiint{\iint\!\!\!\!\!\int}	% triple integral
\def\R#1{{\mathrm{#1}}}		% roman font in math mode
\def\Eq#1{{equation~(\ref{e:#1})}}	% equation reference
\def\Ep#1{{~(\ref{e:#1})}}	% equation reference
\def\Eqs#1#2{{equations~(\ref{e:#1})-(\ref{e:#2})}}
\def\EQN#1{\label{e:#1}}        % eqn labelling a la Texsis
\def\Fig#1{{Fig.~\ref{f:#1}}}	% figure reference
\def\Figs#1{{Figs.~\ref{f:#1}}}	% figure reference
\def\Fip#1{{~\ref{f:#1}}}	% figure reference

\def\M#1{{\mathbf{#1}}}	% matrix notation
\def\T#1{{{#1}^{\bot}}}		% transposition of a matrix
\def\d#1{{\R{d}{#1}}}		% integrant
\def\MG#1{{\mbox{\boldmath $ #1$}}} %bold greek

\begin{document}

\title{Inversion of the Lyman-$\alpha$ forest: \\
 3D investigation of the intergalactic medium
}
\author[   C.~Pichon,       J.L.Vergely,   
	 E.~Rollinde,       S.~Colombi              \&
	 P.~Petitjean ]{   C.~Pichon$^{1,2,5}$,       J.L.Vergely$^{1,3}$,   
	 E.~Rollinde$^{2}$,       S.~Colombi$^{2,5}$              \&
	 P.~Petitjean$^{2,4}$ \\
${}^1$ Observatoire de Strasbourg, 11 rue de l'Universit\'e,
67000 Strasbourg, France. \\
${}^2$ Institut d'Astrophysique de Paris, 98 bis boulevard
       d'Arago, 75014 Paris, France. \\
${}^3$ Institute of Astronomy, Madingley Road CB3 OHA Cambridge, UK.\\
${}^4$ UA CNRS 173 -- DAEC, Observatoire de Paris-Meudon, F-92195 Meudon
Cedex, France \\
${}^5$ Numerical Investigations in Cosmology (N.I.C.), CNRS, France.
}
\date{Accepted for publication in MNRAS}
\maketitle

%\offprints{C.~Pichon}

\begin{abstract}
We  discuss the  implementation of  Bayesian inversion  methods in  order to
recover the properties of the  intergalactic medium from observations of the
neutral hydrogen Lyman-$\alpha$ absorptions  observed in the spectra of high
redshift  quasars  (the  so-called   Lyman-$\alpha$  forest).   We  use  two
complementary  schemes  (i)  a  constrained  Gaussian  random  field  linear
approach and (ii) a  more general non-linear explicit Bayesian deconvolution
method  which  offers  in   particular  the  possibility  to  constrain  the
parameters of the equation of state for the gas.
   
The interpolation ability of the first approach is shown to be equivalent to
the second one in the limit of negligible measurement errors, low-resolution
spectra and null mean prior.

While relying  on prior assumption for the  two-point correlation functions,
we show how to recover, at least qualitatively, the 3D topology of the large
scale  structures in  redshift  space  by inverting  a  suitable network  of
adjacent, low resolution lines of  sight.  The methods are tested on regular
bundles of lines  of sight using $N$-body simulations  specially designed to
tackle this problem.

We also  discuss the  inversion of  single lines of  sight observed  at high
spectral  resolution.   Our  preliminary  investigations  suggest  that  the
explicit Bayesian method  can be used to derive  quantitative information on
the physical  state of the gas  when the effects of  redshift distortion are
negligible.  The information in  the spectra remains degenerate with respect
to two  parameters (the temperature  scale factor and the  polytropic index)
describing the equation of state of the gas.

Redshift distortion is considered by simultaneous constrained reconstruction
of  the  velocity  and  the  density  field in  real  space  while  assuming
statistical correlation  between the two  fields.  The method seems  to work
well in the  strong prior r\'egime where peculiar  velocities are assumed to
be the most likely realization in the density field. Finally, we investigate
the effect  of line of sight separation  and number of lines  of sight.  Our
analyses suggest that  multiple low resolution lines of  sight could be used
to improve most likely velocity  reconstruction on a high resolution line of
sight.
\end{abstract}
%%%%%%%%%%%%%%%%%%%%%%%%%%%%%%%%%%%%%%%%%%%%%%%%%%%%%%%%%%%%%%%%%%%%%%%%%

%    \thesaurus{04  % A&A Section 2: Cosmology
%       	{03.13.2; %Methods: data analysis
% 	 03.13.8; %Methods: N-body simulations
% 	 03.13.6; %Methods: statistical
%          11.09.3; %{\em (Galaxies:)} intergalactic medium
% 	 11.17.1; %{\em (Galaxies:)} quasars: absorption lines
% 	 12.04.1}} %{\em (Cosmology:)} dark matter 


%\markboth{ 3-D Intergalactic Medium reconstruction.}
%{from Lyman $\alpha$ Forest }

%
%________________________________________________________________

%%%%%%%%%%%%%%%%%%%%%%%%%%%%%%%%%%%
\section{Introduction}
It  has been realized  recently that  the cosmological  mass density  of the
baryons  located in  the  intergalactic  medium (IGM)  at  high redshift  is
similar  to the  total cosmological  mass  density of  baryons predicted  by
primordial nucleosynthesis theories (Petitjean et al. 1993; Press \& Rybicki
1993; Meiksin  \& Madau 1993;  Rauch et al.   1997; Valageas et  al.  1999).
Therefore, there is probably a  close interplay between galaxy formation and
IGM evolution.  The IGM acts as the baryonic reservoir for galaxy formation,
while  star formation  activity in  forming galaxies  should  influence the
physical state of the IGM  through metal enrichment and emission of ionizing
radiation. Hence it would be of primary interest to be able to correlate the
spatial distribution of intergalactic gas with that of galaxies.

Neutral hydrogen  in the  IGM is revealed  by the numerous  absorption lines
seen in QSO  spectra (the so-called Lyman-$\alpha$ forest).   The physics of
the  gas   is  remarkably   simple:  its  thermal   state  is   governed  by
photo-ionization heating  and adiabatic cooling  (e.g., Hui \&  Gnedin 1997;
Weinberg 1999) and its dynamics results from the effects of gravity on large
scales   and   pressure   smoothing   on  small   scales   (Reisenegger   \&
Miralda-Escud\'e 1995; Bi  \& Davidsen 1997; Hui et  al. 1997).  Dark matter
and baryons trace each other quite well and the Lyman-$\alpha$ forest is due
to  mildly  over-dense  fluctuations  in  a pervasive  medium  with  density
contrasts  of the order  of 1  to 10.  The gas  should be  distributed along
filaments and/or sheets of significant extension.

This is  supported by observations of multiple  lines-of-sight (LOS) showing
that the  gaseous complexes producing  the Lyman-$\alpha$ forest  have large
sizes.  Indeed,  in the  spectra of multiple  images of lensed  quasars with
separations of  the order  of a few  arcsec (Smette  et al.  1995;  Impey et
al. 1996), the Lyman-$\alpha$ forests appear nearly identical, implying that
the  absorbing  objects  have  sizes  $>$50$h_{75}^{-1}$~kpc.\footnote{where
$h_{75}$ is  the Hubble constant expressed  in units of  75 km/s/Mpc.} Pairs
with separation  up to 500~$h_{75}^{-1}$~kpc  show an excess  of absorptions
common  to both \loss\,  compared to  what is  expected for  an uncorrelated
distribution  of  absorption  lines  (Dinshaw  et  al.  1995;  Petitjean  et
al. 1998; Crotts \& Fang 1998;  D'Odorico et al. 1998). This suggests rather
large dimensions or better coherence  length and a non-spherical geometry of
the absorbing structures (Rauch \& Haehnelt 1995).

Recent $N$-body simulations have provided a consistent theoretical framework
for the description of the  intergalactic medium (Cen et al. 1994; Petitjean
et al. 1995; Hernquist et al. 1996; Zhang et al. 1995; M\"ucket et al. 1996,
Miralda-Escud\'e et  al.  1996; Bond  \& Wadsley 1998). The  simulations are
very   successful   at  reproducing   the   main   characteristics  of   the
Lyman-$\alpha$  forest:   the  column  density   distribution,  the  Doppler
parameter  distribution, the  flux decrement  distribution and  the redshift
evolution of absorption  lines. It has become clear  that the Lyman-$\alpha$
forest is  a powerful tool to  investigate key cosmological  issues such as:
the  re-ionization  of  the  universe  (Abel \&  Haehnelt  1999;  Schaye  et
al.  1999; Ricotti  et al.   2000); the  density  fluctuation power-spectrum
(Croft  et al.   1998; Gnedin  \&  Hui 1998;  Hui 1999;  Nusser \&  Haehnelt
1999a),  the geometry of  the Universe  (Hui et  al.  1999)  or cosmological
parameters (Weinberg et al. 1999).

Applications  to real  data  have  led to  interesting  constraints on  the
fluctuation power-spectrum  (Croft et al.  1999; Nusser  \& Haehnelt 1999b),
cosmological parameters (Weinberg  et al. 1999; Theuns et  al.  2000) or the
physical characteristics  of the gas  (Schaye et al. 1999).   However, these
studies are  presently limited  by the amount  of information  available and
show that it is important to increase current \los\, data sets.

Two  approaches can  be considered:  (i)  increasing the  number of  \loss\,
observed at  intermediate and high  spectral resolution in order  to improve
the precision of the above  measurements; large redshift surveys in progress
or in preparation  such as the Sloan Digital Sky  Survey (SDSS; e.g., Szalay
2000) the Two  degree Field  (2dF; e.g.,  Fokes et al.  1999) or  the VIRMOS
redshift survey (e.g., Le f\`evre  et al. 1998) should dramatically increase
the number  of low spectral  resolution QSO spectra available  for analysis;
(ii) using groups of QSOs to constrain the 3D distribution of the gas and to
study  redshift-space  distortion   effects  taking  into  account  peculiar
velocities in the reconstruction; the ultimate goal would be to increase the
density of \loss\, so that  the reconstructed 3D spatial distribution of the
gas can  be correlated with  galaxies observed in  the same field;  the deep
imaging  surveys planned with  MEGACAM (e.g.,  Boulade et  al. 1998)  at the
Canada-France-Hawaii  Telescope and  follow-up  spectroscopy should  provide
data for such projects.

It  is  thus  of first  importance  to  prepare  the  tools needed  for  the
interpretation of  the wealth of data  that will be provided  by the planned
surveys.   Nusser  \& Haehnelt  (1999a)  have  described  a method  for  the
recovery  of the real  space density  distribution along  one \los.
Using an analytical model of the intergalactic medium, they propose a direct
inversion  of the Lyman-$\alpha$  forest seen  in the  QSO spectra  using an
iterative  scheme based on  Lucy's deconvolution  method (Lucy  1974). {This
method  yields  fields  for  the   density  in  contrast  to  Voigt  profile
decomposition.}

Here we show that these techniques can be generalized to multiple \loss\, to
reconstruct the  3D density  field (see  Vergely et al.  2001 for  a similar
application  to the  3D mapping  of  the local  interstellar medium).   This
should  help  for  characterizing  the  structures  (filaments,  sheets...),
determining physical properties of  the gas (temperature, peculiar velocity)
and discussing the cosmological evolution of the IGM.

This paper is organized as  follows: in section~2 we present basic equations
describing the relationship between  absorption along \loss\, and properties
of  the  IGM.  Section~3  is  concerned with  sketching  the  basis for  the
inversion  technique;  two methods  are  described,  a Bayesian  regularized
inverse method and a constrained random Gaussian field reconstruction, which
can actually be seen as a particular case of the first method.
%(following the ideas of, e.g., Bardeen et al. 1986 and Rice).   
Section~4 describes two N-body simulations from which we construct simulated
data.  Section~5 discusses the  use of inversion techniques implemented here
(i) to  recover the 3D spatial  distribution of the  IGM from Lyman-$\alpha$
forest absorption lines on large scales while neglecting thermal broadening;
(ii) to  address the issue of  thermal broadening on small  scales; (iii) to
take  into  account  peculiar  velocities  and correction  for  the  induced
redshift distortions.

%%%%%%%%%%%%%%%%%%%%%%%%%%%%%%%%%%%%%%%%%%%%%%%%%%%%%%%%%%%%%%%%%%%%%%
\section{The Lyman-{\Large $\alpha$} optical depth along a line of sight}
%%%%%%%%%%%%%%%%%%%%%%%%%%%%%%%%%%%%%%%%%%%%%%%%%%%%%%%%%%%%%%%%%%%%%%
%%%%%%%%%%%%%%%%%%%%%%%%%%%%%%%%%%%%%%%%%%%%%%%%%%%%%%%%%%%%%%%%%%%%%%

\label{s:relation}
The optical  depth, $\tau_{\ell}(w)$, along  the \los \, $\ell$, at
projected  position $  \M{x}_{\perp,\ell}\equiv (y_{\ell},z_{\ell})$  on the
sky, and in velocity space, $w$, is related to neutral hydrogen
density, $n_{\rm HI}$, by :
\begin{equation}
\tau_{\ell}(w)= \frac{c \, \sigma_0 }{H(\overline{z}) \sqrt{\pi}} \int
\!\!\int          \left(   \int_{-\infty}^{+\infty}      \frac{n_{{\rm
HI}}(x,\M{x}_\perp)}{b(x,\xp)}           \exp\left(-         \frac{(w-
x-v_p(x,\xp))^2}{b(x,\xp)^2}      \right)  \d    x  \right)    \delta_{\rm
D}(\M{x}_\perp - \M{x}_{\perp,\ell}) \d{}^2 \xp \,, \quad \ell=1\cdots
L, \EQN{eqfun}
\end{equation} 
where $\sigma_0$  is the  effective  cross-section for  resonant  line
scattering,  $H(\overline{z})$ the   Hubble constant  at mean redshift
$\overline{z}$ and $v_p(x)$ is the projection of the peculiar velocity
along   the \los.   The   double sum  over   $\xp$ corresponds  to the
integration  in the directions  perpendicular to the \loss.  $\delta_D$
is the 2D  Dirac distribution.  The Doppler  parameter $b(\M{x})$ is 
considered a
function of the local temperature of the IGM at point $\M{x}\equiv
(x,\xp)$  where $x$ is the  real  space coordinate  expressed in km/s
[{$={r} {H(\overline{z})}$}].

This work is concerned with
assessing  the inversion of    {\Eq{eqfun}}
with the aim of constraining 
the  3D fields,   $n_{{\rm HI}}(x,\M{x}_\perp) $, $b(x,\M{x}_\perp) $ 
and $v_{p}(x,\M{\xp})$, from the  knowledge of  a  bundle  of  lines of
sight,  $\ell=1\cdots L$.  \\

\subsection{The model}
To relate
the gas density, the dark matter  (DM) density and the temperature,
we follow  the prescriptions of Hui \&  Gnedin (1997). We
refer to this paper for a detailed derivation of  the relations given
below.  We  assume that 
baryons trace dark matter potential (Bi \& Davidsen  1997) and 
are in ionization equilibrium.  Therefore,
\begin{equation}
n_{\rm    HI}\ \propto\ \rho_{\rm DM}^2\ T^{-0.7} \, , \EQN{trac1}
\end{equation}
where $n_{\rm HI}$ is the neutral hydrogen particle density and 
$\rho_{\rm DM}$ the dark matter density.

%%%%\noindnet
Considering that shock heating is  unimportant for the thermal budget of the
intergalactic  gas (Hui  \& Gendin  1997),  an effective  equation of  state
describes the physical state of the gas,
\begin{equation}
T(\M{x})=\overline{T}\left(            \frac{\rho_{\rm           DM}(\M{x})}
{\overline{\rho}_{\rm DM}}\right)^{2\beta} \, .\EQN{EOS}
\end{equation} 
The parameter  $\beta$ is in  the interval $0<\beta<0.31$ (this  upper bound
corresponds  to the  asymptotic  value  at $z  =0$  far from  re-ionization).
Therefore,
\begin{equation}
n_{\rm  HI}(\M{x})=\overline{n}_{\rm HI} \left(  \frac{\rho_{\rm DM}(\M{x})}
 {\overline{\rho}_{\rm DM}}\right)^{\alpha} \quad  \mbox{\rm with a scaling}
 \quad \alpha=2{-} 1.4 \beta \, . \EQN{trac}
\end{equation}
\par\noindent If there is no turbulence then the Doppler parameter $b(x)$ at
each position is due to thermal broadening only,
\begin{equation}
b(\M{x})=13\,  {\rm  km  s^{-1}}\  \sqrt{\frac{\overline{T}}{10^4K}}  \left(
\frac{\rho_{\rm  DM}(\M{x})}  {\overline{\rho}_{\rm DM}}\right)^{\beta}  \,,
\EQN{junkish}
\end{equation} 
and \Eq{eqfun} becomes 
\begin{equation}
                  \tau_\ell(w)=A(\overline{z})c_1                       \iint
\int_{-\infty}^{+\infty}       \left(       \frac{\rho_{\rm      DM}(x,\xp)}
{\overline{\rho}_{\rm  DM}} \right)^{\alpha-\beta}\exp\left(-  c_2 \frac{(w-
x-v_p(x,\xp))^2}{\left[  {\rho_{\rm DM}(x,\xp)}  /{\overline{\rho}_{\rm DM}}
\right]^{2\beta}}   \right)   \d   x   \,   \delta_{\rm   D}(\M{x}_\perp   -
\M{x}_{\perp,\ell}) \, \d{}^{2} \xp \,. \EQN{sp1}
\end{equation}
 The parameters $c_1$ and $c_2$  depend on the characteristic temperature of
the IGM:
\begin{equation}
c_1=\left(13\sqrt{\pi}   \sqrt{\frac{\overline{T}}{10^4}}\right)^{-1}  \quad
{\rm  ,} \quad  c_2= \left(13^2  \frac{\overline{T}}{10^4} \right)^{-1}\quad
{\rm   and}  \quad   A(\overline{z})=  \overline{n}_{\rm   HI}   \frac{c  \,
\sigma_0}{H(\overline{z})}\    \propto    \    \frac{{\bar{T}}^{-0.7}}{J}\,,
\EQN{defc1}
\end{equation}
\par\noindent where $J$ is the ionizing flux assumed to be uniform. Here the
temperatures are given  in Kelvin.  The value of  $A(\overline{z})$ is fixed
by  matching the  observed average  optical  depth ($\simeq$  0.2 at  ${\bar
z}=2$)

\subsection{The r\'egimes of interest for the reconstruction}
\label{s:totor1}
Several r\'egimes will be considered in \S~5 when performing the inversion:
\begin{enumerate}
\item[(i)] {\em  Small scales or high  resolution} ($\ell \la  0.1$ Mpc): in
this  r\'egime, and although  it   might  not  necessarily be  a  good  approximation
(e.g. Hui, Gnedin \& Zhang  1997), we simply assume that redshift distortion
is negligible  [$v_p=0$ in  \Eq{sp1}] and reconstruct  the density  field in
redshift space while constraining the equation of state.
\item[(ii)] {\em Large scales or low-resolution} ($\ell \ga 1$ Mpc): in this
r\'egime, applicable  to low resolution  spectra, thermal broadening  can be
neglected and \Eq{eqfun} simply becomes:
\begin{equation}
\tau_{\ell}(w)=  A(\overline{z})  \int  \!\!  \int  {\left(  \frac{\rho_{\rm
DM}(w-v_p(x(w,\xp)),\xp)}   {\overline{\rho}_{\rm   DM}}   \right)^{\alpha}}
\delta_{\rm D}(\M{x}_\perp - \M{x}_{\perp,\ell})  \d{}^2 \xp \,, \qquad {\rm
for } \quad \ell=1\cdots L \,, \EQN{eqfun2}
\end{equation} 
where  $x(w,\xp)$ is  defined implicitly  by the  equation $x=w-v_p(x,\xp)$.
Our efforts in this r\'egime will  focus on 3D reconstruction of the density
in redshift  space, i.e. with $v_p=0$  in \Eq{eqfun2} and  known equation of
state  for  the  gas.   In  principle, redshift  distortion  should  not  be
neglected,  but this  does not  change significantly  the topology  of large
scale structures,  at least  at weakly non-linear  scales, making  thus such
simplified analysis still relevant.
\item[(iii)] {\em Intermediate scales  or intermediate resolution} ($0.1 \la
\ell  \la 1$ Mpc):  Redshift distortion  will not  be neglected  anymore and
\Eq{sp1} will be  used to determine simultaneously the  density and velocity
fields, assuming that the effective equation of state is known.
\end{enumerate}
Note that we neglect here the  statistical scatter away from \Eq{EOS} and in
particular the departure from a unique power law for larger over-densities.
%%%%%%%%%%%%%%%%%%%%%%%%%%%%%%%%%%%%%%%%%%%%%%%%%%%%%%%%%%%%%%%
%%%%%%%%%%%%%%%%%%%%%%%%%%%%%%%%%%%
\section{Deconvolution of the IGM }
\label{s:totor4}
%%%%%%%%%%%%%%%%%%%%%%%%%%%%%%%%%%%%%%%%%%%%%%%%%%%%%%%%%%%%%%%
%%%%%%%%%%%%%%%%%%%%%%%%%%%%%%%%%%%%%%%%%%%%%%%%%%%%%%%%%%%%%%%
The basic idea  is to interpolate between adjacent  \loss\, the fields which
are measured along the \loss.  This first requires assumptions on the nature
of the fields.  In fact, strictly speaking, our ability to say anything away
from the  \loss \, could  be questioned, since  to the best of  our unbiased
knowledge, space  between the \loss \,  could well be  empty.  Moreover, the
inversion of  \Eq{eqfun} is obviously not unique  and additional assumptions
must  be made  in order  to reduce  the parameter  space.  For  example, the
Doppler  parameter and/or  the  peculiar  velocity fields  are  taken to  be
described by  a simple function of  the sought density  field, $n_{\rm HI}$.
Indeed, dynamical considerations  supported by numerical simulations suggest
there  exists  a statistical  relationship  between  over-densities and  the
corresponding projected  velocity field,  while temperature and  density are
also statistically related by an equation of state.

This paper addresses these issues via two techniques:
\begin{enumerate}
\item[(i)]    a   general,    explicit    Bayesian   deconvolution    method
(\S~\ref{s:taran}),  capable of  dealing with  fields and  priors such  as a
given equation of state.  This method should allow one to deconvolve thermal
broadening  non  linearly  while  accounting  for  peculiar  velocities  and
therefore  to reconstruct  the  density/velocity field  along  a \los\,  and
constrain the  equation of state of  the gas. With several  \loss, it should
simultaneously be possible to obtain the three dimensional density field.
\item[(ii)]   a   constrained   Gaussian   random  field   linear   approach
(\S~\ref{s:constrain}),  which  relates  the peculiar  velocities  projected
along the \los\, to the 3D density field or directly the 3D density field to
the \los\,  density in redshift space.   It requires prior  knowledge of the
logarithm of  density in redshift space along  each \los \, but  can be used
after applying method (i) to each \los.
\end{enumerate}
In fact, method (i)  is very general and can be applied  in many ways, which
mainly  differ in the  priors taken  for the  statistical properties  of the
density and velocity  fields.  Method (ii) corresponds to  a given choice of
strategy  for  the  3D  density/velocity reconstruction  step:  like  Wiener
filtering, it is a particular case of method (i) (\S~\ref{s:overlap}).

%%%%%%%%%%%%%%%%%%%%%%%%%%%%%%%%%%%%%%%%%%%%%%%%%%%%%%%%%%%%%%%
\subsection{A non-parametric  explicit Bayesian regularized inverse method}
\label{s:taran}
We aim to invert \Eq{eqfun}, i.e. reconstruct the density field $n_{\rm HI}$
and the  velocity field $v_p(x,\xp)$.  To  that end, we take  a {\em model},
$g$,  such  as  \Eqs{EOS}{junkish},   which  basically  relate  the  Doppler
parameter $b$ and  the gas density $n_{\rm HI}$ to  the dark matter density,
$\rho_{\rm DM}$, and obtain \Eq{sp1}.  In this equation, there are a certain
number of {\em parameters} to  be determined, which can be continuous fields
such  as  the  dark  matter  density  or the  velocity  field,  or  discrete
parameters  such as  $\alpha$ and  $\beta$. This  set of  parameters  can be
formally  described as a  vector, $\M{M}$.   The goal  here is  to determine
$\M{M}$ by fitting the data,  $\M{D}$, i.e. the absorption spectra along the
$N$ \loss.

Since the problem is under-determined, we use a Bayesian technique described
in Tarantola \& Valette (1982a; see also e.g. Craig \& Brown 1986; Pichon \&
Thi\'ebaut 1998).  In order to achieve regularization,  this method requires
prior guess for  the parameters, or in statistical  terms, their probability
distribution function, $f_{\rm prior}(\M{M})$.

Using Bayes' theorem, the conditional  probability density
$f_{\rm post}(\M{M}|\M{D})$  for the realization $\M{M}$  given the observed
data $\M{D}$ then writes:
\begin{equation}
f_{\rm post}(\M{M}|\M{D})={\cal L}(\M{D}|\M{M})f_{\rm prior}(\M{M}) \,,
\end{equation}
where $\cal L$ is the likelihood function of the data given
the model.

If we assume that both functions  $\cal L$ and $f_{\rm prior}$ are Gaussian,
we can write:
\begin{equation}
f_{\rm
post}(\M{M}|\M{D})={\cal A}\exp\left(-\frac{1}{2}(\M{D}-g(\M{M}))^{\perp}
\cdot\M{C}_d^{-1}\cdot(\M{D}-g(\M{M}))
-\frac{1}{2}(\M{M}-\M{M}_0)^{\perp}\cdot\M{C}_0^{-1}\cdot(\M{M}-\M{M}_0)\right) 
\,,
\EQN{fpost}
\end{equation}
with   $\M{C}_d$   and   $\M{C}_0$   being   respectively   the   covariance
``matrix''\footnote{Formally defined  on continuous $+$  discrete fields, 
as is the 
vector  $\M{M}$.}  of  the observed  data  and of  the prior  guess for  the
parameters, $\M{M}_0$.   $\cal A$ is a normalization  constant.  The superscript,
$\perp$, stands for transposition.  The first argument of the exponential in
\Eq{fpost} corresponds to the likelihood of the data given the model and the
parameters\footnote{Note  that the  model  $g$ taken  here would  correspond
to \Eq{sp1}  instead of \Eqs{EOS}{junkish}  as said  earlier.}, while  the last
correspond to  the likelihood of  the parameters given the  prior $\M{M}_0$.
Note that  the assumption of a  Gaussian field for $f_{\rm  prior}$ could be
lifted, in  particular to account  for the presence of  contrasted filaments
(i.e.  we could  introduce 3  point correlation  functions, or  higher order
statistics  to account  for  the fact  that,  say, the  prior likelihood  of
aligned  overdensities is  higher).  A  possible method  for  maximizing the
posterior    probability    given    in    \Eq{fpost}   is    sketched    in
Appendix~\ref{s:minim}.
In a nutshell, the minimum,  $\langle{\M{M}}\rangle$, of the argument of the
exponential  in  \Eq{fpost}  is  shown  by  a  simple  variational  argument
(Tarantola and Valette, 1982a ; 1982b) to obey the implicit equation:
\begin{equation}
\langle{\M{M}}\rangle=\M{M}_0 + \M{C}_0 \cdot \M{G}^{\perp} \cdot (\M{C}_d +
\M{G}  \cdot \M{C}_0 \cdot  \M{G}^{\perp})^{-1} \cdot  (\M{D} +  \M{G} \cdot
(\langle{\M{M}}\rangle -\M{M}_0)-g(\langle{\M{M}}\rangle)) \,, \EQN{eq0}
\end{equation}
where $\M{G}$ is the matrix (or more rigorously, the functional operator) of
partial  derivatives   of  the  model  $g({\M{M}})$  with   respect  to  the
parameters.  Note  that, under the  assumption of Gaussianity,  the extremum
$\langle{\M{M}}\rangle$  is at  the same  time the  most  likely constrained
value of the parameters vector and its mean value. The posterior covariances
of the parameters, $\M{C}_{\M{M}}$, can be computed from \Eq{covpost}.

The method can in principle  be iterated, taking in \Eq{eq0} $\M{M}_0=\M{M}$
and $\M{C}_0=\M{C}_{\M{M}}$ to compute a new value of $\M{M}$ until possible
convergence. However, in this paper, we did not test this procedure.
We  might then  wonder  how the  choice  of the  prior  for the  parameters,
$\M{M}_0$ and  their covariance matrix, $\M{C}_0$, affect  the final result,
$\langle{\M{M}}\rangle$. 

We  will show in \S~\ref{s:overlap}  that for null
prior,  $\M{M}_0=0$,  the  method  proposed  here is  equivalent  to  Wiener
filtering if the model  is linear [$g(\M{M})=\M{G}.\M{M}$].  However, we may
include more prior information when possible.  For instance, if in the field
of interest, redshifts of galaxies and clusters, gravitational lensing or SZ
data, etc.,  are available, we  may explicitly incorporate  these additional
constraints  in the  prior $\M{M}_{0}$  instead of  extending the  data set,
$\M{D}$.  More realistic expressions  accounting for the statistical scatter
around \Eq{EOS}  and a possible  slope break are also  possible.  Additional
information about our  prejudice on the evolution of  large scale structures
can  also  be incorporated  in  the  description  of the  prior  probability
distribution  function   to  account  for,  say,   dynamically  induced  non
Gaussianity.
                                       
%%%%%%%%%%%%%%%%%%%%%%%%%%%%%%%%%%%%%%%%%%%%%%%%%%%%%%%%%
\subsection{ Constrained random field reconstruction}
\label{s:constrain}
The explicit Bayesian  method described above can be applied
to the data to reconstruct along  each \los \, the density field in redshift
space  while   constraining  the  equation  of  state,   as  illustrated  in
\S~\ref{s:T}.  When dealing    with the  large scale
r\'egime  of \S~\ref{s:totor1},  \Eq{eqfun2}  applies and the  density
contrast, defined by
\begin{equation}
\delta (x) \equiv \log\left[{\rho}_{\rm DM}/ \overline{\rho}_{\rm DM}\right]
\approx(\rho_{\rm DM}-\overline{\rho}_{\rm  DM})/\overline{\rho}_{\rm
DM} \, , \EQN{defdelta}
\end{equation}
reads, along each \los\, and in redshift space ($x=w$), 
\begin{equation}
{\delta_\ell(x  ,\M{x}_{\perp})} 
= \frac{1}{\alpha} \log\left(\frac{ \tau_{\ell}(x)}{A(\overline{z})}
 \right) \ .
\EQN{eqfun22}
\end{equation}  
This section focuses on  recovering
the 3D density field in redshift space or in real space, the latter
case requiring treatment of peculiar velocities. 
To achieve  that, we   use a constrained
random field method (e.g., Hoffman \&  Ribak 1992). 
Broadly speaking, such a method 
assumes that part of a model (here, the density in redshift space
along the \loss) is fixed by the observations. It then provides the relation
between  these  ``data'' and  the  most  likely  value of the
remaining part of the parameters (here, the density between the
\loss\, and the full 3D velocity field). This method requires some
assumptions on the statistical properties of the searched fields.
The idea is to consider large enough scales
so that non-linear effects have not driven dynamically the system 
too far away from its initial conditions which we assume to be
Gaussian distributed.\footnote{Hence, we do not address here 
possible non Gaussianity due to topological defects.}
The theory of constrained random Gaussian fields is well known (e.g., 
Rice 1944, 1945; Longuet-Higgins 1957; Adler 1981;
Bardeen et al. 1986 and references therein) and
application to our problem is detailed in Appendix~\ref{a:CRF}.

We assume that the constraints are distributed along a bundle of $L$
\loss, i.e. that the density contrast [defined above in
\Eq{defdelta}] takes the values $[{\delta}_\ell(x)]_{\ell=1\cdots L}$ 
along the \loss. Then, using linear perturbation theory and 
the Gaussian nature of underlying fields,
we can write the probability distribution function of the 
3D velocity or density field in redshift space in terms of these
constraints and of the 3D power-spectrum of the density field, $P_{3D}({\bf
k})$. A prior is thus required for $P_{3D}({\bf
k})$, but an iterative procedure can in principle 
be implemented, using the $P_{3D}({\bf
k})$ measured in the reconstructed data after redshift distortion
deconvolution as a new prior.

We demonstrate that the most likely velocity 
${\left\langle { v_p} \right\rangle}_\ell  $ along the
line  of sight $\ell$ is given by the linear relationship 
[\Eq{defKv}]
\begin{equation}
{\left\langle     {v_p}     \right\rangle}_\ell(x)    =     \sum_{\ell'}\int
K_{\ell\ell'}  (x,x') \delta_{\ell'}  (x')  \d x'  \,, \quad  \mbox{or
discretely}  \quad \left\langle \M{v_{p}}  \right\rangle =  \M{C}_{v \delta }
\cdot \M{C}_{\delta\delta}^{-1} \cdot \MG{\delta} \,, \EQN{defvconv2}
\end{equation}
where the kernel, $K_{\ell\ell'}(x,x')$,  is a simple function of the
assumed 3D power spectrum given by \Eq{defKv},
while  $  \M{C_{\delta\delta}}$ and  $  \M{C_{  v
\delta}}$ are respectively  the log density auto correlation,  and the mixed
log density-velocity correlation given by
\begin{equation}
     \M{C}_{\delta\delta} \equiv  \left( {\left\langle {  \delta_i\delta _j}
\right\rangle} \right)_{i=1\cdots n,j=1\cdots  n} \,,\quad \M{C}_{ v \delta}
\equiv\left( {\left\langle {v_i \delta _j} \right\rangle} \right)_{i=1\cdots
p,j=1\cdots n} \, , \EQN{defCdd}
\end{equation}
assuming we  know the  log-density at $n$  points in  space ($p$ stands  for the
number  of points at  which we  seek the  velocity).  

To obtain the density in real space along one \los, 
it is possible to rely on the explicit Bayesian method once
more, by using for the model, $g$, \Eq{sp1} or \Eq{eqfun2} 
with $v_p$ given by \Eq{defvconv2}.
This ``strong prior'' r\'egime will be tested against simulations in 
\S~\ref{s:velSP}. Of course, the Bayesian method could as well
allow us to perform the simultaneous 3D reconstruction of the density field.

The constrained
random field machinery can also  be used
to reconstruct the  3D density field  in redshift
space (or in real space once the density along
each \los \, is deconvolved from redshift distortion), 
${\left\langle {  \delta^{(\rm 3  D)}} \right\rangle}(\M{x}) $.
This is particularly relevant at low  spectral resolution 
which corresponds to the large scale r\'egime,
where \Eq{eqfun22} can be directly used for ${\delta}_\ell(x)$.  One obtains
[\Eq{defK}]
\begin{equation}
{\left\langle {\delta^{\rm   ({\rm 3D})}}     \right\rangle}(\M{x}_\lambda) =
  \sum_{\ell}\int  K^{\rm (3D)}_{\lambda        \ell}  (\M{x}_{\lambda},\M{x'}_\ell)
  \delta_{\ell}  (x')   \d   x' \,,   \quad 
  \mbox{or} \quad
   \left\langle  
\MG{\delta}^{({\rm 3D})}
\right\rangle
= 
\M{C}_{\delta^{({\rm 3D})}\delta} \cdot \M{C}_{\delta\delta}^{-1}
 \cdot \MG{\delta} \,, \EQN{defvconv22} 
\end{equation}
where the  kernel, $K^{\rm (3D)}_{\lambda \ell}(\M{x}_\lambda,\M{x'}_\ell)$,
is  also  a function  of  the assumed  3D power  spectrum  given by  \Eq{defK}.
$\M{C_{\delta\delta}} $ is  given by \Eq{defCdd}, $\M{C_{\delta^{({\rm
3D})}\delta}}  $  is  the  mixed  \los-3D  over-density  correlation  given  by  $
\M{C_{\delta^{({\rm       3D})}\delta}}      \equiv\left(      {\left\langle
{\delta_i^{({\rm   3D})}   \delta   _j}  \right\rangle}   \right)_{i=1\cdots
p,j=1\cdots n} $.
%%%%%%%%%%%%%%%%%%%%%%%%%%%%%%%%%%%%%%%%%%%%%%%%%%%%%%%%%
\subsection{ Overlap between the two methods and connection with Wiener filtering}
%%%%%%%%%%%%%%%%%%%%%%%%%%%%%%%%%%%%%%%%%%%%%%%%%%%%%%%%%
\label{s:overlap}
The above extrapolation technique  
is restricted to  quasi linear
analysis in redshift space and unsaturated absorption lines, 
since it assumes {\em a priori} that
the density is {\em known}  along each  \los\, and that
it is Gaussian distributed.    
As  such,  constrained random
fields methods cannot be  applied directly  to \Eq{eqfun}  which involves  a double
non-linear  convolution  over  the  underlying density  both  explicit  (via
$n_{\rm HI}$) and  implicit (via $v_p$).  The Bayesian  approach sketched in
\S~\ref{s:taran} is more general  and makes less stringent assumptions.
In  particular it should provide means of applying redshift distortion
correction on the fly while accounting for temperature induced
blending. We nonetheless show that, for linear models, when the  prior  dominates, 
the extrapolation ability of \Eq{fpost} reduces  
to  constrained  random  field extrapolation, while, in contrast, 
in the zero prior limit,  it reduces to
Wiener filtering. We  also show how the covariance of the prior
log-density and velocity can be adjusted to fix a unique linear relationship
between the sought density field and its redshift distortion.

Let us start from the explicit Bayesian method.
If the prior is null, $\M{M}_{0} \equiv 0$, the error in the
measurements   negligible,  $\M{C}_d  \approx  0$, the   model linear,
$g(\M{M})= \M{G} \cdot {\M{M}}$, \Eq{eq0} becomes
\begin{equation}
\langle{\M{M}}\rangle= \M{C}_0  \cdot \T{\M{G}} \cdot (  \M{G} \cdot \M{C}_0
\cdot \T{ \M{G}})^{-1} \cdot \M{D} \, . \EQN{eq0a}
\end{equation}
When recovering the  3D density field  from the
measured  density along the  \loss, $\M{C}_{0}  \equiv \M{C}_{\delta^{{({\rm
3D})}}  \delta^{({\rm 3D})}}$,  the  linear operator  $\M{G}$ operates  then
simply like a Dirac comb on a field $\eta$:
\begin{equation}
    \M{G}_{\ell}\cdot \eta\equiv  \int
    \delta_{D}(\xp-\M{x}_{\perp\ell}) \eta(\M{x})\, \d \xp 
    \,,
    \EQN{defGdirac}
\end{equation}
so that 
\begin{equation}
    \M{C}_0 \cdot\T{\M{G}} = \M{C}_{\delta^{({\rm 3D})} \delta} \quad {\rm 
    and }\quad \M{G} \cdot \M{C}_0   \cdot   \T{\M{G}}  
    = \M{C}_{\delta \delta}\,, \quad {\mbox{which implies for 
    \Eq{eq0a}:}} \quad 
   \left\langle  
\MG{\delta}^{({\rm 3D})}
\right\rangle=   \M{C}_{\delta^{({\rm 3D})} \delta}\cdot 
    (\M{C}_{\delta \delta})^{-1} \cdot \M{\delta}  \, .
    \EQN{overlap}
\end{equation}
Equation\Ep{overlap} is identical to \Eq{defvconv22}.  Note incidentally that if the
prior is null and the model linear but if the errors in the measurements are
accounted for, \Eq{eq0} becomes
\begin{equation}
\langle{\M{M}}\rangle=  \M{C}_0 \cdot  \T{\M{G}}  \cdot (  \M{G} \cdot
\M{C}_0 \cdot \T{ \M{G}}+\M{C}_d)^{-1} \cdot \M{D}
=   (  \T{\M{G}} \cdot
\M{C}_d^{-1} \cdot { \M{G}}+\M{C}_0^{-1})^{-1} \cdot  
\T{\M{G}} \cdot \M{C}_d^{-1}  \cdot \M{D}  \, , \EQN{weiner}
\end{equation} 
which  corresponds to  Wiener   filtering  (Wiener 1949; Zaroubi et al.~1995).
In other 
words, when the model is linear, our method is equivalent to Wiener filtering applied to  ${\M{M}}-{\M{M}}_{0}$.
When  we  seek to invert  for  both $\delta$ and  $v_p$
(hence imposing a weak prior on the field),
\begin{equation}
    \M{M} \equiv \left( \MG{\delta}, \M{v_p} \right) \,,
\end{equation}
The penalty function  [corresponding to the log of  the prior in \Eq{fpost}]
 can be re-arranged [cf.~\Eq{rest}]
\begin{equation}
 (\M{M}-\M{M}_0)^{\perp}\cdot\M{C}_0^{-1}\cdot(\M{M}-\M{M}_0)=
{\T{\left(\M{v_p}-\M{C}_{v\delta }    \cdot{\M{C}_{\delta \delta}^{-1}}
\cdot    \M{\delta}  \right)}} \cdot{{\left(\M{C}_{v v}-{\M{C}_{v \delta
}}\cdot{\M{C}_{\delta              \delta}^{-1}}\cdot\T{\M{C}_{v \delta
}}\right)^{-1}}\cdot  \left(\M{v_p}-\M{C}_{v\delta }\cdot{\M{C}_{\delta
\delta}^{-1}}\cdot\MG{\delta} \right)} \,.  \EQN{penalty}
\end{equation}
The strong prior r\'egime, mentioned in \S~\ref{s:constrain} and
tested in \S~\ref{s:velSP}, is  therefore a sub-case of \Eq{penalty} where
\[
\M{C}_{v        v}        \approx{\M{C}_{v       \delta}}\cdot{\M{C}_{\delta
\delta}^{-1}}\cdot\T{\M{C}_{v  \delta  }}  \,  \quad  \mbox{implying}  \quad
\M{v_p}  \approx \M{C}_{v \delta  } \cdot{\M{C}_{\delta  \delta}^{-1}} \cdot
\MG{\delta} \,, \EQN{weakprior}
\]
i.e. $\M{v_p}$ will take its most likely value as was assumed in
\Eq{defvconv2}.  

Both  the  explicit  Bayesian   method  and  the  constrained  random  field
reconstruction require detailed  description of a {\em prior}  model for the
large-scale structure of  the IGM in order to  fix $\M{M}_{0}$, $\M{C}_{0}$,
$P_{3D}(\M{k})$,  plus additional  relationships such  as those  sketched in
\S~\ref{s:relation}.  As mentioned  earlier, these  methods can  be iterated
with new priors  measured in the reconstructed data, but  we have not tested
the convergence of such a scheme and leave that to future work.

%-------------------------------------------------------
\begin{figure}
\centerline{\psfig{figure=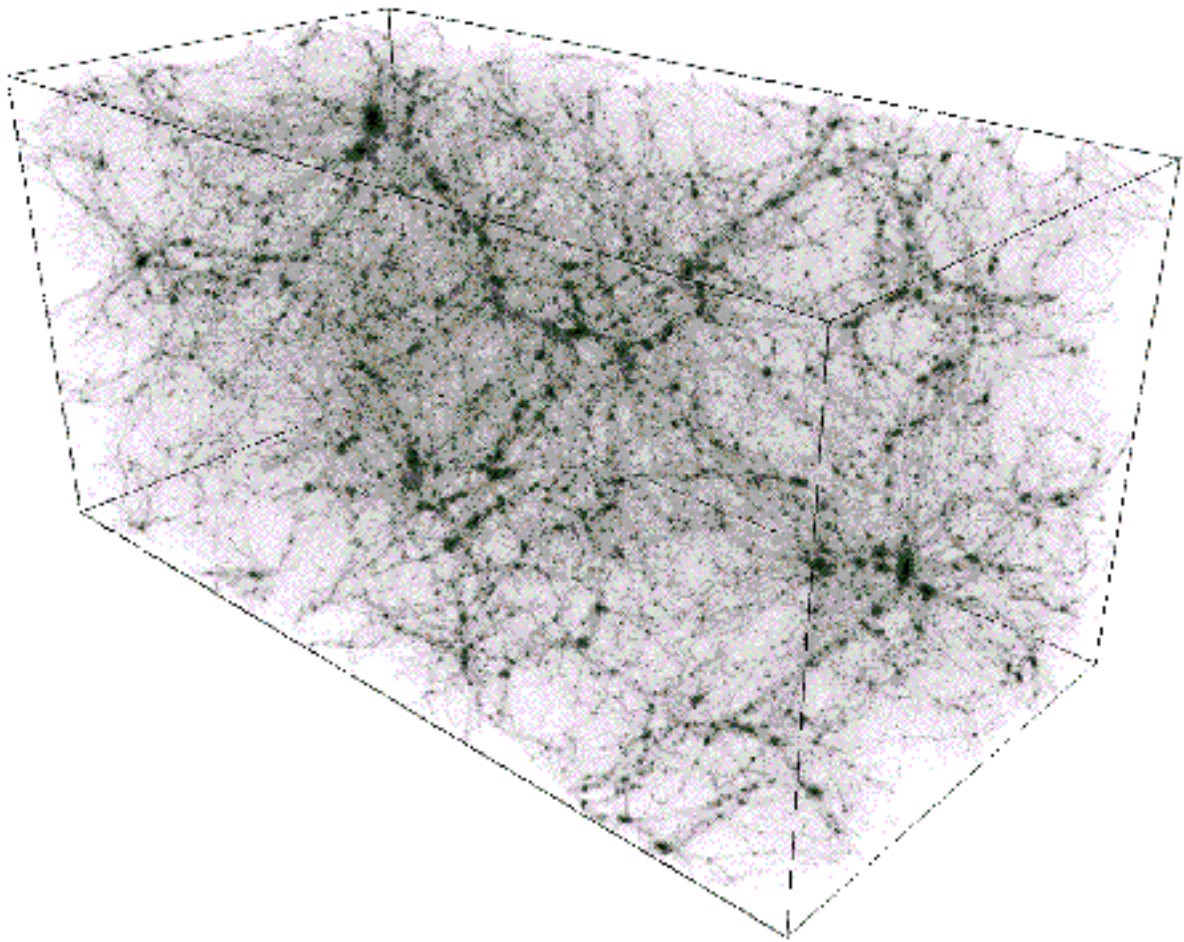,bbllx=57pt,bblly=109pt,bburx=526pt,bbury=472pt,width=17cm}}
\caption{ The  dark matter distribution in  the small simulation  box, S, at
$z=2$  (see Table~\ref{table:carasim}  and text).  The color  scales roughly
logarithmically with the projected density. Darker regions are denser.}
\label{f:figure_S}
\end{figure}
\begin{figure}
\centerline{\psfig{figure=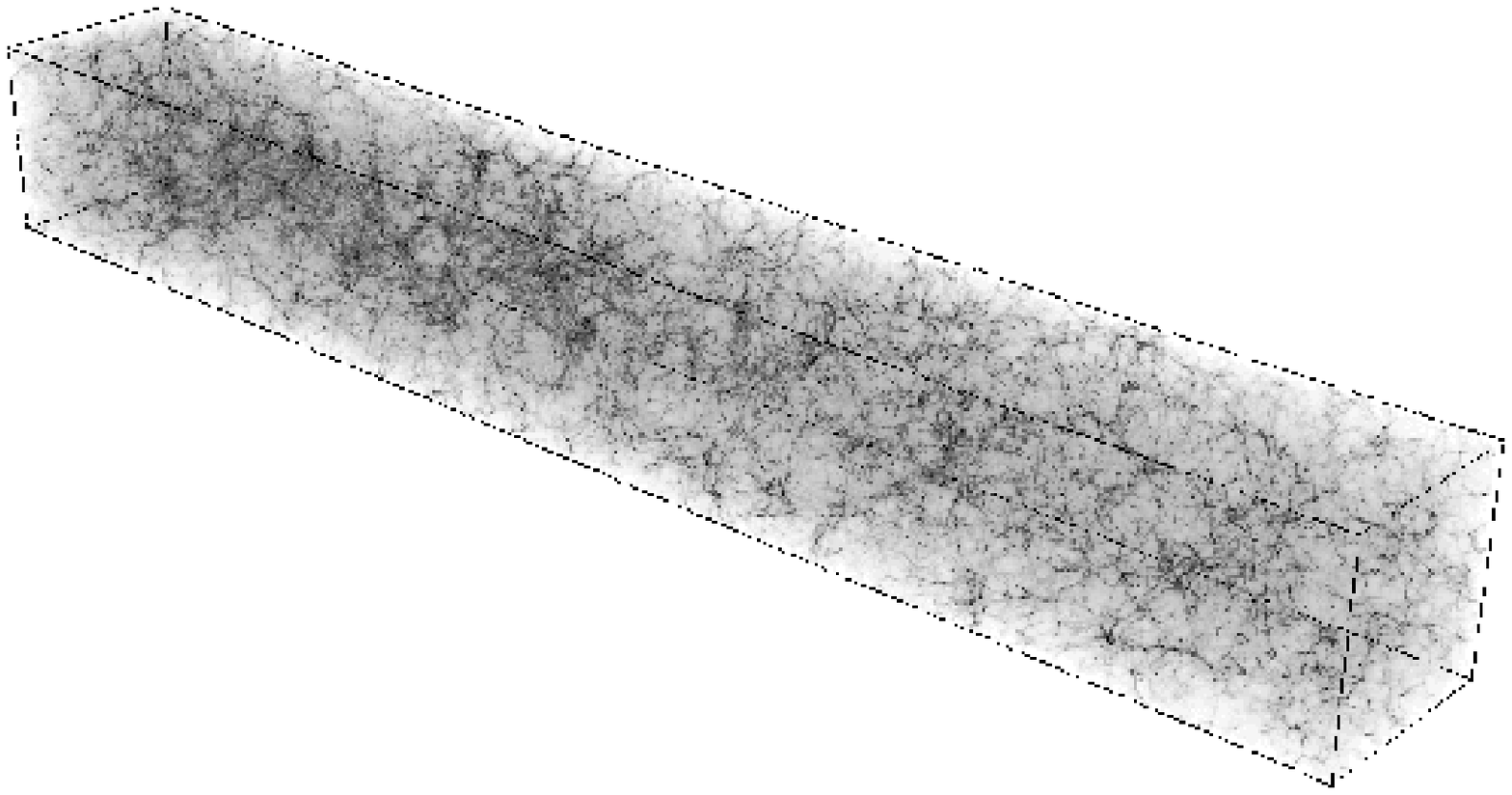,bbllx=57pt,bblly=113pt,bburx=562pt,bbury=377pt,width=17cm}}
\caption{Same as \Fig{figure_S} but for the large simulation
box, B.}
\label{f:figure_B}
\end{figure}
%-------------------------------------------------------

%%%%%%%%%%%%%%%%%%%%%%%%%%%%%%%%%%%
\section{Numerical simulations}
\label{s:simu}
%%%%%%%%%%%%%%%%%%%%%%%%%%%%%%%%%%%

To  test  our  methods  we  use  two  standard  Cold  Dark  Matter  $N$-body
simulations. The gas distribution is derived
from  the  dark  matter  distribution,  using simple  recipes  described  in
\S~2 and based  on previous  works (e.g.,  Hui \&  Gnedin 1997;  Nusser \&
Haehnelt 1999a). As discussed in   Analysis of more realistic  numerical simulations, taking
fully into account the details of  the gas dynamics is left for future work.
Many aspects  of the  reconstruction problem do  not strongly depend  on the
detail of the gas dynamics.

%%%%%%%%%%%%%%%%%%%%%%%%%%%%%%%%%%%
%\subsection{ Description of the $N$-body CDM simulations}
%
The simulations  were run with  a Particle-Mesh (PM) code,  fully vectorized
and parallelized on  SGI-CRAY architecture with shared memory.\footnote{This
program is  an improved version  of an older  code (Bouchet, Adam  \& Pellat
1985; Alimi  et al.~1990;  Moutarde et al.~1991;  Hivon 1995).  It  uses for
better   performances   a   ``predictor-corrector''  (e.g.,   Rahman   1964)
implementation of  the time-step  (instead of the  traditional ``leapfrog'',
e.g., Hockney \& Eastwood 1981).   It is still in construction but available
on request by email at  nic@iap.fr.} The characteristics of the simulations,
S  and  B, which  involve  respectively $\sim  32$  and  $\sim 16$  millions
particles,   are  given   in  Table~\ref{table:carasim}.   The  cosmological
parameters are inspired from Jenkins et al.~(1998).  The particles were laid
down on a mesh  with the same shape as the grid  used to compute the forces.
Then the Zel'dovich  (1970) approximation was used to  perturb the positions
of  the  particles  and to  set  up  Gaussian  initial conditions  with  the
appropriate  power-spectrum for standard  Cold Dark  Matter (CDM).  This was
done in a  similar way as in the COSMICS package  of Berstchinger (1995). To
avoid effects  of transients (e.g., Scoccimarro 1998),  the simulations were
started at  high redshift  $z=255$ and evolved  until the  desired redshift,
$z=2$.   \Figs{figure_S}   and  \Fip{figure_B}  display   the  corresponding
dark-matter distribution. A detailed  analysis of the power-spectrum and the
variance of  the density field measured  in the simulations  is presented in
appendix~\ref{a:simul}.

% CONSIDERATIONS SUR LA RESOLUTION
%From   Table~\ref{table:carasim},  
The   spatial  comoving   resolutions  of
simulations S and  B are $\lambda_{\rm g}\simeq 4.9$ and $40$  km~s$^{-1}$ 
respectively,  which corresponds to  physical resolutions
$\sim 8.5$ and $68$ km~s$^{-1}$ at $z=2$.
\begin{table}%[htp]
\caption[]{Characteristics of the $N$-body experiments.}
\vskip 0.4cm
\begin{tabular}{lllllllll}
\hline
Model & $\Omega_0$ & $\Lambda$ & $h$ & $\Gamma$  & $\sigma_8$  & $N_{\rm p}$ & $
L$ \\
\hline
S     &   $1.0$    &   $0.0$   & $0.5$ & $0.5$ & $0.51$ & $512\times256\times256
$ & $50 \times 25 \times 25$ \\
B     &   $1.0$    &   $0.0$   & $0.5$ & $0.5$ & $0.51$ & $1024 \times 128 \times
 128$ & $800 \times 100 \times 100$ \\
\hline
\end{tabular}
{\small
\vskip 0.3cm
Model: ``S'' and ``B'' stand for ``small'' and ``big'' respectively.

%\vskip 0.3cm
$\Omega_0$: value of the density parameter of the universe.

%\vskip 0.3cm
$\Lambda$: value of the cosmological constant.

%\vskip 0.3cm
$h$: parameterizes Hubble constant, $H_0=100 h$ km/s/Mpc.

%\vskip 0.3cm
$\Gamma$: shape parameter of the initial power-spectrum (see, e.g.,
Jenkins et al.~1998 for details).

%\vskip 0.3cm
$\sigma_8^2$: the linear variance in the dark matter
at present time in a sphere of radius $8h^{-1}$ Mpc (to fix the
normalization).

%\vskip 0.3cm
$N_{\rm p}$: size of the grid used to compute the potential and the
forces; also the number of particles.

%\vskip 0.3cm
$L$: dimensions of the rectangular periodic box in comoving Mpc.
\normalsize}
\label{table:carasim}
\end{table}
This is  to be compared with  the maximum possible pixel  resolutions of the
instruments available on  the VLT: UVES, $\lambda \simeq  3$ km/s, and FORS,
$\lambda \simeq 100$ km/s.  However, the actual resolution of the simulation
depends on the  physical parameter of interest and is  always worse than the
mesh  resolution.  For density  related  processes,  we  can expect  the  PM
simulation  to  be sufficiently  accurate  at scales  as  small  as $\sim  2
\lambda_{\rm  g}$,  although  dynamics   can  actually  be  contaminated  by
softening of the forces on scales as large as $6\lambda_{\rm g}$ (Bouchet et
al.~1985).   For  velocities,  which  are  quite  sensitive  to  resolution,
numerical  comparisons between  PM simulations  and higher  resolution codes
show that results  are correct within $\sim 25$ per cent  at scales close to
$\lambda_{\rm  g}$  (e.g.,  Colombi  1996).  Concerning  the  gas  dynamics,
density fluctuations are  expected to be damped out  below the Jeans length,
and therefore it  is not necessary to have a  spatial resolution much better
than this  cut-off scale.  For example,  the thorough analysis  of Gnedin \&
Hui  (1998)  shows that  this  scale is  of  the  order of  $50-100\,h^{-1}$
comoving kpc, i.e.~$5-10$ comoving km~s$^{-1}$.  This roughly corresponds to
the spatial  resolution of  the S simulation  (at least  for density-related
quantities).  In  this respect,  the resolution of  the B simulation  is not
high  enough, and this  simulation is  only used  to test  reconstruction of
weakly non-linear structures.

%LE SMOOTHING ADAPTATIF
In addition  to small scale  softening and limited  resolution, discreteness
effects  represent another  source of  concern, particularly  in under-dense
regions.    We  apply   adaptive  Gaussian   smoothing  to   the  particle
distribution as  follows.  The mean quadratic distance,  $d_i$, between each
particle, $i$,  and its  six nearest neighbours  is computed.  This  
sets a
smoothing  length, $\ell_i=d_i$,  i.e.   the Gaussian  filter associated  to
particle   $i$  is   $W_{\ell_i}(r)  \propto   \exp(-r^2/2\ell_i^2)$  within
$3\ell_i$  after  appropriate renormalization.   In  practice, the  smoothed
density (or mass weighted velocity) is  computed on a grid chosen here to be
the same  as the simulation  grid.  Each cell,  $j$, is subdivided  in $N^3$
sub-pixels, $k_j$,  corresponding to  positions $x_{k_j}$, with  $N=3$.  The
contribution of particle $i$ to the grid site $j$ writes
\begin{equation}
   C_{j,i} \propto \sum_{{k_j},|r-x_{k_j}| \leq 3\ell_i} W_{\ell_i}(|r-x_{k_j}|)
,
\end{equation}
with the appropriate normalization $\sum_j C_{j,i}=m_i$ where $m_i$ is
the mass of particle $i$.

%%%%%%%%%%%%%%%%%%%%%%%%%%%%%%%%%%%
\section{Application}
%%%%%%%%%%%%%%%%%%%%%%%%%%%%%%%%%%%
%%%%%%%%%%%%%%%%%%%%%%%%%%%%%%%%%%%
\label{s:tomo}

In this section, we apply the methods discussed in
\S~\ref{s:totor4} to simulated Lyman $\alpha$ spectra extracted
from the $N$-body simulations [using \Eq{sp1}].

Our preliminary analyses are organized as follows.  In \S~\ref{s:prostotor},
we give some details on the models and the priors used for both the Bayesian
method     and    the     constrained    random     field    reconstruction.
Section~\ref{s:prostotor2}  deals  with  3D  reconstruction of  the  density
field. We first test the constrained random field method in a r\'egime where
the density along each \los \, is supposed to be known.
Next, we  test the Bayesian approach.   The latter method does  not rely on
such a strong prior for the density, and is first applied to the large scale
r\'egime  discussed in  \S~\ref{s:totor1}, where  thermal broadening  can be
neglected.  Moreover, redshift  distortion is  not taken  into  account.  In
section~\ref{s:T}, we apply the Bayesian method to constrain the equation of
state  of the  gas. We  consider the  small scale  r\'egime as  discussed in
\S~\ref{s:totor1}  but neglect  redshift distortion  again for  the  sake of
simplicity,  although  peculiar  velocity  effects should  realistically  be
accounted  for. These  velocities are  dealt with  in  \S~\ref{s:vel}, which
assume in turn that the equation of state of the IGM is well constrained. We
analyse the  efficiency of velocity  reconstruction versus number  of \loss,
and test  Bayesian reconstruction in  the frameworks of strong  and floating
priors.

The reader will notice that for  each problem considered, we neglect in turn
either redshift distortion  or thermal broadening. Accounting simultaneously
for both  effects can  in principle be  achieved with the  explicit Bayesian
method or  a combination with  the constrained random  field reconstruction.
However, our  main goal here  was to illustrate  the method and to  pin down
various effects  at each  step of the  reconstruction, concentrating  on one
particular property  of the IGM,  such as the  structures of the  3D density
field,  the  equation  of  state,  or  redshift  distorsion.   More  general
applications will be developed in future work.

%-------------------------------------------------------
\begin{figure}
\centerline{\psfig{figure=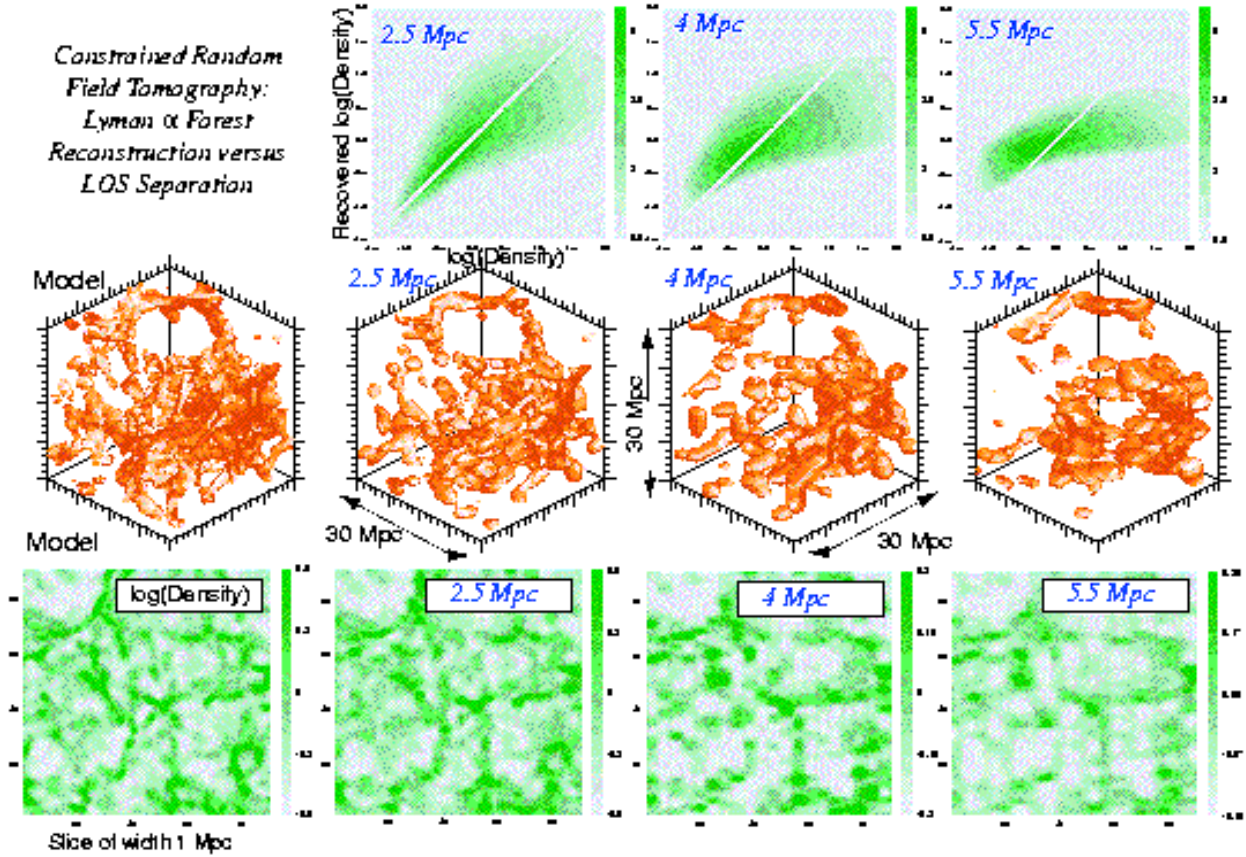,width=17cm}}
\caption{ {\em  Top panels,  from left to  right:} The recovered  log density
versus  the real  (simulated)  log density  as  a function  of the  distance
between  the \loss,  $L_{\rm  LOS}$, as  labeled:  as expected  the
bias  increases with $L_{\rm  LOS}$;  {\em Middle  panels,  from left  to
right:} the model and the  reconstructed density for $L_{\rm LOS}=2.5,4$ and
$5.5$  Mpc comoving;  {\em Bottom  panels, from  left to  right:} a  slice of
1x80x80 Mpc across the simulation and the reconstructed fields (the
scale on the panels is in pixels). Most
of the small scale structures are lost in the reconstructed field. The large
scale  topology   is  however recovered.  The rounded features in  the
reconstructed density are an artifact of the interpolation method.  }
\label{f:CRF}
\end{figure}
%-------------------------------------------------------

%%%%%%%%%%%%%%%%%%%%%%%%%%%%
\subsection{The priors}
\label{s:prostotor}
%%%%%%%%%%%%%%%%%%%%%%%%%%%%
%%%%%%%%%%%%%%%%%%%%%%%%%%%%
\subsubsection{Explicit Bayesian method}
\label{s:priorbayes}
%%%%%%%%%%%%%%%%%%%%%%%%%%%%
The Gaussian  Bayesian prior [\Eq{fpost}]  is fully described by  the first
two  moments: the  prior choice for the parameters  of  the model,  
$\M{M}_0$, and its covariance, $\M{C}_0$.  

For the model we choose the following combination of fields and
discrete
parameters:
\begin{equation}
\M{M}= \left[\gamma(x,\xp),  v_p(x,\xp), \overline{T}, \beta \right]\, .
\EQN{totorcustom1}
\end{equation}
Function $\gamma(x,\xp)$ is defined as
\begin{equation}
\frac{\rho_{\rm DM}(x,\xp)}{{\overline{\rho}}_{\rm DM}}=D_{0}(x,\xp)
\exp\left[\gamma(x,\xp)\right] \,, \EQN{defD0}
\end{equation}
so that positivity of density is insured.
Here, $D_0(x,\xp)$ is an arbitrary function (specified
later) which fixes
the value of the prior for $\rho_{\rm
DM}(x,\xp)/{\overline{\rho}}_{\rm DM}$, when
$\gamma(x,\xp)=\gamma_0\equiv 0$. 
Note that $A(\overline{z})$ is assumed to be known throughout the 
paper.
%%%%%%%%%%%%%%%%%%%%%%%%%%%%%%%%%%%%%%%%%%%%%%%%%%%%%%%%%%%%%%%%%%%%%%%%%%%%%%
%                                                                            %
%   Rigorously, in \Eq{totorcustom1}, we                                     %
%   should have $A(\overline{z})$ as a parameter                             %
%   [\Eq{defc1}], but in our preliminary                                     %
%   analyses, we supposed it fixed to $A(\overline{z})=$ \Steph{egal \`a     %
%   quoi? C'est $A(\overline{z})$ ou $J$ qui est fix\'e? C'est diff\'erent   %
%   dans les 2 cas, puisqu'il y a une d\'ependance en temp\'erature de $A$   %
%   et donc \c{c}a peut changer les r\'esultats dans la d\'etermination      %
%   de l'\'equation d'\'etat}.                                               %
%                                                                            %
%%%%%%%%%%%%%%%%%%%%%%%%%%%%%%%%%%%%%%%%%%%%%%%%%%%%%%%%%%%%%%%%%%%%%%%%%%%%%%

For the prior, we take
\begin{equation}
 \M{M}_{0}= \left[0, 0, \overline{T}_0, \beta_0 \right]\,,
\end{equation}
where the values of  $\overline{T}_0$ and $\beta_0$ will be
given in  \S~\ref{s:T}.

We derive the prior covariance operator $\M{C}_0$ either in an {\em
ad hoc} manner (\S~\ref{s:0T}, \ref{s:T} and \ref{s:velSP})
or from the simulations (\S~\ref{s:velFP}).  In the first case,
$\M{C}_{\gamma\gamma}$, is chosen to obey :
\begin{equation}
\M{C}_{\gamma\gamma}(x,x',\xp,\xp') \equiv
\sigma_\gamma^2
\exp\left(-\frac{|x-x'|}{\xi_{x}}\right)
\exp\left(-\frac{|\xp-\xp'|}{\xi_{T}}\right)\,, \EQN{cov}
\end{equation}
where $\xi_x$ and $\xi_{T}$ are  natural lengths in the inversion and govern
the level of smoothness of the reconstruction.  Typically, $\xi_{T}$ will be
of order  of the mean transverse  distance between two lines  of sight.  The
optimal choice for  $\xi_x$ depends on the problem  considered.  If peculiar
velocity effects are neglected, $\xi_x$ can be taken as small as the maximum
scale  between  spectral  resolution  and Jeans  length  (\S~\ref{s:0T}  and
\S~\ref{s:T}). In  that case, no small  scale information is  lost along the
\loss.   However, when  redshift  distortion  is to  be  taken into  account
(e.g.  \S~\ref{s:velSP}),  it is  necessary  to  have  a smoother  prior  to
stabilize the inversion, typically  the length marking the transition toward
the non-linear r\'egime (in other words, the typical size of clumps).

The  parameter $\sigma_\gamma$  may,  if required,  depend  on position.  On
average, it corresponds  roughly to the variance of  $\gamma$ in a rectangle
of  volume  $\xi_x  \xi_{T}^2$.   It  governs indirectly  by  how  much  the
reconstructed field, $\langle \M{M} \rangle$, is allowed to float around the
prior $\M{M}_0$ while solving \Eq{eq0} with the iterative method detailed in
Appendix A. When peculiar velocity effects are neglected, this parameter can
be taken to be rather  large, of the order of $0.2$. Otherwise,  the inversion process is
more complicated: details will be given in \S~\ref{s:velSP}.
Exponential correlation  functions
turned out  to be more  appropriate than Gaussian  ones in order  to recover
filamentary  structures:   the  covariance  kernel  given   in  \Eq{cov}  is
steeper,  which  allows us  to  take  into  account high  density
fluctuations.   
%%%%%%%%%%%%%%%%%%%%%%%%%%%%
\subsubsection{Constrained random field reconstruction priors}
%%%%%%%%%%%%%%%%%%%%%%%%%%%%
The constrained random field reconstruction method,
applied in \S~\ref{s:CRF}, \S~\ref{s:NLOS} and \S~\ref{s:velSP}, 
 also requires  values for the prior covariance matrix $\M{C}_0$, which
is taken to be those measured  in the simulations, as detailed in
Appendix~\ref{a:CRF}. 
Some of the biases involved in this choice are discussed in \S~\ref{s:totor2}.

%%%%%%%%%%%%%%%%%%%%%%%%%%%%%%%%%%%
\subsection{Large scales structures: tomography of the IGM}
\label{s:prostotor2}
%%%%%%%%%%%%%%%%%%%%%%%%%%%%%%%%%%%
We apply the  two methods described in \S~3 to  recover the large scale
structures in simulation B. For this purpose,  we use a network of equally separated
\loss\, along which  we simulate spectra in  accordance with
\Eq{sp1} ( as shown in \Fig{figtmp}) while varying the
separation. We
proceed in two steps: we first ignore all issues related to finite signal to
noise ratios, thermal broadening  or line  saturation, and use 
constrained random  fields to
extrapolate the density away from the \loss \, assuming
that this latter is fully determined along the \loss \, (\S~\ref{s:CRF}); 
we then illustrate the Bayesian 
technique, which does not suppose that the density along the \loss \, is 
known (\S~\ref{s:0T}). In the latter case,
only the large scale r\'egime is considered [i.e., the r\'egime (ii)
discussed in \S~\ref{s:totor1}]
and redshift distortion is neglected ($v_p=0$). Section~\ref{s:totor2}
discusses shortcomings of the two methods and realistic extensions.

%%%%%%%%%%%%%%%%%%%%%%%%%%%%%%%%%%%
\subsubsection{Constrained random field}
\label{s:CRF}
%%%%%%%%%%%%%%%%%%%%%%%%%%%%%%%%%%%
Let us  first consider redshift  space and assume  that we have  derived the
density on  each \los  \, using for  example \Eq{eqfun22}.  Recall  that the
most likely 3D  density away from the lines  of sight obeys \Eq{defvconv22}.
The covariance matrix of the prior, $\M{C}_0=\M{C}_{\delta\delta}$, is shown
on the  top of the  bottom right panel  of \Fig{figrecvel}.  We  present the
results of a reconstruction of part  of simulation B in \Fig{CRF}.  For this
figure, we used  the discrete form of \Eq{defvconv22},  on a regular network
of overlapping sub-grids of size $20\times 20\times 20$ pixels such that the
centers  of  adjacent  sub-grids  are   separated  from  each  other  by  10
pixels. The value of the reconstructed density on one pixel is obtained by a
weighted interpolation of the  recovered density on each sub-grid containing
this pixel, the  weight being inversely proportional to  the distance of the
pixel from  the center of  the sub-grid considered.  This  procedure insures
smoothness  of the  reconstruction while  keeping the  size of  the matrices
reasonable.   The  top  panels  of  \Fig{CRF} illustrate  the  bias  in  the
extrapolation procedure as  we vary the distances between  \loss, the middle
panels  display  the 3D  reconstructed  iso-log  densities corresponding  to
$\delta=0.2$, while the bottom panels  show a slice through this field.  The
large scale  filaments are recovered  for all separations  investigated, but
small scale  structures disappear beyond  $2.5$ Mpc comoving  of separation.
The topography of the structures is well described.  As expected the density
is poorly recovered for the largest separations.
%
%-------------------------------------------------------
\begin{figure} %[tbp]
\unitlength=1cm
\begin{picture}(10,11)
\centerline{\mbox{\psfig{file=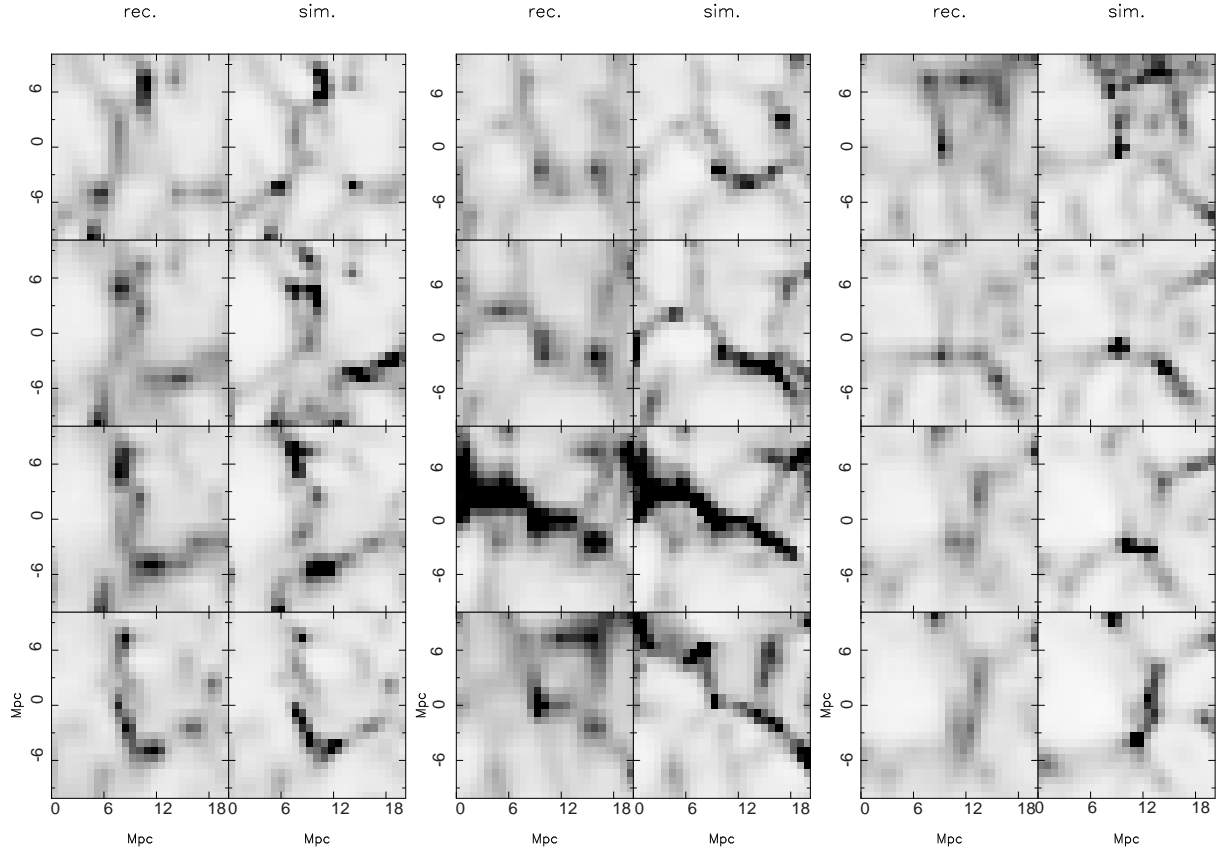,bbllx=50pt,bblly=35pt,bburx=560pt,bbury=757pt,angle=-90.0,width=16cm}}}
\end{picture}
\caption[]{  Density contrast  reconstruction using  the  Bayesian algorithm
from a  set of  9$\times$9 lines of  sight taken through  simulation~B.  The
distance between two adjacent lines of sight is equal to $2.4$ Mpc comoving.
Each panel represents respectively on the left the reconstruction and on the
right the  simulation.  Dark regions correspond to  over-dense regions.  The
filaments are well recovered.}
\label{f:figrecb1}
\end{figure}
%-------------------------------------------------------

%%%%%%%%%%%%%%%%%%%%%%%%%%%%%%%%%%%
\subsubsection{Bayesian reconstruction: line saturation and finite
signal-to-noise ratio (SNR)}
%%%%%%%%%%%%%%%%%%%%%%%%%%%%%%%%%%%
\label{s:0T}
Choosing simply  $D_0\equiv 1$  in \Eq{defD0}, our  model $g$,  on pixelized
data, reads [\Eq{eqfun2} with $v_p=0$, see also Appendix~\ref{s:Azero}]
\begin{equation}
  g_{i\ell}(\gamma) = 
A(\overline{z}) \exp[\alpha \gamma(w_{i\ell},x_{\perp\ell})]  \,,
\end{equation}
with $\alpha$ fixed equal to $1.7$  Here, $w_{i\ell}$ is the velocity at bin
$i$ corresponding to  the \los\, labelled $\ell$ and  $\gamma(x,\xp)$ is the
only parameter,  for which  the prior covariance  is given by  \Eq{cov}. The
parameters  $\sigma_{\gamma}$, $\xi_x$ are  $\xi_T$ are  respectively chosen
equal to 1,  twice the resolution and 1.5 times  the distance between \loss.
The matrix $\M{G}$ is given in Appendix~\ref{s:Azero}.
Errors in the simulated data are modeled as follows.  We
assume that  they are  uncorrelated,  so that the  covariance
error matrix $\M{C}_d$  is diagonal, with elements given
by
\begin{equation}
\sigma_{\tau}^2   \equiv   \frac{\sigma_{F}^2}{F^2}   \simeq   \frac{1}{(S/N)^2}   +
\frac{\sigma^2_0}{F^2}  =  \frac{1}{(S/N)^2}   +
{\sigma^2_0}\,{\exp(2 \tau)}  \,,  \EQN{deferror}
\end{equation}
since  the  observed flux  is  simply: $F(w)=\exp\left({-\tau(w)}\right)  $.
Equation  (\ref{e:deferror}) states  that  the  error on  the  flux has  two
origins:  a  constant  signal  to  noise  ratio  component  and  a  residual
instrumental noise, $\sigma_0$, which  dominates at large optical depth.  In
the  inversion illustrated  in \Fig{figrecb1},  we  use a  SNR of  25 and  a
residual error of magnitude 0.01.

The  reconstruction of filamentary structures is only
effective in  the r\'egime where the  distance between lines of  sight is of
the  order of  1-3 Mpc  comoving. Beyond  this limit,  the  isotropic method
presented here  is insufficient  to recover the  structure of the  IGM [such
anisotropic features may be  described by higher order correlation functions
and  stronger assumptions  relying on  a   prior different from \Eq{fpost}].
Inherent to the method is the limitation that density fluctuations at scales
smaller than  the separation between \loss\, are damped out  by the
reconstruction.   Also,  the  probability   to  intersect  a  given  strong
over-density is inversely proportional to the amplitude of the over-density.
In other words the information  regarding rare high over-densities is simply
not sampled enough by the lines of  sight.  { A related effect is induced by
flux saturation in the spectra  depending on the spectral resolution and the
SNR.  For  instance optical depths of  $\tau =5$ or $10$  will correspond to
very different  over-densities but very similar ($\approx  0$) fluxes.  Note
finally that for simplicity we have  made use of Gaussian line profiles when
Lorentzian would have been more appropriate.  }

%%%%%%%%%%%%%%%%%%%%%%%%%%%%%%%%%%%%
\subsubsection{Discussion}
%%%%%%%%%%%%%%%%%%%%%%%%%%%%%%%%%%%%
\label{s:totor2}
In the reconstruction of \S~\ref{s:CRF},  the density is assumed to be known
along the \loss,  together with the covariance matrix  of the 3D log-density
field.  At low  spectral resolution, we may neglect  both thermal broadening
and  peculiar velocities  and  use \Eq{eqfun22}  to  determine directly  the
density in redshift space from the Lyman-$\alpha$ forest along each \los. At
high  spectral  resolution,  thermal  debroadening and  redshift  distortion
deconvolution  could  in  principle  be  achieved  simultaneously  with  the
explicit  Bayesian method  or a combination  of the  Bayesian method  with the
constrained    random     field    reconstruction,    as     discussed    in
\S~\ref{s:constrain} and  shown below.

Note also that  our prior for the 3D covariance  matrix in \S~\ref{s:CRF} is
optimal: it is  measured directly in the  simulation. In that
sense, our reconstruction is biased since  we use part of the correct answer
in advance. Moreover,  we go beyond Gaussian linear  approximation, since we
work  on {\em  log}-density,  which  contributes to  improve  even more  the
reconstruction.  In real observations, we would  not have a prior as good as
that  chosen here at  our disposal. 
% Moreover  we would use  linear theory
% which applies to  density and not log-density. In fact  it might be possible
% to extend  linear theory  to log-density but  we did  not test that  in this
% paper. 
However,  as shown  in \S~\ref{s:0T}, the  results from  the explicit
Bayesian reconstruction, which  rely on a much weaker  prior, \Eq{cov}, give
very similar  results to the  constrained random field  reconstruction. This
shows that the  non linear features present in  the measured correlations do
not play an important role in our  ability to carry out the inversion on the
scales  explored here.   Finally,  it  may be worth  mentioning  again that  the
methods should be iterated, using  for new priors and covariance matrixes the
measured  ones  in  the  reconstructed  field.  

%-------------------------------------------------------
%
%%%%%%%%%%%%%%%%%%%%%%%%%%%%%%%%%%%%
\subsection{ Small scales: the IGM temperature}
%%%%%%%%%%%%%%%%%%%%%%%%%%%%%%%%%%%
\label{s:T}

% %-------------------------------------------------------
% \begin{figure*}
% \unitlength=1cm
% \begin{picture}(12,8)
% \put(0,0)\centerline{\psfig{figure=TARAN-composite-2.5.ps,width=9cm}}
% \put(0,7)\centerline{\psfig{figure=TARAN-composite-4.ps,width=9cm}}
% \end{picture}
% \caption{
% {\em  On the left: top panel, from   left  to right:}  the model 
% and  the reconstructed density $L_{\rm  LOS}=2.5$ Mpc; {\em  On the left: 
% bottom panel, from left to  right:} a slice through the  simulation and the 
% reconstructed density; Note that the number of LOS through the cube is kept 
% constant. { \em On the right:} same figure but for  $L_{\rm  LOS}=4$ 
% Mpc }
% \label{f:TARAN}
% \end{figure*}
% %-------------------------------------------------------

%-------------------------------------------------------
\begin{figure} % [tbp]
\unitlength=1cm
\begin{picture}(12,7.5)
\put(0,0){\psfig{figure=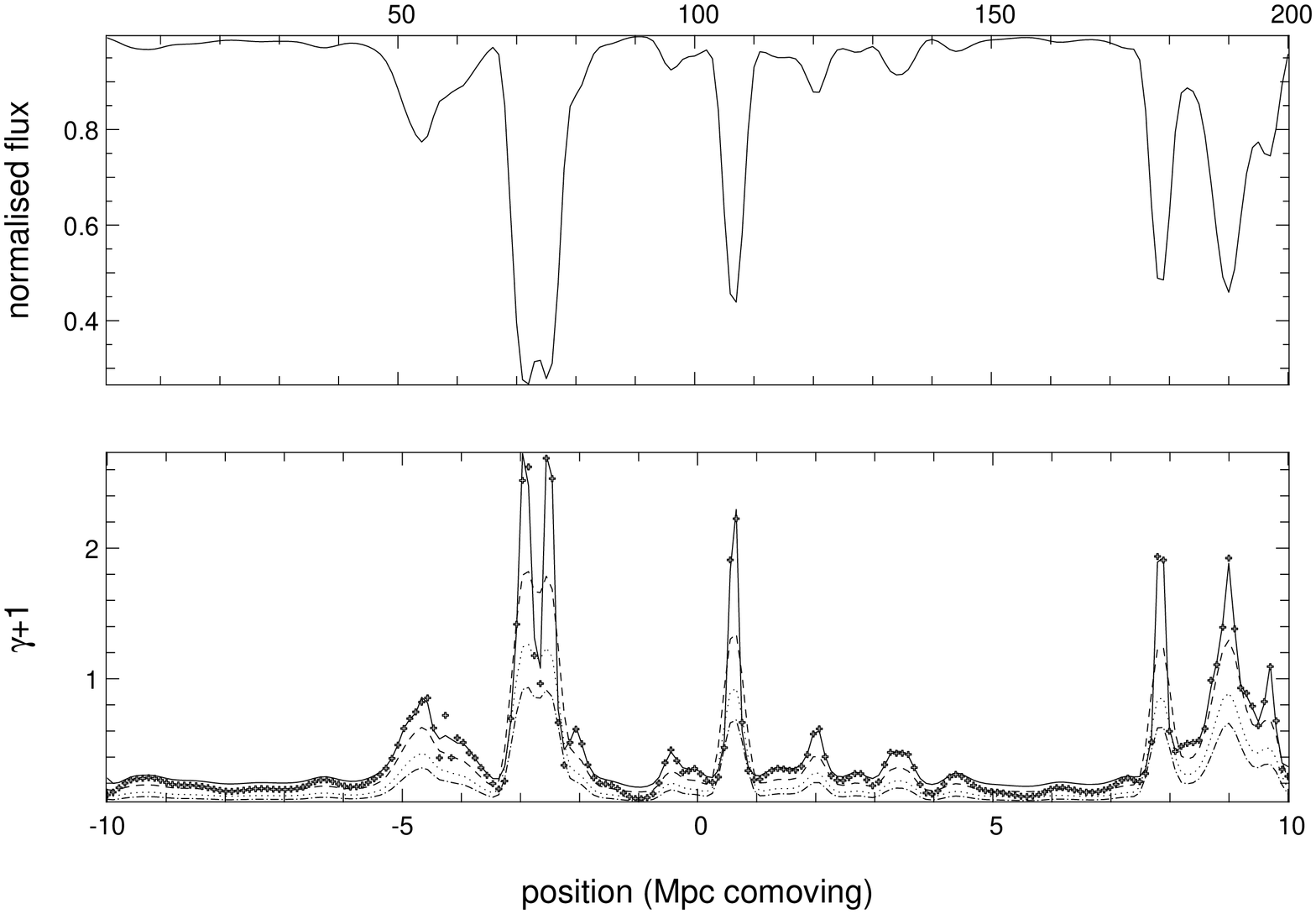,bbllx=0pt,bblly=0pt,bburx=564pt,bbury=749pt,angle=-90,width=10cm,height=8.18182cm}}
\put(11.5,0){\psfig{figure=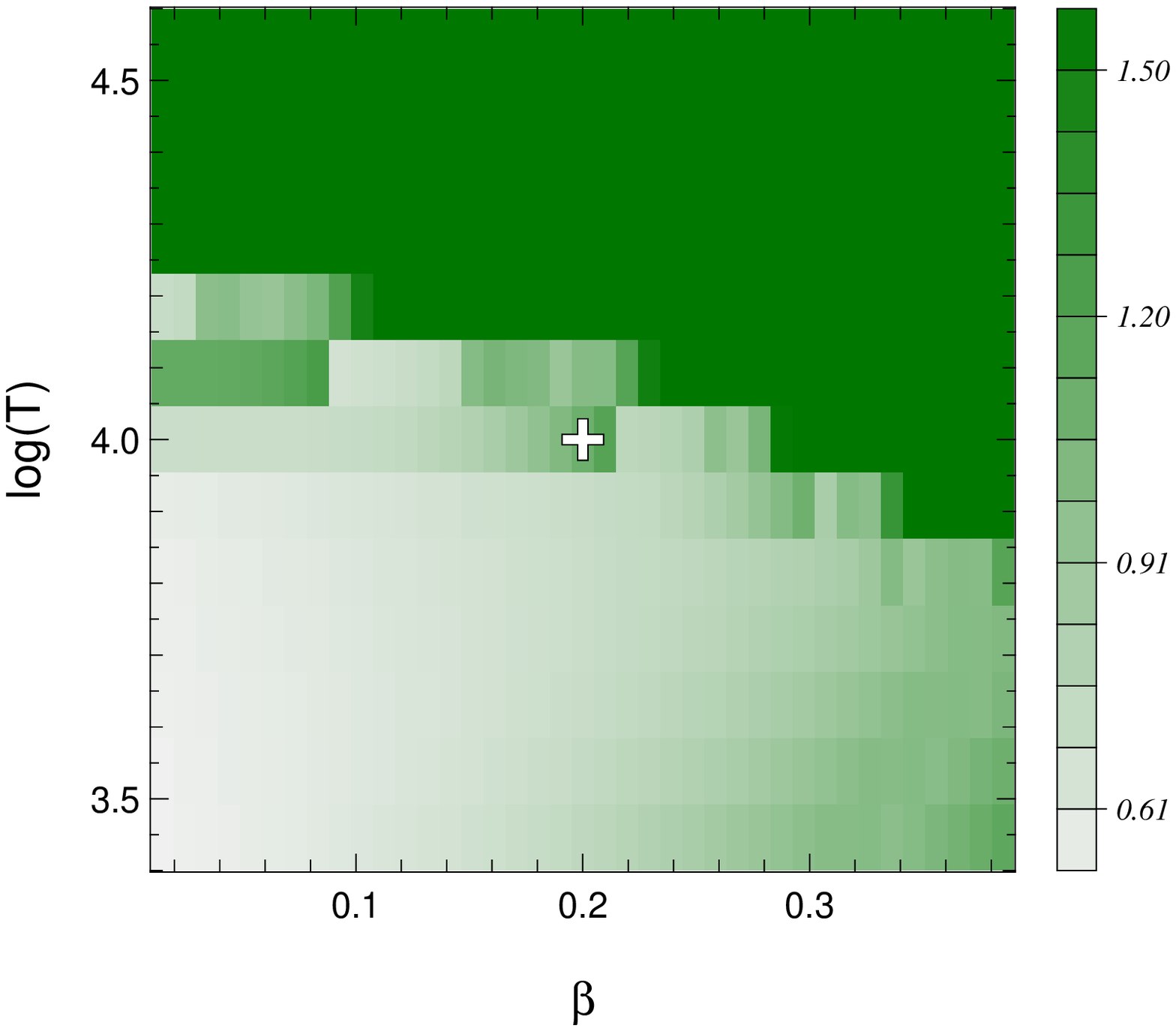,bbllx=70pt,bblly=260pt,bburx=600pt,bbury=740pt,width=7cm,height=7cm}}
\end{picture}
\caption{  {\em  Left  panels:}   Inversion  using  different  equations  of
state. The upper panel shows a  portion of simulated spectrum through S. The
equation of state used corresponds to \Eq{EOS} with $\bar T={\bar T}_{\rm t}
\equiv 10^4\  K$, $\beta=\beta_{\rm t}\equiv 0.2$.   Peculiar velocities are
not  considered.  The  lower  panel  shows the  simulated  density as  black
dots. The  density recovered using the same equation of state is plotted as a
solid line;  it is apparent that  even the internal  structure of absorption
blends is recovered.   Other curves correspond to the  results of inversions
using various lower values of $\bar  T$ at fixed $\beta=0.2$.  The effect of
lowering $\bar T$ is to give smaller values for reconstructed density with a
reduced $\chi^2  < 1$.   If, on the  contrary, ${\bar  T} > T_{\rm  t}$, one
obtains $\chi^2  \gg 1$.  {\em Right  panel:} Map of  convergence ($\chi^2 <
1$) or  divergence ($\chi^2 \gg  1$) for inversions using  \Eq{defg1MT} with
different values  of ${\bar T}$ and $\beta$.   The \los\, is the  same as in
left panels.
%A  threshold of  2  was applied  to the  $\chi^2$ value. 
}
\label{f:figtmp}
\end{figure}
%-------------------------------------------------------
%
We now aim to determine the equation  of state of the IGM by considering the
inversion of a single \los\,  observed at high spectral resolution [r\'egime
(i)  in \S~\ref{s:totor1}].   The  inversion of  the  density, velocity  and
temperature fields from a single  \los\, is not unique (Theuns et al.~1999a;
Hui \& Rutledge 1999).  Indeed,  the same spectrum can be reconstructed with
different  equations of state  and density  distributions as  illustrated by
\Fig{figtmp}.   Neglecting peculiar  velocities for  the sake  of simplicity
($v_p=0$), the problem reduces to  the determination of two parameters $\bar
T$  and   $\beta$  and  one   unknown  field,  $\gamma$.   The  simultaneous
determination of these  parameters {\em and} the field  remains a degenerate
problem. As  detailed in appendix~\ref{s:AT},  our model, $g$,  on pixelized
data reads, from \Eq{sp1},
\begin{equation}
g_{i\ell}(\gamma)=A(\overline{z})c_1   \iint  \left(\int_{-\infty}^{+\infty}
(D_{0}(x,\xp)\exp[\gamma(x,\xp)])^{\alpha-\beta}\exp\left(-               c_2
\frac{(w_{i\ell}-  x)^2}{(D_{0}(x,\xp)\exp[\gamma(x,\xp)])^{2\beta}} \right)
\d x \right) \delta_{D}(\xp-x_{\perp\ell})\d{}^{2} \xp.\EQN{defg1MT}
\end{equation}
Here, $A({\bar z})$ is arbitrarily fixed to  $A({\bar z})=0.7$ as
explained in \S~\ref{s:priorbayes},  $\alpha=2-1.4\beta$ 
[\Eq{trac}] and $c_1$ and  $c_2$ are functions of
$\bar T$ [\Eq{defc1}]. The function $D_0(x,\xp)$ is chosen to be 
\begin{equation}
D_0(x,x_{\perp})=\left(
\frac{\tau_\ell(w \equiv x)}{A(\overline{z})} \right)^{1/\alpha} 
\,.
\EQN{defD1}
\end{equation}
The  prior covariance matrix  $C_{\gamma\gamma}$ is  given by  \Eq{cov} with
$\xi_{T} \rightarrow \infty$.  Here $\xi_x$ and $\sigma_{\gamma}$ are chosen
equal to 0.2 Mpc comoving and 0.2.

We conduct our  analyses as follows. We first simulate  a spectrum along one
\los\ with a given {\em real} pair ($\beta_{\rm t}$, $\bar{T}_{\rm t}$). The
noise   matrix  $\M{C}_d$   is  the   same  as   in  \S~\ref{s:0T}   with  a
$(S/N,\sigma_0)=(50,0.05)$.  We  then invert this \los\  for $\gamma(x)$ while
varying ($\beta$, $\bar{T}$) over a given range of realistic values. In that
sense,  the only {\em  effective} parameter  in the  inversion is  the field
$\gamma$.  For each  value of  $(\beta,{\bar  T})$, we  compute the  reduced
$\chi^2$,  i.e.  $(\M{D}-g(\M{M}))^{\perp}.\M{C}_d^{-1}.(\M{D}-g(\M{M}))$ in
\Eq{fpost},  as  shown  in  right  panel  of  \Fig{figtmp}.   The  value  of
($\beta_{\rm  t}$,  $\bar{T}_{\rm  t}$)  is  shown by  a  white  cross.  The
($\beta$,  $\bar{T}$) plane  is  divided  into two  regions  separated by  a
straight borderline, one with $\chi^2  \gg 1$ (corresponding to large values
of ${\bar T}$)  and the other one with $\chi^2 \le  1$.  This arises because
the  absorption lines  are indeed  thermally broadened  and  resolved.  When
${\bar  T} >  {\bar T}_{\rm  t}$, the  absorption features  in the  data are
narrower than the model and cannot be fitted anymore.

As expected, the real parameters stand on the borderline between convergence
and  divergence: these  parameters  correspond  to a  good  fit.  We  cannot
however  distinguish  --using  a   $\chi^2$  criterion--  between  pairs  of
$(\beta,{\bar T})$  on this borderline.   Even though the degeneracy  is not
completely  lifted, this  analysis provides  a complementary  method  to the
standard  techniques of  Voigh  profile fitting  (see  Ricotti et  al.~2000;
Schaye  et al.~1999)  to measure  the  mean properties  of the  IGM and  its
cosmological  evolution.  The  application of  our  method to  real data  is
developed to a companion paper (Rollinde, Petitjean \& Pichon, submitted).

Note finally that, for close enough lines of sight (e.g. multiple lensed QSO
images)  we might  in  theory be  able  to investigate  the  small scale  3D
properties of the IGM while accounting for thermal broadening.

%
%-------------------------------------------------------
\begin{figure} %[tbp]
\unitlength=1cm
\begin{picture}(12,12.5)
\put(9,7.5){\psfig{file=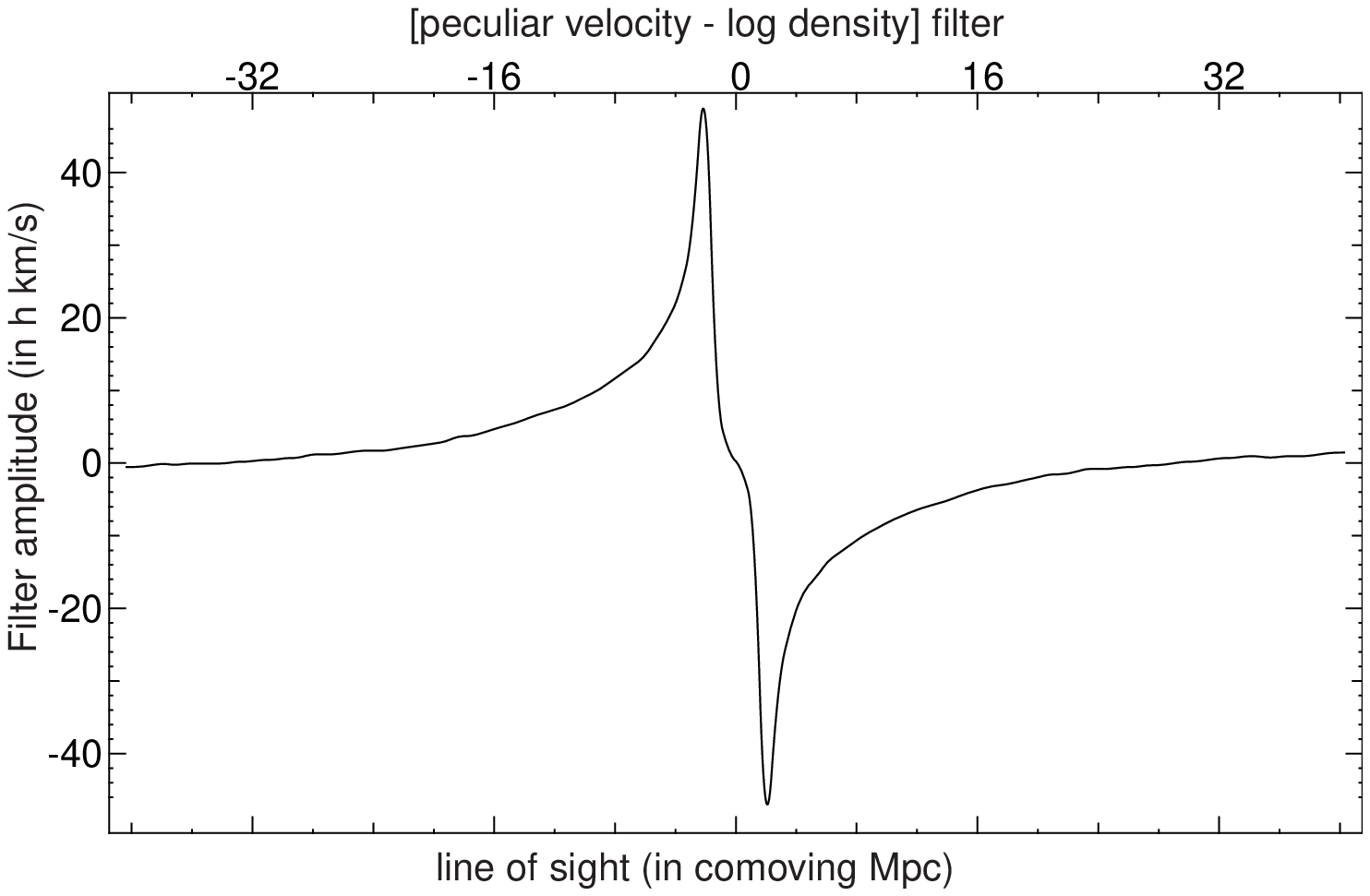,angle=0.0,width=8cm}}
\put(9,0){\psfig{file=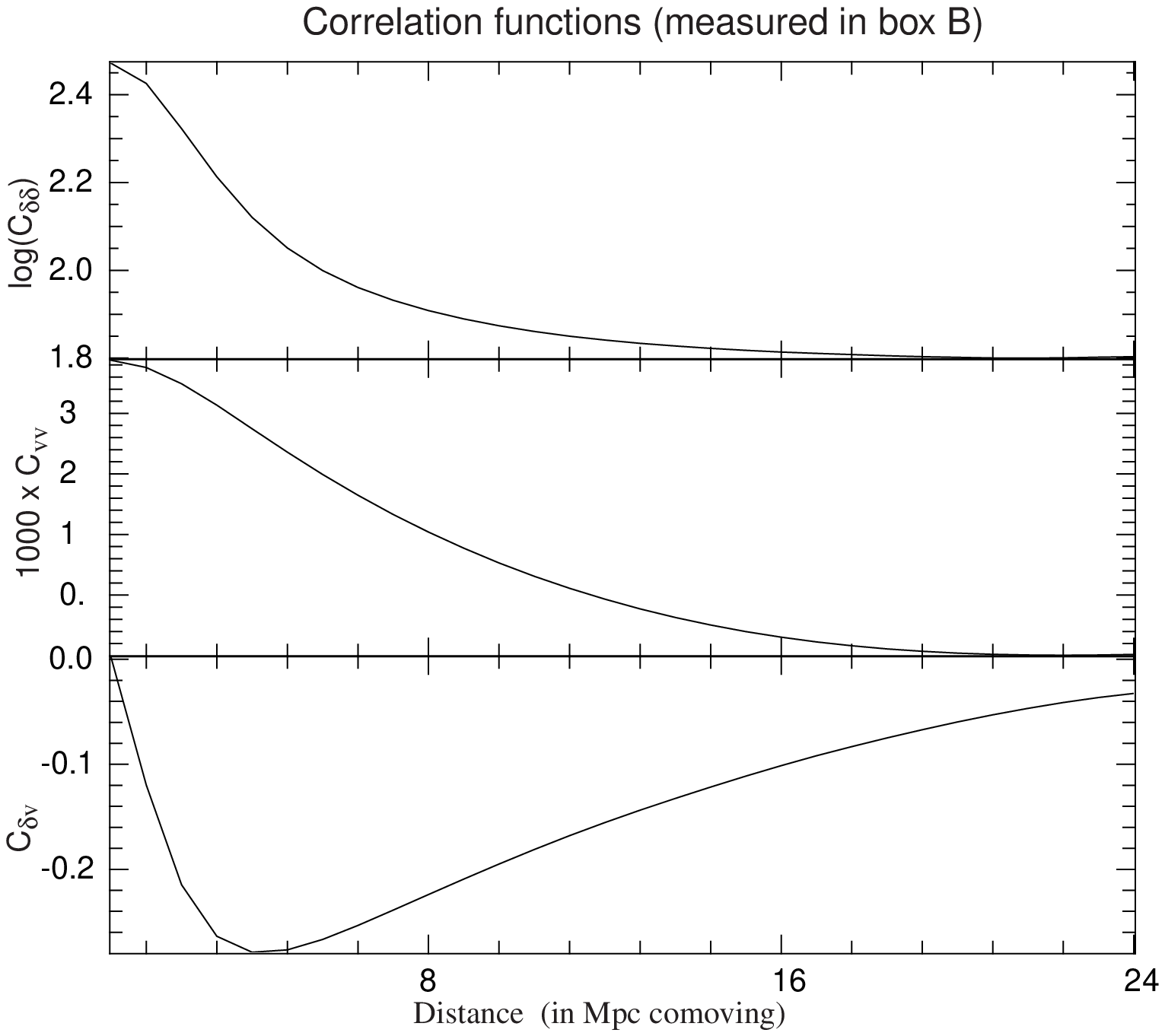,angle=0.0,width=8cm}}
\put(0,0){\psfig{file=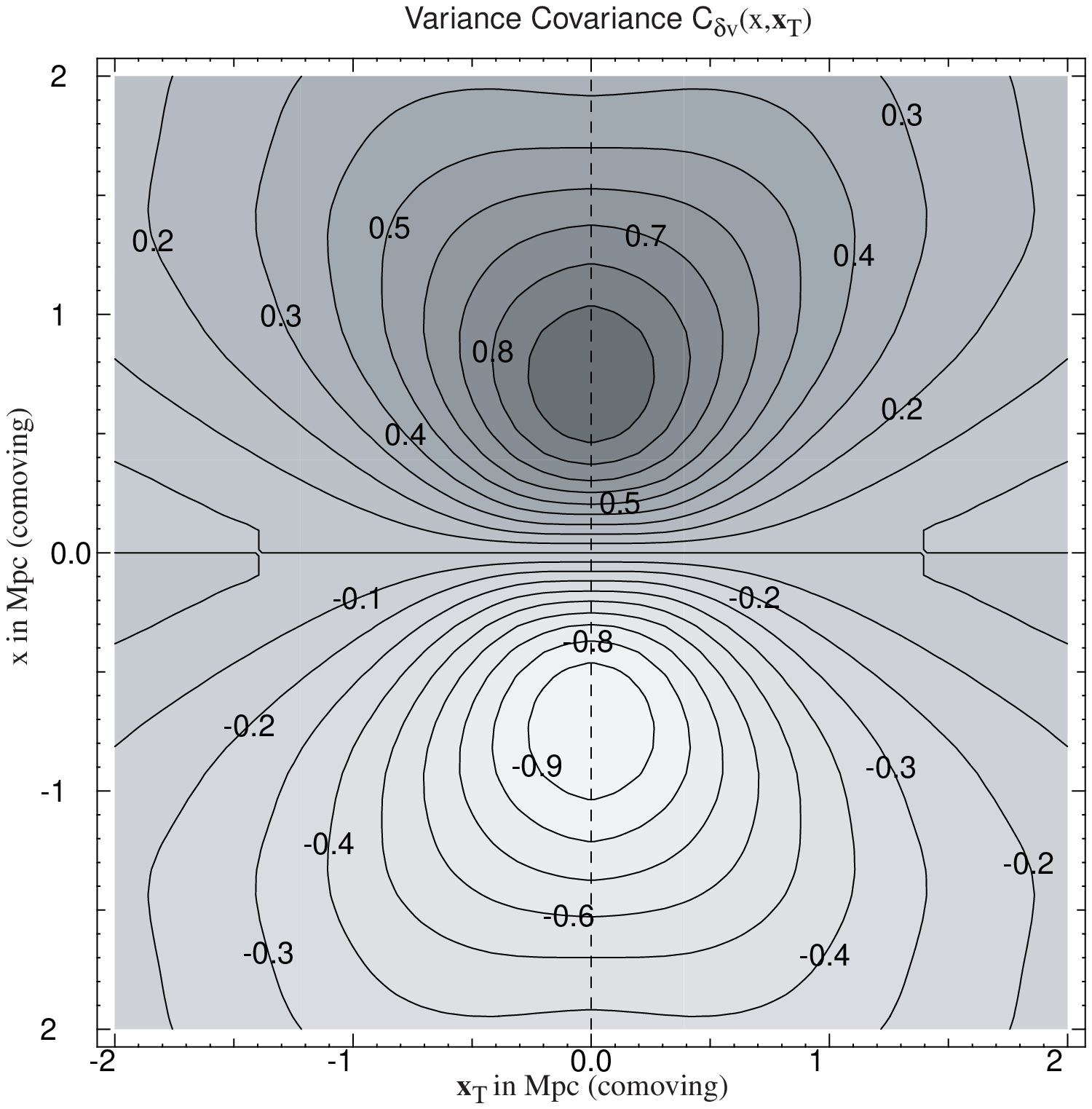,angle=0.0,width=8cm,height=8cm}}
\end{picture}
\caption{  {\em   Left  panel:}  the  3D   correlation  function,  $\M{C}_{v
\delta}(x,\xp)$,  measured in  simulation  S.  {\em  Top  right panel:}  the
filter $K^{(v)}$  required to compute  the most likely velocities  along one
\los\, [\Eq{defvconv2} with $\ell=\ell'=L=1$]. The width of the filter shows
that the peculiar  velocity has two natural scales as  discussed in the main
text.  {\em Bottom  right panel:} the 1D \los  \, correlation functions: top
sub-panel: $\log(\M{C}_{\delta \delta })$ ; middle sub-panel: $\M{C}_{v v}$;
bottom sub-panel: $\M{C}_{\delta v}$.  }
\label{f:figrecvel}
\end{figure}
%-------------------------------------------------------
%

%%%%%%%%%%%%%%%%%%%%%%%%%%%%%%%%%%%%%%%%%%%%%%%%%%%%%%%%%%%%%%%%%%%%%%
\subsection{Redshift distortion}
%%%%%%%%%%%%%%%%%%%%%%%%%%%%%%%%%%%%%%%%%%%%%%%%%%%%%%%%%%%%%%%%%%%%%%
\label{s:vel}
Recall that in this section, for the sake of simplicity, we assume that the equation of state
of the IGM is known.

There are several issues to address here. The optical depth along
a bundle of  \loss\, does not constrain uniquely  the corresponding velocity
field. This would require the  knowledge of the full 3D density distribution
together with  the assumption that  linear dynamic applies.  Thus,  we first
investigate how  increasing the number  of measured \loss\, or  changing the
mean separation  between them improves  the likelihood of  the corresponding
realization  of the  constrained velocity  field for  a given  density field
along  the  bundle  (\S~\ref{s:NLOS}).   We  then turn  to  the  problem  of
deconvolving  the optical  depth in  real space,  but conduct  a preliminary
analysis on  {\em a single} \los. We  test two approaches. The first 
approach is a
strong prior inversion (\S~\ref{s:velSP}),  i.e.~it relies on  the Bayesian 
formalism while
assuming that  the velocity  field takes its  most likely value.  The second
method  allows the velocity  field to  float around  this most  likely value
(\S~\ref{s:velFP}).  Finally, we discuss the limitations of the present work
and possible improvements (\S~\ref{s:totor3}).

Let us briefly  describe the filters and  correlation function
involved.   \Fig{figrecvel}   (left  panel)  displays   the  3D  correlation
function,  $\M{C}_{v  \delta}(x,\xp)$,  measured  in simulation  S.   It  is
antisymmetric  along the \los,  and symmetric  orthogonally.  The  top right
panel   shows   the   1D   filter,   $K^{(v)}(x,y)$   [\Eq{defvconv2}   with
$\ell=\ell'=L=1$],  which   was  in  practice  computed   according  to  the
prescription  sketched in  Appendix~\ref{a:CRF}.  This  antisymmetric filter
presents  two  characteristic scales:  a  strong  peak  at $\approx  2$  Mpc
(comoving) and  broad wings up  to $\approx 20$  Mpc. This implies  that the
most likely velocity at a given  point will depend on the local density {\em
and  also significantly} on  the density  further away  (up to  $\approx 20$
Mpc).  Transversally  the  shape  of  the  3D  cross  correlation  function,
$\M{C}_{v \delta}(x,\xp)$, which vanishes near the line ${x}=0$, implies that
the density away from a given point will dominate the local velocity field.

%%%%%%%%%%%%%%%%%%%%%%%%%%%%%%%%%%%
\subsubsection{Most likely velocity versus \los \, separation \& number of \loss}
%%%%%%%%%%%%%%%%%%%%%%%%%%%%%%%%%%%
\label{s:NLOS}
In this subsection we assume temporarily that the log-density field is known
along  a bundle  of \loss.   In the  framework of  constrained  random field
(\S~\ref{s:constrain}),  \Eq{defvconv2} gives  the relationship  between the
most likely velocity  along a given bundle of  \loss\, and the corresponding
log-density.

\begin{figure} %[tbp]
\unitlength=1cm
\begin{picture}(12,12.5)
\put(0,6.5){\psfig{file=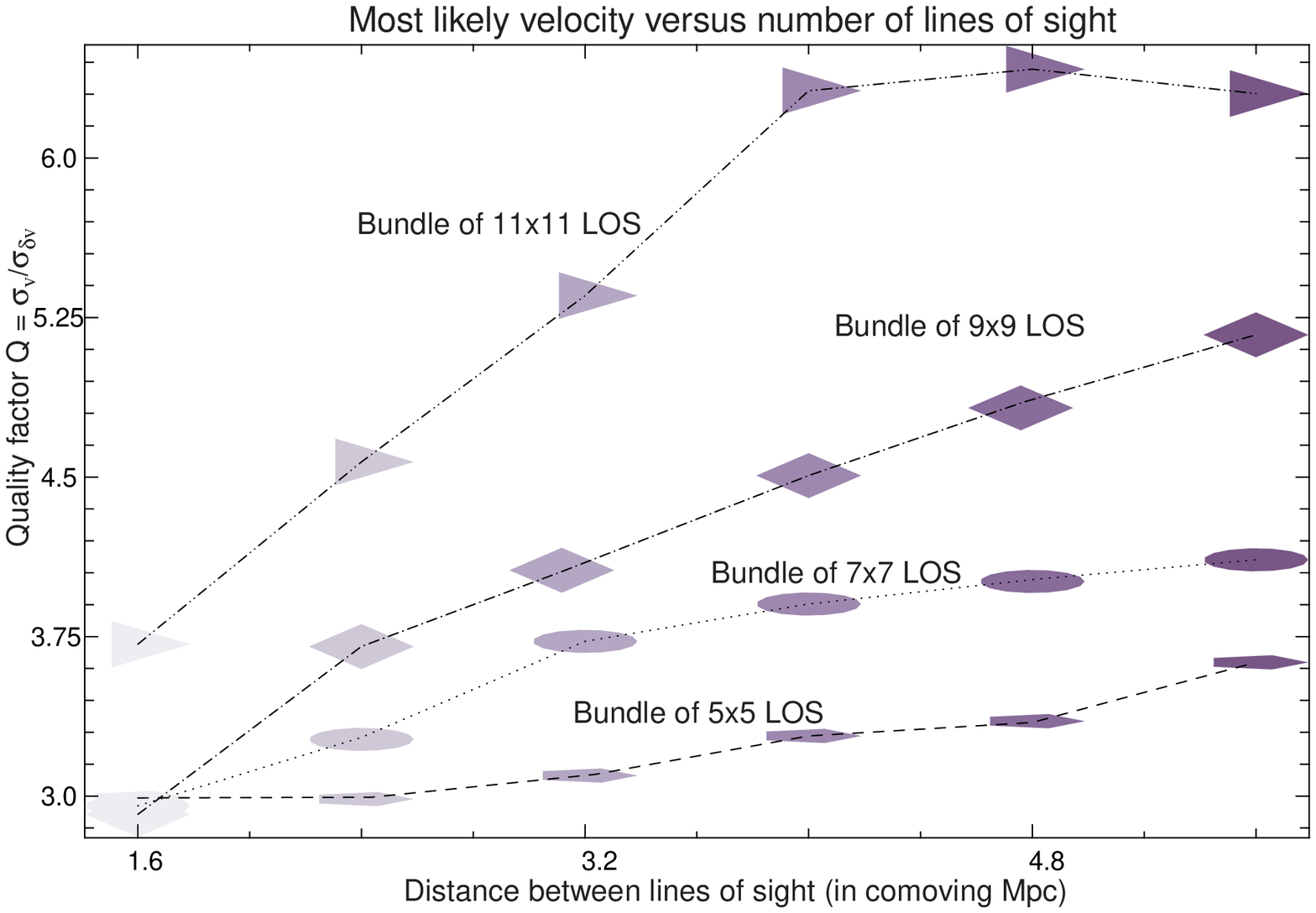,angle=0.0,width=7cm,height=6cm}}
\put(8.,6.){\psfig{file=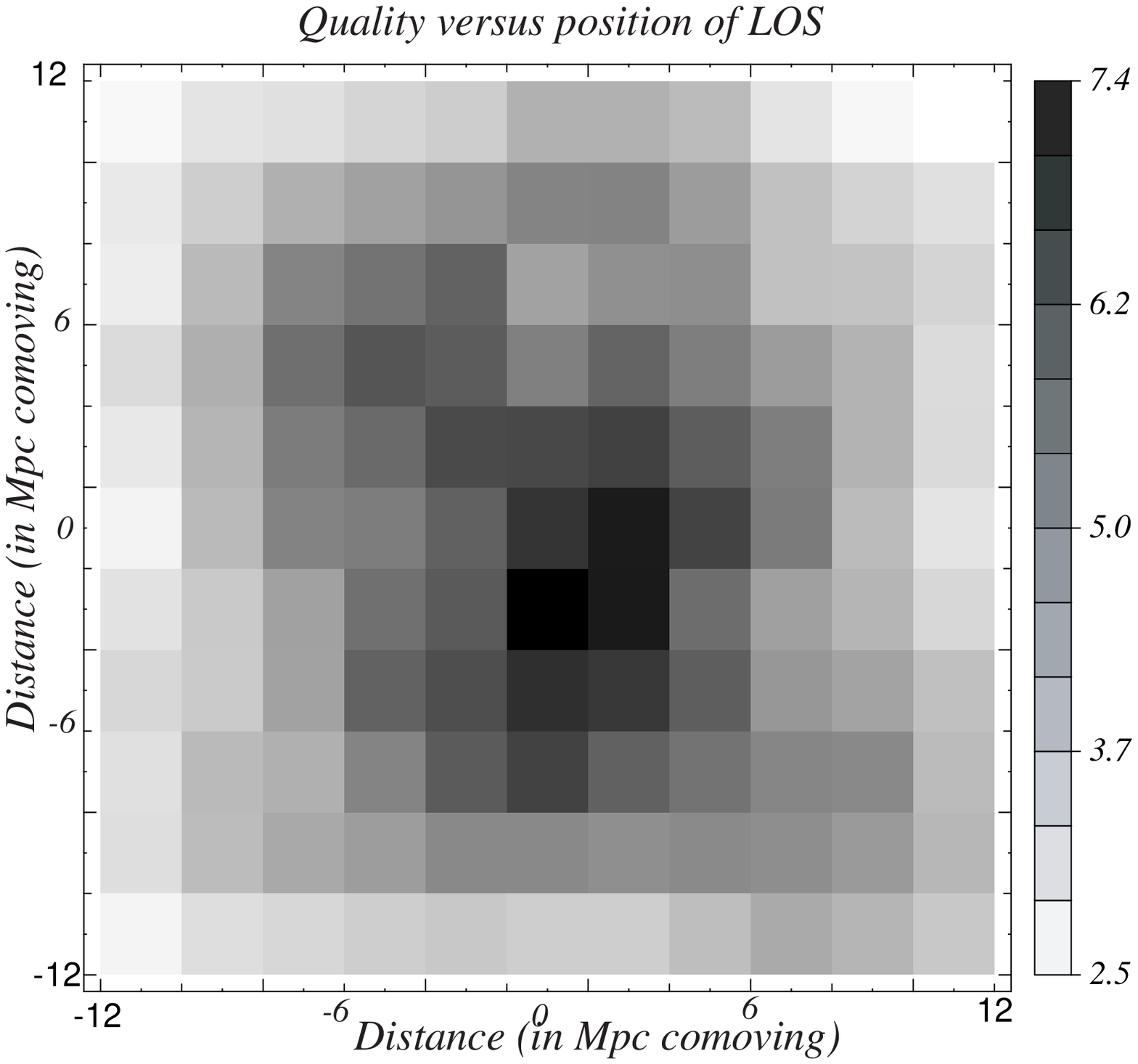,angle=0.0,width=7cm}}
\put(0,0){\psfig{file=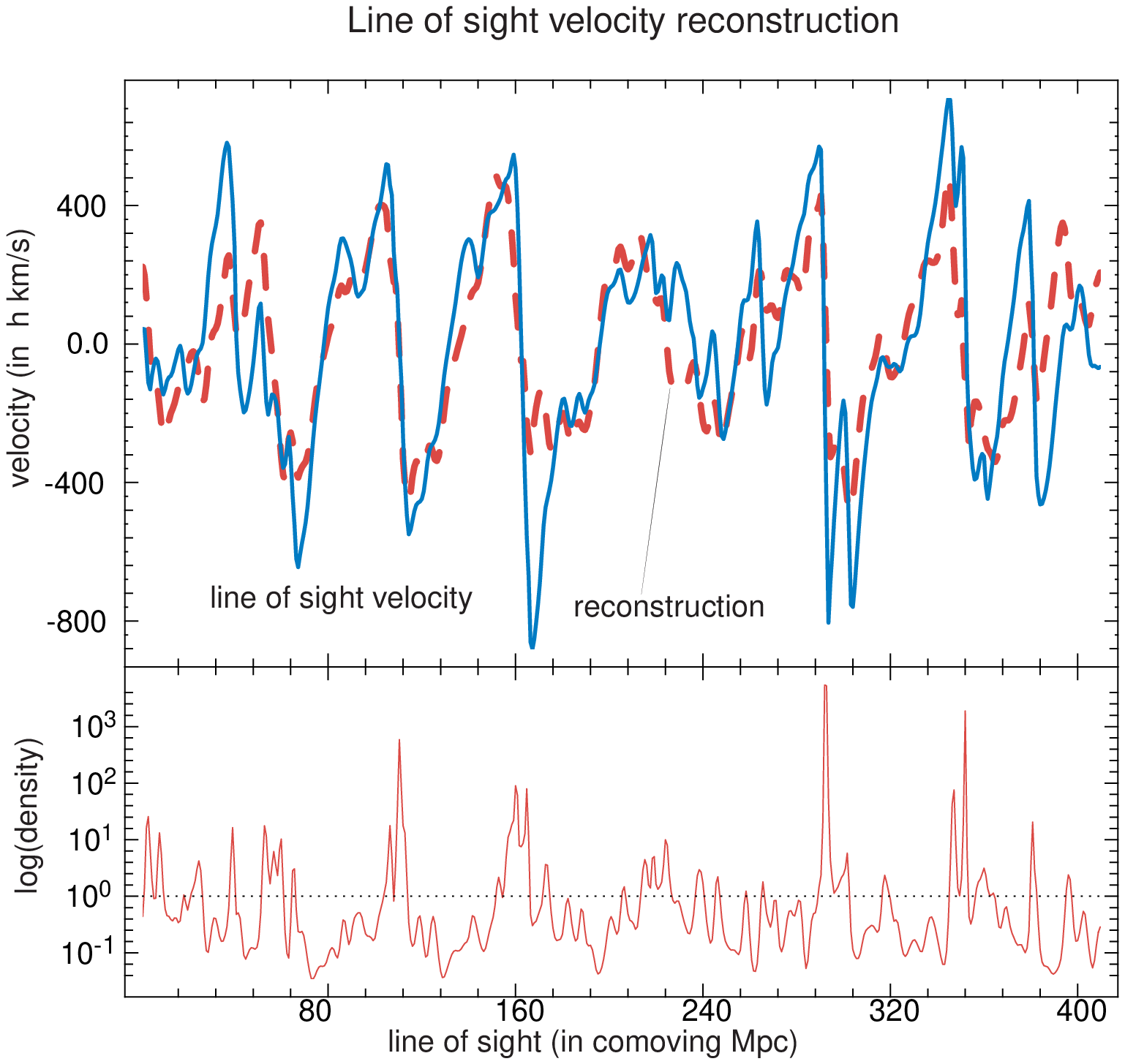,angle=0.0,width=7cm,height=6cm}}
\put(8.,0){\psfig{file=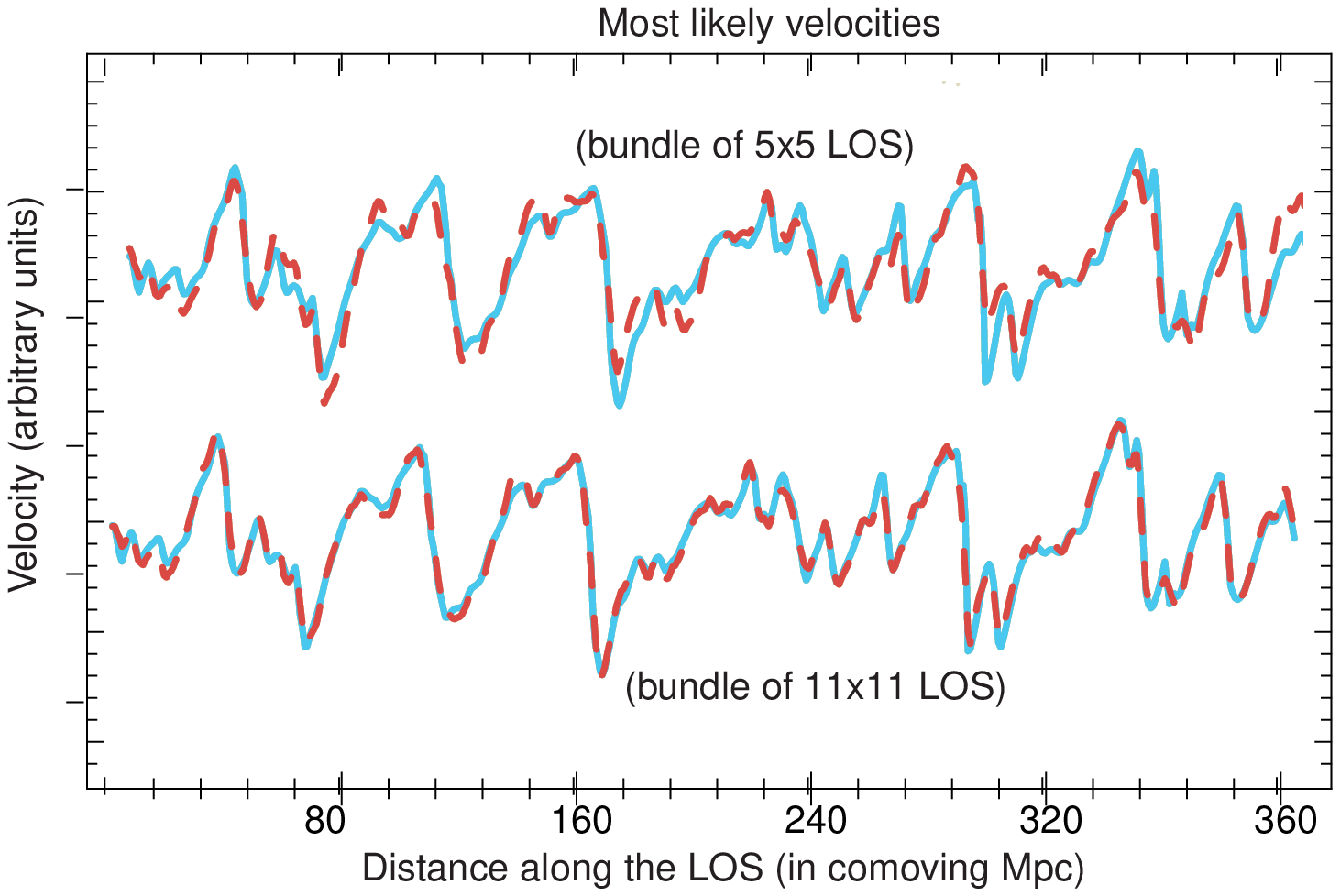,angle=0.0,width=8cm}}
\end{picture}
\caption{ {\em  Top left panel:}  quality of the  reconstruction [\Eq{defQ}]
versus \los \,  separation and number of \loss.   Increasing the sampling on
the  sky  decreases  the  dispersion  between the  constrained  most  likely
velocity  and the  measured velocity  as discussed  in the  text.   Note the
saturation  for 11$\times$11 \loss  \, at  a separation  of $\simeq$  5 Mpc.
{\em  Top right  panel:} isocontour  for the  quality of  the reconstruction
projected on  the sky  for a bundle  of $11  \times 11$ \loss,  separated by
$2.4$ Mpc comoving.  Note that the reconstruction obviously works better for
the central  \loss.  {\em Bottom left  panel:} in simulation  B, most likely
velocity  constrained  by  a single  \los.   The  solid  line on  the  upper
sub-panel  corresponds to  simulated  velocity  and the  dashed  one to  the
reconstructed  velocity.   The  simulated  density  is  displayed  on  lower
sub-panel.   {\em Bottom  right  panel:} solid  lines: simulated  velocities
along the  center of a  bundle of $5  \times 5$ \loss  \, or $11  \times 11$
\loss; dashed lines: corresponding recovered velocities.}
\label{f:figrecvel2}
\end{figure}

Let us define the
quality factor, $Q$, as
\begin{equation}
 Q \equiv
 \frac{
\sigma_{v_p}}{\sigma_{\delta v_p}}
= \sqrt{ \frac{\langle v_p^2\rangle }{\left\langle \left(v_p-v_{\rm rec}\right)^2
\right\rangle}}\, ,\EQN{defQ}
\end{equation}
where $v_{\rm  rec}$ is the reconstructed velocity.   Parameter $Q$ measures
the  inverse  residual  misfit in  units  of  the   variance  for  the
velocity.  We  show in  \Fig{figrecvel2} (top left  panel) that  this number
increases with the number of \loss\, sampling the sky, as expected. However,
$Q$ increases as  well with the distance between \loss\,  until it reaches a
maximum,  which might  sound confusing.   This can  be easily  understood by
examining  left panel  of \Fig{figrecvel}.  In fact,  a bundle  of  \loss \,
constrains  the transverse  3D velocity  distribution {\em  at intermediate
scales}, as  a result of  a competition between  short range and  long range
correlations:
\begin{enumerate}
\item[(i)] High frequency  structures are read from the  \los \, through the
two strong peaks along the x coordinate axis  on the left panel
of \Fig{figrecvel} (at  approximately $\pm 0.8$ Mpc).  Other  \loss\, can in
principle contribute to small scales, but  only if  they are found very close to the \los \,
of interest (i.e. with $\M{x}_{\perp} \simeq 0$).
\item[(ii)] Low  3D frequency  features are mainly  sampled by  \loss\, away
from  the  \los\, of  interest,  due to  the  significant  tails present  on
$\M{C}_{v\delta}$ at scales as large as $\sim$ 20 Mpc, as illustrated by top
right panel of \Fig{figrecvel}.  This effect  is three-dimensional,
i.e. in {\em all directions}: it thus provides information on the structures
transverse to the \los.
\end{enumerate}
(Note that in this discussion, we implicitly  assumed that $\M{C}_{\delta\delta} \simeq$ identity
in  \Eq{defvconv2}. Taking  into  account the  real  contribution of  matrix
$\M{C}_{\delta\delta}^{-1}$  would simply  boil down  to smoothing  the density
with an isotropic filter, which  has no effect on our 
qualitatives conclusions).

The competition  between effects  (i) and (ii)  fixes an  optimal separation
between the  \loss\, as a function of  their number. From the top  left panel of
\Fig{figrecvel2}, we see for example  that the optimal separation is $5$ Mpc
for a  bundle of 11$\times$  11 \loss. For  a bundle with  a smaller  number of
\loss, the optimal separation becomes  larger so that the  tails of  
$\M{C}_{v\delta}$  are  still fully sampled (but with  a  sparser binning  and thus  a
smaller quality factor).

The bottom right and left panels  of \Fig{figrecvel2} compare the velocity along
one \los \, measured in the  simulation to the reconstructed one by applying
\Eq{defvconv2}  to  bundles of  various  sizes  (1$\times$1, 5$\times$5  and
11$\times$11) distributed  uniformly on the sky (from  simulation~B), with a
mean separation of $2.5$ Mpc. With only one \los, the reconstructed velocity
does not   account in  detail for small  structures although it  seems to
match well large  scale flows in the example  studied here.  Increasing the
number of \loss \, significantly  improves the reconstruction: with a bundle
of  11$\times$11  \loss, the  reconstruction  almost  perfectly matches  the
simulation.

An  important  outcome of  this analysis  is  that  since the  optimal
separation between  \loss \, is  rather large (a  few Mpc), the  small scale
information  in the  reconstruction  is only  contained  in the  \los \,  of
interest. Therefore, having  high resolution spectra on all  the \loss \, is
not  required:  a survey  dedicated  to  real  space reconstruction  should
provide a  high resolution  spectrum together with  a set of  low resolution
spectra separated by distances smaller than or of the order of $\approx 4-5$ Mpc
comoving.  Note that $Q$ was computed while averaging over the whole bundle:
the quality  of the reconstruction  in fact depends  on the position  of the
\los  \,  in  the  bundle,  as   illustrated  in  the  top  right  panel  of
\Fig{figrecvel2}. Obviously, the quality factor  is optimal at the center of
the bundle: the high resolution spectrum should be located there.

%-------------------------------------------------------
%
We assumed here the 3D  covariance matrices needed for the reconstruction 
were known. In fact,  we used the best possible guess for  them since they 
were
derived  from direct measurement in  the simulation.  In reality
we would have  to proceed iteratively: for a given  power spectrum, we could
recover the 3D density, compute perturbatively the corresponding 3D velocity
field and derive a new  covariance matrix until convergence is achieved.  We
have  not demonstrated  here  that  this procedure  is  convergent. This  is
certainly a possible shortcoming of the procedure.

%%%%%%%%%%%%%%%%%%%%%%%%%%%%%%%%%%%
\subsubsection{Strong prior inversion}
%%%%%%%%%%%%%%%%%%%%%%%%%%%%%%%%%%%
\label{s:velSP}

Let  us now try  to deconvolve the  density in  real space along  {\em one}
\los.   A combination  of the  general Bayesian  method and  the constrained
random field  technique is implemented: the constrained  random field method
allows us to  relate the unknown field $v_p$ to  $\gamma$, imposing that the
peculiar velocity takes its most  likely value, but the recovery of $\gamma$
is still based on the Bayesian  method.  Our model, $ g_{i}(\gamma)$, is now
:
\begin{equation}
  g_{i}(\gamma)=A(\overline{z})c_1                   \int_{-\infty}^{+\infty}
(D_{0}(x)\exp[\gamma(x)])^{\alpha-\beta}\exp\left(-   c_2   \frac{(w_{i}-  x
-v_{p}(x)  )^2}{(D_{0}(x)\exp[\gamma(x)])^{2\beta}}  \right)  \,  \d  x  \,,
\EQN{defgvMT}
\end{equation}
with the supplementary assumption that the peculiar velocity in \Eq{defgvMT}
equals the most likely velocity (Appendix~\ref{a:CRF}):
\begin{equation}
v_{p}(x)= \langle v \rangle \equiv \int K^{(v)}(x,y) \gamma{(y)} \d  y \,, \quad \mbox{ where } \quad
  K^{(v)}(x,y)  \equiv \frac{1}{2\pi}  \int  e^{i k_x  (x-y)} \,  \frac{P_{v
  \delta,{\rm  1D}}(k_x)}{P_{\delta   \delta,{\rm  1D}}(k_x)}  \d   k_x  \,.
  \EQN{defvpMT1}
\end{equation}
The  unknown parameter  remains the  density  contrast.  The  prior for  the
density is chosen as $D_0 \equiv 1$ so that $\gamma=\delta$.  For the filter
$K^{(v)}(x,y)$  we use  a  simple analytic  fit  of  the function  $K^{(v)}(x,y)$
measured  in the  simulation  as explained  in Appendix~\ref{a:aponl}.   The
derivation of  the different vectors and  matrixes involved in  this case is
sketched in  Appendix~\ref{s:AvelSP}.  The practicalities involves   
fixing appropriately
the  parameters $(\sigma_{\gamma},\xi_x)$  in  \Eq{cov} ($\xi_T\equiv\infty$
for   a  single   \los)   for  the   minimization   procedure  detailed   in
Appendix~\ref{s:minim}   to    converge   while   providing    as   accurate
reconstruction as possible. To stabilize  the inversion, we need to take for
$\xi_x$  a value  close to  the correlation  length, $\xi_x=1$  Mpc.  With a
larger  value  of $\xi_x$,  the  inversion is  still  stable  but makes  the
reconstructed density  field too  smooth, while a  smaller value  of $\xi_x$
makes the inversion unstable.   The choice of $\sigma_{\gamma}$, which fixes
the amount of  variations allowed around the prior,  is more delicate.  
A small
value  of $\sigma_{\gamma}$  makes convergence  easier, but  does  not leave
enough  freedom for  the reconstructed  density to  float around  the prior:
voids tend  to be filled, and high  density peaks are not  saturated. On the
contrary,  a large value  for $\sigma_{\gamma}$  allows  significant deviations
from the prior but makes the  iteration procedure less stable. 
For this reason,
the reconstruction  is carried in two steps.   We first take a  small value for
$\sigma_{\gamma}=0.0175$,  and  reconstruct the  density
while  using   \Eq{defvpMT1}  to   determine  accurately  the   most  likely
velocity.  Because of  our  choice of  $\sigma_{\gamma}$, the  reconstructed
density  is not as  contrasted as  it should  be, but  this does  not affect
significantly  the corresponding  most  likely velocity:  it  just makes  it
smoother.  In the second step, we  {\em fix} the most likely velocity 
at the
value obtained from  the first step. Thus \Eq{defvpMT1}  is 
disregarded,
and  we  iterate   once  more  on  the  density  with   a  larger  value  of
$\sigma_{\gamma}$,  $\sigma_{\gamma}=0.2$, allowing  more variations  of the
density around the  new prior --the reconstructed density  obtained from the
first step.   The fact that the  most likely velocity is  fixed indeed makes
the inversion more stable and allows larger values of $\sigma_{\gamma}$.

\Fig{figrecvel2}  illustrates how  the  method performs  on two  unsaturated
\loss: the first isolated and the latter  nearby a cluster.  The simulated spectra assume
$A=0.39$,  $\beta=0.4$,  ${\bar  T}=10^4$   K,  and  were  calculated  after
smoothing the density and velocity fields with a cube of size $\sim 200$ kpc
(2 cells). The errors in the data are modelled as described in \S~\ref{s:0T}
with  $(S/N,\sigma_0)=(100,0.05)$   in  \Eq{deferror}.   As   expected,  the
reconstructed  velocity matchs the original  only  when  there  is  no  significant
structure  close  to the  \los,  likely  to  induce large-scale  infall
contamination. Bottom panels of \Fig{figrecvel2} show that the reconstructed
density reproduces well  the shape of most structures,  except that they are
not correctly located along the velocity axis on bottom right panel.

Note that  our two-step procedure is  similar in spirit to  that proposed by
Nusser \& Haehnelt (1999a), although we use same smoothing length $\xi_x$ in
both  steps, which  allows more  small scale  features on  the reconstructed
density.  Also, our  method is not yet able to  deal with spectra containing
significantly  saturated absorption lines:  in that  case, the  inversion is
much  less  stable and  the  reconstructed  most  likely velocity  is  often
unrealistic, even if  the \los\, is isolated. Finally,  we assumed the kernel
function  $K^{(v)}(x,y)$  was  known,  which should  not  be  the case  in
reality: a more  detailed study of the  effects of the assumed  shape for this
function will be needed in the future to fully qualify the method.

%%%%%%%%%%%%%%%%%%%%%%%%%%%%%%%%%%
\subsubsection{Floating prior for the velocities.}
\label{s:velFP}
%%%%%%%%%%%%%%%%%%%%%%%%%%%%%%%%%%

A less biased representation of the underlying field would be to assume that
$\gamma$ and $v_{p}$  are two fields which are  statistically correlated (by
the dynamics) but whose realizations are independent.  The model is formally
identical to  \Eq{defgvMT} with  the restriction that  $v_p$ does  {\em not}
obey  \Eq{defvpMT1}  anymore.  The  vector  of  the  model parameters  is  :
$\M{M}=\left( \gamma(x) , v_p(x) \right)$.  The correlation between $\gamma$
and  $v_p$, $\M{C}_{  v\gamma}$, is  considered to be  linear. Recall  that the
prior variance-covariance matrix,  $\M{C}_{0}$, has three independent terms,
shown in bottom right panel of \Fig{figrecvel}:
\begin{equation} 
   \M{C}_0=\left(\begin{array}{cc}  \M{C}_{\gamma  \gamma} &  \M{C}_{  v \gamma}  \\
\T{\M{C}_{v \gamma }} & \M{C}_{v v} \\ \end{array} \right) \,.  \EQN{defCgv}
\end{equation} 
The  penalty  function then  obeys  \Eq{penalty},  and  realizations of  the
velocity  field are  entitled to  float around  their most likely
values,  \Eq{defvconv2}.   The  corresponding  model, $g$,  is  sketched  in
Appendix~\ref{s:AvelWP}.  
%
% An illustration  is given  in \Fig{figweak}  in a
% somewhat unphysical r\'egime where  thermal broadening is effectively larger
% than redshift distortion. Note also  that the correlations functions used in
% this  inversion  were  {\em  measured}  from the  selected  \loss\,  in  the
% simulations (and truncated beyond  5 Mpc). 
%
The  iterative  procedure  presented  in Appendix~\ref{s:minim}  brings  the
reduced $\chi^2$ down from values of  about a $100$ to $1\pm \sqrt{2/N}$ in
a  few iterations,  but does  not  converge if  peculiar velocities  induce
displacements larger than the effective  width of the absorption lines. Even
though the weak prior inversion is more elegant and easier to implement than
the strong prior approach (cf.~Appendix~\ref{s:AvelWP}), it seems to fail to
constrain sufficiently our  model when  redshift distortion  is important.
This arises because the effective correlation in \Eq{penalty} is too weak to
induce convergence.

%%%%%%%%%%%%%%%%%%%%%%%%%%%%%%%%%%
\subsubsection{Discussion}
\label{s:totor3}
%%%%%%%%%%%%%%%%%%%%%%%%%%%%%%%%%%
A priori, the  best approach for reconstructing the  density in redshift
space would be to use the explicit Bayesian method with a floating prior for
the  velocity described in  \S~\ref{s:velFP}.  However,  our preliminary
analyses show that  this method fails to converge when  applied to {\em one}
\los  \,  if redshift  distortion  becomes  of the  order  of  the width  of
absorption lines,  which is unfortunately the case  in realistic situations.
The strong  prior inversion of  \S~\ref{s:velSP}, tested again on  one \los,
seems to  be more  reliable, but gives  accurate reconstruction only  if the
considered \los\, is unsaturated and  is isolated from large structures.   The only 
reliable way to improve the reconstruction  is therefore to have
more information  on the  3D structure of  the intergalactic  medium through
bundles of  \loss, as studied in  \S~\ref{s:NLOS}. 
The diference between \S~\ref{s:velFP} and \S~\ref{s:velSP} would then 
vanish, since the discrepency between the most likely velocity and the 
actual field becomes smaller and smaller, while  the correlation 
between the density and the velocity becomes  simultaneously tighter and tighter.
However, we have not explicitely tested
the methods of \S~\ref{s:velSP} and \S~\ref{s:velFP} on several 
\loss\ : this is  left  for   future  work.
%
%-------------------------------------------------------
\begin{figure}%[tbp]  

%%[tbp]
\centerline{\mbox{\psfig{file=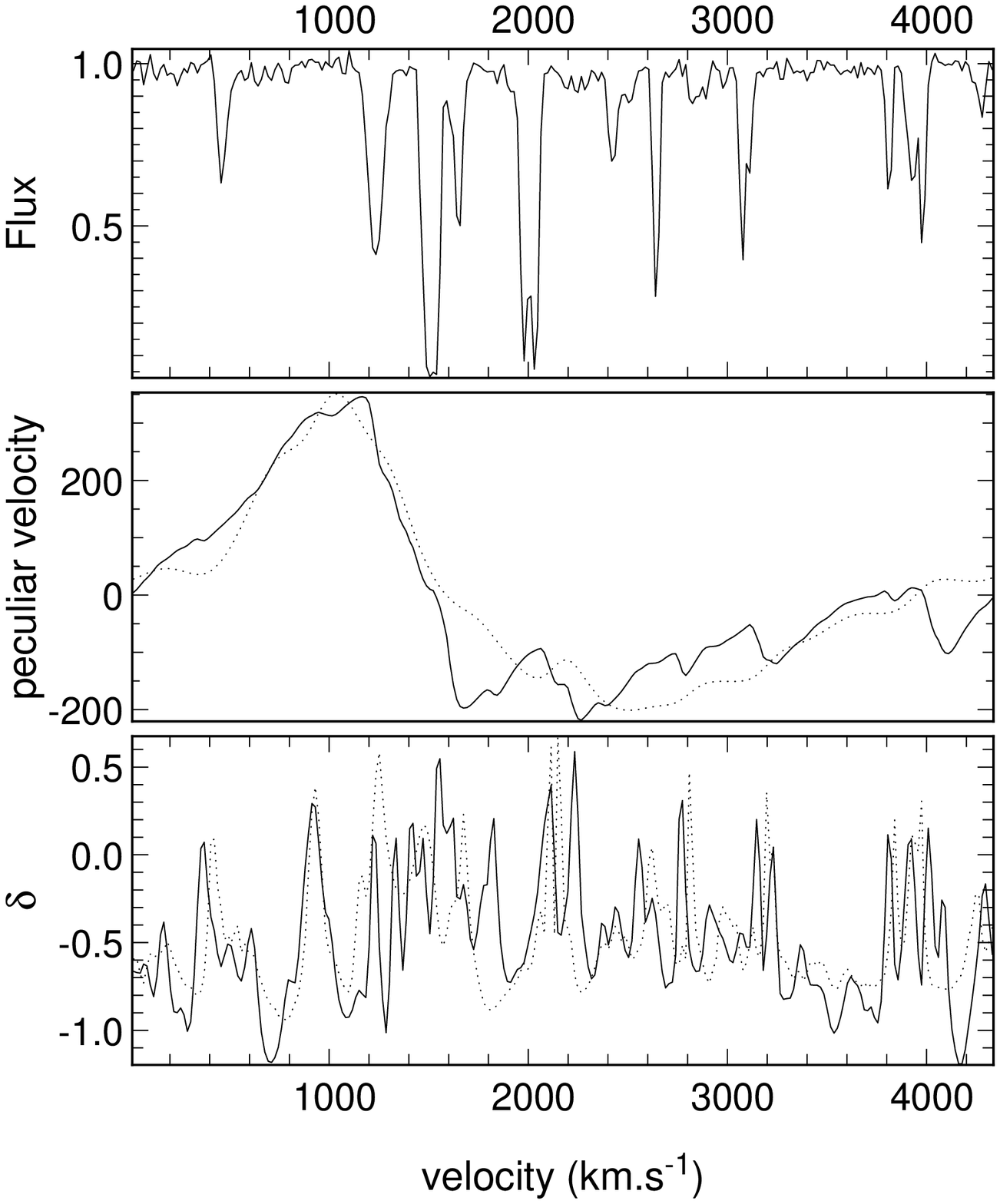,bbllx=55pt,bblly=200pt,bburx=530pt,bbury=739pt,width=8cm}\psfig{file=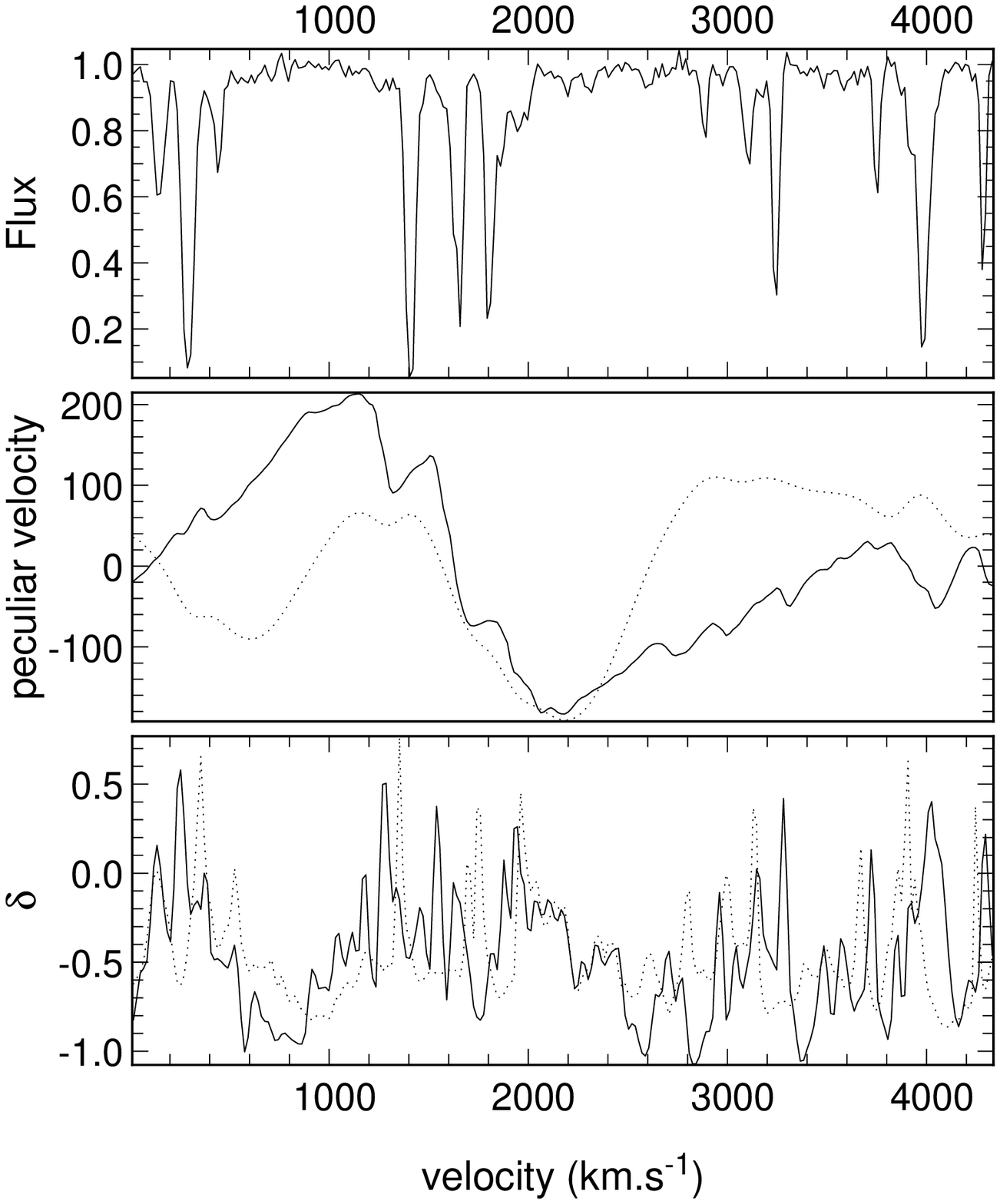,bbllx=55pt,bblly=200pt,bburx=530pt,bbury=739pt,width=8cm}}}
\caption{{\em Inversion  while accounting for peculiar  velocity with strong
prior.}  Simulation  S  is  used  to  test the  method.   Two  examples  are
considered,  according to  whether there  is  a large  structure nearby  the
\los\, or not  (respectively right and left panels).   {\em Top panels:} the
simulated spectra. {\em Middle panels:}  the simulated (solid line) and most
likely (dotted line) peculiar velocity  along the \los. {\em Bottom panels:}
the simulation (solid line)  and reconstructed (dotted line) log-density (in
$\log_{10}$ units).}
\label{f:figstrong}
\end{figure}
%-------------------------------------------------------
%

% %-------------------------------------------------------
% \begin{figure}%[tbp]
% %%[tbp]
% \unitlength=1cm
% \begin{picture}(15,10)
% \centerline{\psfig{file=weak.ps,width=15cm,height=10cm}}
% \end{picture}
% \caption{
% {\em  Inversion while accounting for peculiar  velocities with weak
% prior. Simulation S.} The upper panel shows 
% Solid line:  simulated density (lower panel) and most
% likely velocity (upper  panel). Dotted line: upper panel  shows the spectrum
% (computed with  a fixed equation  of state  and the  most likely  
% velocity, arbitrary units)  and  the  smoothed  reconstructed  density; 
% lower  panel  shows  the
% reconstructed  peculiar  velocity  ($v_{\rm  rec}$).  Dashed  line:  density
% inverted from the spectra and $v_{\rm rec}$ (see main text for details).}
% \label{f:figweak}
% \end{figure}
% %-------------------------------------------------------
% %

%%%%%%%%%%%%%%%%%%%%%%%%%%%%%%%%%%%
%%%%%%%%%%%%%%%%%%%%%%%%%%%%%%%%%%%
\section{Conclusion}
\label{s:conclusion}

In this paper, an explicit Bayesian technique and a constrained random field
method have been proposed to recover various properties of the intergalactic
medium  from observations  of  the Lyman-$\alpha$  forest  along \loss\,  to
quasars.  In particular, our   preliminary  analyses suggest  that these
methods may be  used (i) to recover  the large scale 3D topology  of the IGM
from inversion  of a  network of adjacent  \loss\, observed at  low spectral
resolution; (ii) to  constrain the physical characteristics of  the gas from
inversion of single  \loss\, observed at high spectral  resolution; (iii) to
investigate how the  number of and the distance between  \loss \, constrain the
projected  peculiar velocities;  (iv)  to correct  in part for redshift  distortions
induced by these velocities using either strong or weak priors.

Both approaches rely on prior assumptions on the covariance of the log-density
field and the cross-correlation between the log-density field and the peculiar
velocity field.  

These  methods are  used in  various  r\'egimes: as  extrapolation tools  to
recover the  3D structure of the  IGM; as non-linear  deconvolution tools to
correct  for  blending; as  non-parametric  field  extractors  and as  model
fitting routines to constrain the parameters of the equation of state.

We have  demonstrated (\S~\ref{s:overlap}) that  as far as  extrapolation is
concerned the  standard constrained random field  interpolation scheme could
be viewed  as a  specific linear sub-case  of the Bayesian  inversion scheme
presented in \S~\ref{s:taran}.  The  method presented in \S~\ref{s:taran} is
therefore  complementary  to, and  more  general  than standard  constrained
random field techniques: it can also cope with thermal broadening and finite
signal to  noise, in a  manner similar to  Wiener filtering, but  allows for
non-linear  models  and non-zero  mean  priors.   The correlation  functions
required for the  prior need not be measured in the  simulations, and can be
postulated.  It is more flexible since some level of redshift distortion can
in principle be corrected for using the full 3D information along the bundle
(although we did not demonstrate it explicitly in this paper).  It is well suited
for this  kind of problems since  it deals directly  with unknown continuous
fields  [i.e.   the  parameter   space  is  the   Hilbert  space   L2;  see,
e.g.~\Eq{excont}].   In  contrast   with   the  Lucy-Richardson   algorithm,
regularization is built in.

We have shown
that  temperature inversion  is degenerate  with respect  to  two parameters
describing the equation  of state of the gas,  the temperature scale factor,
$\bar T$ and the effective polytropic index $\beta$.  

Recall that  we have assumed in  this paper the correlation  matrices of the
log density to be fixed a priori, together with the cross-correlation of the
log density and the velocities  when dealing with peculiar velocities.  When
the  method  is  applied  to real  data,  we  will proceed
iteratively  and   recompute  these   (cross-)  correlations  once   the  3D
reconstruction is  achieved. We expect  this procedure to converge  and that
the convergence limit will not depend too strongly on the initial prior.

A thorough analysis of the  various biases involved in the methods presented
here is postponed to a  companion paper which will investigate statistically
the properties of the reconstructed  fields and the degeneracies involved in
recovering  the  density and  the  temperature  while  relying on  numerical
hydrodynamical simulations. Since this inversion method relies
on existing cross  correlation between the density and  the velocity fields,
it should still  be applicable on scales where dark  matter dynamics is less
relevant, so  long as such correlations  exist.  We have left  aside for now
the simultaneous  true 3D deconvolution  of both the temperature  and the
peculiar velocities.

\section*{acknowledgments}
We thank  F.~Bernardeau, E.~Thi\'ebaut and D.~Pogosyan  for many discussions
and an  introduction to  constrained random field  theory. JLV and  PPJ also
thank Bob Carswell for useful discussions.  JLV was supported in part by the
EC  TMR  network  ``Galaxy  Formation  and Evolution''  and  the  Centre  de
Donn\'ees  astronomique  de Strasbourg.   This  work  was  supported by  the
Programme National de Cosmologie.
%

%%%%%%%%%%%%%%%%%%%%%%%%%%%%%%%%%%%
%%%%%%%%%%%%%%%%%%%%%%%%%%%%%%%%%%%
%%\vskip 1cm
%%\par\noindent
%%{\bf REFERENCES}

\appendix

%%%%%%%%%%%%%%%%%%%%%%%%%%%%%%%%%%%
%%%%%%%%%%%%%%%%%%%%%%%%%%%%%%%%%%%
\section{Minimization procedure.}
\label{s:minim}

In this section we sketch an iterative procedure leading to the optimization
of  the  posterior  probability  of  the  model for  a  given  data  set  in
\Eq{fpost}.  The minimum of the argument of the exponential in \Eq{fpost} is
shown by a simple variational argument (Tarantola and Valette, 1982a; 1982b)
to obey the implicit equation:
\begin{equation}
\langle{\M{M}}\rangle=\M{M}_0 + \M{C}_0 \cdot \T{\M{G}} \cdot (\M{C}_d + \M{G}
\cdot \M{C}_0   \cdot   \T{\M{G}})^{-1} \cdot (\M{D}    +  \M{G} \cdot
(\langle{\M{M}}\rangle -\M{M}_0)-g(\langle{\M{M}}\rangle)) \,,
\EQN{eq1}
\end{equation}
with $\M{G}_{}$, the matrix of partial derivatives:
\begin{equation}
\M{G}_{}= \left(\frac{\partial g_{}}{\partial \M{M}_{}}\right) \, .
\end{equation}
 This minimum is found using an iterative procedure:
\begin{equation}
{\M{M}}_{[k+1]}=\M{M}_0 + \M{C}_0  \cdot \T{ \M{G}_{[k]} } \cdot
(\M{C}_d + \M{G}_{[k]} \cdot \M{C}_0 \cdot \T{\M{G}_{[k]}})^{-1} \cdot
(\M{D}       +     \M{G}_{[k]}          \cdot     ({\M{M}}_{[k]}
-\M{M}_0)-g({\M{M}}_{[k]})) \,,
\EQN{eq11}
\end{equation}
where subscript  $[k]$ refers  to the iteration  order.  In this  scheme the
minimum corresponds to ${\tilde \M{M}} =\M{M}_{[\infty]}$ and in practice is
found via a convergence criterion  on the relative changes between iteration
$[k]$  and $[k+1]$.   For  the  sake of  numerical  efficiency, rather  than
inverting $(\M{C}_d + \M{G}_{[k]} \cdot \M{C}_0 \cdot \T{ \M{G}_{[k]}})$, we
solve for the vector $\M{W}_{[k]}$ satisfying
\begin{equation}
 \M{S}_{[k]} \cdot \M{W}_{[k]} =
(\M{D}       +     \M{G}_{[k]}          \cdot     ({\M{M}}_{[k]}
-\M{M}_0)-g({\M{M}}_{[k]})),
\quad {\rm
where}
\quad
\M{S}_{[k]}=\M{C}_d + \M{G}_{[k]} \cdot \M{C}_0 \cdot \T{ \M{G}_{[k]}} \,,
\EQN{defS}
\end{equation}
and iterate:
\begin{equation}
{\M{M}}_{[k+1]}=\M{M}_0 + \M{C}_0  \cdot  \T{\M{G}_{[k]} } \cdot
\M{W}_{[k]} \,.
\EQN{eqit}
\end{equation}
From now on, we drop  the subscript $[k]$.  
Once  the maximum of \Eq{fpost} has been
reached,  an approximation  of  the  internal error  made  on the  parameter
estimation is  derived from a second  order development of  the posterior
distribution function in the vicinity of the solution :
\begin{equation}
\M{C}_{{\M{M}}}=\M{C}_0-\M{C}_0  \cdot \T{\M{G}}  \cdot  \M{S}^{-1} \cdot  \M{G}
 \cdot \M{C}_0 \,. \EQN{covpost}
\end{equation}
The high  spatial frequency  fluctuations are lost  in the  inverse process
because of limited resolution and finite signal to noise ratio.  The
{\em prior} correlation function therefore plays an important role to transform
an ill-posed problem into an invertible one.  How is the density information
degraded in the  spectra? This question can be  addressed via the resolving
kernel, $\M{R}$, introduced for the  first time by Backus and Gilbert (1970)
and which  gives the  spread  of the  density  estimation at  a given  position.
Suppose that  we know the  true model, $\M{M}_{\rm  true}$. The data  can be
written:  \( \M{D}=\M{g}(\M{M}_{\rm true})  \,  \).   Approximating locally
operator $g$ near its minimum as a linear operator, \Eq{eq1} yields:
\begin{equation}
\label{eqkernel}
\langle{\M{M}}\rangle-\M{M}_0=\M{C}_0 \cdot  \T{\M{G}} \cdot \M{S}^{-1}\cdot
\M{G}       \cdot       (\M{M}_{\rm true}-\M{M}_0)\equiv       \M{R}       \cdot
(\M{M}_{\rm true}-\M{M}_0)\,, \EQN{defR}
\end{equation}
which defines the resolving kernel $\M{R}(x,x')$  as a low band pass
filter.  
%%%%%%%%%%%%%%%%%%%%%%%%%%%%%%%%%%%
%%%%%%%%%%%%%%%%%%%%%%%%%%%%%%%%%%%
\section{Constraints random fields \& Multiple line of sights}
\label{a:CRF}

As a  though experiment,  let us  assume that we  know the  density contrast
$\delta$ on n points and ask what the corresponding most likely velocity (or
density)  at points labeled $k=1\cdots p$, $\varpi_k$ is.
We shall  not assume  that the densities  $\delta _1,\cdots ,\delta  _n$ are
necessarily along the same \los\, nor that the quantity $\varpi_k$ is
sought  along  any of  these.   Let  $ X=[  {\varpi_1,\cdots,\varpi_p,\delta
_1,\cdots ,\delta _n}] $. We define 
\begin{equation}
\M{C}\equiv \left[  {\matrix{  \langle  {\varpi_1}{\varpi_1}\rangle  &\cdots  &\langle
	{\varpi_1}{\varpi_p}\rangle  &\langle {\varpi_1}{{\delta }_1}\rangle
	&\cdots &\langle  {\varpi_1}{{\delta }_n}\rangle \cr  \vdots &\ddots
	&\vdots      &\vdots      &\ddots      &\vdots      \cr      \langle
	{\varpi_1}{\varpi_p}\rangle             &\cdots             &\langle
	{\varpi_p}{\varpi_p}\rangle  &\langle {\varpi_p}{{\delta }_1}\rangle
	&\cdots   &\langle   {\varpi_p}{{\delta   }_n}\rangle  \cr   \langle
	{\varpi_1}{{\delta  }_1}\rangle &\cdots  &\langle {\varpi_p}{{\delta
	}_1}\rangle  &\langle   {{\delta  }_1}{{\delta  }_1}\rangle  &\cdots
	&\langle  {{\delta  }_1}{{\delta   }_n}\rangle  \cr  \vdots  &\ddots
	&\vdots  &\vdots  &\ddots  &\vdots  \cr  \langle  {\varpi_1}{{\delta
	}_n}\rangle &\cdots &\langle {\varpi_p}{{\delta }_n}\rangle &\langle
	{{\delta   }_1}{{\delta   }_n}\rangle   &\cdots  &\langle   {{\delta
	}_n}{{\delta }_n}\rangle }} \right] \equiv {\left[
\begin{array}{cc}
  \M{C}_{w  w}  &  \M{C}_{  w  \delta}  \\  {\M{C}}^{\perp}_{  w  \delta}  &
    \M{C}_{\delta \delta}
\end{array}
 \right]} \, ,
\end{equation}  
so that  $ \M{C}_{w w}$  is the $p\times  p$ autocorrelation matrix  of the
sought field,  $\M{C}_{\delta \delta} $  is the $n\times  n$ autocorrelation
matrix of the known density field,  and $\M{C}_{ w \delta} $ is the $p\times
n$ cross-correlation matrix  of the sought field with  the density field. The
joint probability  of achieving velocity  $\varpi_k$ and density  profile $\delta
_1,\cdots ,\delta _n$ is given by
\[ 
p(\M{X}) \d{}^{n+p}  X = p(\varpi_1,\cdots, \varpi_p  , \delta _1,\cdots  ,\delta _n )
\d{\varpi_1} \cdots  \d{\varpi_p} \d  \delta_1 \cdots \d  \delta_n = \exp  \left[ {-{1
\over   2}\left\{   {\sum\limits_{a,b=   1\cdots   n+p}   {\left(   {C^{-1}}
\right)_{a,b}X_a  X_b}} \right\}} \right]  \,\frac{ \d{}^{n+p}  X}{\sqrt{ (2
\pi)^{n+p} \rm{det}{|C|}}} \,.
\]
The argument of the exponential can be rearranged as
\begin{equation}
{\T{\left( \MG{\varpi},\M{\delta} \right)}}\cdot{{\left[
\begin{array}{cc}
 \M{C}_{w  w}   &  \M{C}_{w \delta  }   \\  {\M{C}}^{\perp}_{  w   \delta}  &
    \M{C}_{\delta \delta}
\end{array}
 \right]}^{-1}}\cdot\left(\MG{\varpi},\M{\delta} \right)= {\T{\left(\MG{\varpi}-\M{C}_{w
\delta  }  \cdot{\M{C}_{\delta   \delta}^{-1}}  \cdot  \M{\delta}  \right)}}
\cdot{{\left(\M{C}_{w        w}-{\M{C}_{w        \delta}}\cdot{\M{C}_{\delta
\delta}^{-1}}\cdot\T{\M{C}_{w \delta }}\right)^{-1}}\cdot
\left(\MG{\varpi}-\M{C}_{w               \delta               }\cdot{\M{C}_{\delta
\delta}^{-1}}\cdot\M{\delta} \right)}+\quad {\rm rest} \EQN{rest}
\end{equation}
where ``rest'' stands for terms independent of $\MG{\varpi}\equiv(\varpi_1\,
\cdots \varpi_p)$.   Applying Bayes' theorem, the  conditional probability of
$\MG{\varpi}$, given  the density  profile $(\delta _1,\cdots  ,\delta _n)$,
obeys
\[ p(\varpi_1,\cdots, \varpi_p | \delta _1,\cdots ,\delta _n ) 
 \d{\varpi_1} \cdots \d{\varpi_p} = p(\varpi_1,\cdots  \varpi_p , \delta_1, \cdots ,\delta _n )/
p( \delta _1,\cdots ,\delta _n ) \d{\varpi_1} \cdots \d{\varpi_p} \, ,
\]
which in turns implies that 
\[  p(\varpi_1,\cdots, \varpi_p | \delta _1,\cdots ,\delta _n ) \propto
\exp   \left[  {-{1  \over   2}\left\{  {\T{\left(\MG{\varpi}-\M{C}_{w   \delta  }
\cdot{\M{C}_{\delta     \delta}^{-1}}     \cdot     \M{\delta}     \right)}}
\cdot{{\left(\M{C}_{w      w}-{\M{C}_{w     \delta     }}\cdot{\M{C}_{\delta
\delta}^{-1}}\cdot\T{\M{C}_{w \delta                      }}\right)^{-1}}\cdot
\left(\MG{\varpi}-\M{C}_{w               \delta               }\cdot{\M{C}_{\delta
\delta}^{-1}}\cdot\M{\delta} \right)} \right\}} \right] \,,
\]
since $p(  \delta _1,\cdots ,\delta _n  ) $ is  independent of $\MG{\varpi}$.
The  maximum  of  the  conditional  probability  is  therefore  reached  for
$\MG{{\left\langle {\varpi} \right\rangle}}$ given by 
\begin{equation}
 \MG{{\left\langle   {\varpi}   \right\rangle}}    =   \M{C}_{w   \delta   }
 \cdot{\M{C}_{\delta \delta}^{-1}} \cdot \M{\delta} \,. \EQN{sol}
\end{equation}

%%%%%%%%%%%%%%%%%%%%%%%%%%%%%%%%%%%%%%%%%%%%%%
\subsection{Peculiar velocity-density relation.}

Let us now be more specific  about $\varpi_k$ and assume, in this subsection,
that we are seeking the most likely peculiar velocity field, $v_{k}$, {where
we dropped the subscript $p$ referring to ``peculiar''}.

%%%%%%%%%%%%%%%%%%%%%%%%%%%%%%%%%%%
\subsubsection{One line of sight}
\label{a:aponl}
Recall  that  nothing has  been  said about  the  relative  position of  the
$\delta_{i}$ and the  $v_{k}$ at this stage.  Let us now  assume for a while
that the subscript $i$ refers to a regular ordering along  the \los,
so that $\delta_{i}=  \delta( i \Delta x)  $, and $v_{i}= v( i  \Delta x) $.
Let     us     also    introduce     the     intermediate    field,     $\M{
u}=\left({u_i}\right)_{i=1\cdots  n}   \equiv  {\M{C}_{\delta  \delta}^{-1}}
\cdot \MG{\delta} $, so that \Eq{sol} reads
\begin{equation}
 \M{{\left\langle {v} \right\rangle}} = \M{C}_{v \delta }\cdot \M{u} \,
, \quad \MG{\delta}=\quad \M{C}_{\delta \delta} \cdot \M{u} \,. \EQN{sol2}
\end{equation}
Multiplying both sides of \Eq{sol2} by $\Delta x $, we get
\begin{eqnarray}
  \sum_{j} \left(C_{v \delta   }\right)_{i,j}{ {u_j}   } \Delta x   &=&
  \sum_{j}{ {u[    j \Delta x]    } } {\left\langle   {v[j \Delta
  x] \delta[ (i-j)  \Delta x] } \right\rangle} \Delta x =
  \left\langle  {v[ i \Delta  x]   } \right\rangle  \Delta x \,, \cr
   \sum_{j}
  \left(C_{\delta   \delta}\right)_{i,j}   u_{j}  \Delta   x  &=&
  \sum_{j}{  {u[  j     \Delta x] }   }
  {\left\langle { \delta [j  \Delta x]\delta[ (i-j) \Delta  x ] }
  \right\rangle} \Delta x = {\delta[ i \Delta x] } \Delta x \,. \label{e:sol1}
\end{eqnarray}
In the limit of $\Delta x$ going to zero, \Eq{sol1} reads
\begin{equation}
 \int  {\left\langle {\delta(x-x') v(x')}
\right\rangle} {u(x')} \, \d x' =  {\left\langle
{v(x)}  \right\rangle} \Delta x  \, \quad {\rm and} \quad 
\int {\left\langle {\delta(x-x') \delta(x')} \right\rangle}  u(x')  \, \d  x' = 
{
{\delta(x)}  } \Delta x \, . \EQN{solcont}
\end{equation}
Transforming  \Eq{solcont} in Fourier space leads to
\begin{equation}
{\left\langle   {\tilde   v}   \right\rangle}(k_x)  =\frac{P_{v  \delta,{\rm
1D}}(k_x)}{P_{\delta \delta,{\rm  1D}}(k_x)}  \, {\tilde  \delta}  (k_x) \,,
\EQN{solfour}
\end{equation}
where ${P_{\delta \delta,{\rm 1D}}(k_x)}$ and ${P_{v \delta,{\rm 1D}}(k_x)}$
are respectively the {\rm 1D} density power spectrum  and the {\rm 1D} mixed
velocity density  power  spectrum,  while    ${\tilde \delta} (k_x)$     and
${\left\langle {\tilde v} \right\rangle}(k_x) $ are the Fourier transform of
${   \delta}   (x)$  and    ${\left\langle    {  v}   \right\rangle}(x)    $
respectively. Here the {\rm 1D} power spectra satisfy
\begin{equation}
{P_{\delta \delta,{\rm 1D}}(k_x)}= \int P_{{\rm 3D}}(\M{k}) W_{\rm J}(\M{k})
\, \d{}^2  k_\perp  \quad  {\rm  and} \quad   {P_{v   \delta,{\rm
1D}}(k_x)}=  \int \frac{P_{{\rm    3D}}(\M{k})   k_x}{k_x^2+k_\perp^2}
W_{\rm J}(\M{k}) \, \d{}^2 k_\perp \, ,\EQN{p1d}
\end{equation}
where  $P_{{\rm  3D}}(\M{k})$  is  the  {\rm  3D}  power  spectrum  of
the density contrast while  $W_{\rm J}(\M{k})$ is a window function whose
characteristic scale $R_{\rm J}$ should be the Jeans length, but is chosen here to
be the maximum of the Jeans length and the sampling scale. Indeed, below
this latter scale  no
information is to be derived from  the data.  Note that the direct inversion
of  \Eq{sol} may  lead to  significant aliasing  if the  power  spectrum has
energy beyond the cutoff frequency  $1/R_{\rm J}$.  The power spectrum ratio
in \Eq{solfour}  is an  antisymmetric filter which  relates the  most likely
velocity field to a given density field in linear theory.

Equation\Ep{solfour} can be transformed back into real space as
\begin{equation}
\langle v \rangle(x)= \int K^{(v)}(x,x') \delta(x') \d x'\,, \EQN{defvconv}
\quad \mbox{ where } \quad  
K^{(v)}(x,x')  \equiv \frac{1}{2\pi} \int   e^{i  k_x (x-x')} \,  \frac{P_{v
\delta,{\rm      1D}}(k_x)}{P_{\delta     \delta,{\rm   1D}}(k_x)}    \d k_x
\,. \EQN{defK0}
\end{equation}
This filter is illustrated  in \Fig{figrecvel}. Equation\Ep{defK0} could be
used  to derive  $K^{(v)}(x,x') $  from  perturbation theory  in the  weakly
non-linear  r\'egime given  an initial  power spectrum.   In  practice, this
filter is constructed here from  the simulation in the following manner: for
each \los \, in the simulation, we compute the FFT of the over-density and of
the velocity;  we multiply  one by  the complex conjugate  of the  other and
repeat the operation on the whole box; we then average over the box (using a
bundle of  $60\times 60$ \loss) and  FFT transform back in  real space: this
yields \Eq{defK0}.

%%%%%%%%%%%%%%%%%%%%%%%%%%%%%%%%%%%
\subsubsection{Multiple lines of sight}
Let us now turn to the more general problem of multiple \loss.  How
can we  take advantage  of larger scale  information on multiple  \loss\, to
constrain  the velocity {\em  along} the  measured \loss  ?  

To conduct the calculation which follows, we
order the $\delta_1,\cdots ,\delta_n$ where $n=L \, p$ so that the first $p$
corresponds to the first line  of sight,  the next $p$  to the second  line of
sight and so  on for the $\ell=1\cdots L$ line of  sights.  Our purpose here
is to account for the fact that in realistic situations, the \loss\, distribution 
on the sky is not necessarily uniform and that the 
volume covered by all \loss\, is rather elongated (i.e.~$L \ll p$).  For the sake
of  numerical  efficiency,  we  Fourier  transform  along  the  longitudinal
direction and are  left with a matrix representation  for the two transverse
dimensions.  We write each block  in Fourier space in terms of the corresponding {\rm
1D} power spectra (this is  possible since both Fourier transform and matrix
multiplication are  linear operations which therefore  commute when applied
on different directions); following the derivation of \Eq{solfour} we find
\begin{equation}
\M{{\left\langle { \tilde  v}  \right\rangle}} = \M{\tilde  \Xi} \cdot
   \M{\tilde \Delta}^{-1} \cdot \MG{\tilde \delta} \,, \EQN{solfour2}
\end{equation}
where
\begin{equation}
\M{\tilde    \Delta}    \equiv  \left[   {\matrix{{P_{_{\delta  \delta
}}^{11}(k_x)}&\cdots  &{P_{_{\delta  \delta   }}^{1L}(k_x)}\cr  \vdots
&\ddots  &\vdots   \cr {P_{_{\delta    \delta     }}^{1L}(k_x)}&\cdots
&{P_{_{\delta \delta }}^{LL}(k_x)}\cr  }} \right] \, ,\,\,\, \M{\tilde
\Xi}  \equiv  \left[    {\matrix{{P_{_{v \delta   }}^{11}(k_x)}&\cdots
&{P_{_{v \delta }}^{1L}(k_x)}\cr  \vdots  &\ddots &\vdots \cr  {P_{_{v
\delta    }}^{1L}(k_x)}&\cdots   &{P_{_{v \delta   }}^{LL}(k_x)}\cr }}
\right] \, , \EQN{ptilde}
\end{equation}
and  $  \M{{\left\langle  { \tilde v}    \right\rangle}}  = \left[  { \tilde
v}^1(k_x),  \cdots  {\tilde  v}^L(k_x) \right], $   $  \M{ \tilde  \delta} =
\left[{\tilde \delta}^1(k_x), \cdots {\tilde \delta}^L(k_x) \right], $ where
the superscript refers to the $L$ \loss.  Here
\begin{equation}
P_{_{\delta       \delta         }}^{\ell{\ell'}}(k_x)=\int     {\exp \left(
{i\,{\M{k}_\perp}\cdot\left\{       {\M{x}_{\perp,\ell}-\M{x}_{\perp,\ell'}}
\right\}}   \right)P_{{\rm 3D}}(\M{k})\,W_{{\rm J},{\bar R}}(\M{k})}\,\d{}^2
\M{k}_\perp\,, \EQN{pijk}
\end{equation}
\begin{equation}
P_{_{v         \delta     }}^{\ell{\ell'}}(k_x)=\int     {\exp        \left(
{i\,{\M{k}_\perp}\cdot\left\{       {\M{x}_{\perp,\ell}-\M{x}_{\perp,\ell'}}
\right\}} \right)\,W_{\rm  J,{\bar R}}(\M{k})}\frac{P_{{\rm  3D}}(\M{k}) k_x
}{k_x^2+\M{k}_\perp^2}\,\d{}^2 \M{k}_\perp\,. \EQN{pijk2}
\end{equation}
The window function,   $W_{\rm J,{\bar   R}}(k_{x},\M{k}_{\perp})$
involves two scales: the longitudinal  Jeans length and the transverse
mean   inter-\los\, separation, $\bar   R$.   The latter  filtering is
required to apodise the  inversion.    Note that $P_{_{\delta   \delta
}}^{\ell\ell}(k_x)   =   {P_{\delta  \delta,{\rm   1D}}(k_x)}   $  and
$P_{_{\delta v }}^{\ell\ell}(k_x) = {P_{\delta v,{\rm 1D}}(k_x)} $ are
given by \Eq{p1d}.  Equation\Ep{solfour2} reads back into real space:
\begin{equation}
v_{\ell'} (x)= \sum_\ell\int K_{\ell'\ell} (x,x') \delta_\ell (x') \d x' \,,
\quad \mbox{where} \quad K_{\ell'\ell}(x,x') \equiv \frac{1}{2\pi} \int e^{i
k_x   (x-x')}   \,    \left(\M{\tilde   \Xi}   \cdot\M{\tilde   \Delta}^{-1}
\right)_{\ell' \ell} \d k_x \,, \EQN{defKv}
\end{equation}
where the matrix $ \M{\tilde  \Xi} \cdot\M{\tilde \Delta}^{-1} $ is given in
\Eq{ptilde}.  In  practice, this  filter is also  constructed here  from the
simulation  following the prescription  sketched above:  for each  bundle of
\loss \, in the simulation, we compute the FFT of the log density and of the
velocity; we multiply  one bundle by the complex conjugate  of the other and
repeat the operation on the whole box; we then average over the box (using a
bundle of  $20\times 20$  \loss): this yields  the matrix\Ep{ptilde}.  The
matrix multiplication in \Eq{defKv} is carried Fourier mode by Fourier mode,
while  the inverse  Fourier transform  is done  by FFT.   

%%%%%%%%%%%%%%%%%%%%%%%%%%%%%%%%%%%%%%%%%%%%%%
\subsection{3D density-\loss\, density relation.}
Let us now assume that $\varpi_k$ refers to the 3D density on a grid of $P$
points                   at                 the                  point
$\M{x}_\lambda=(\M{x}_{\perp,\lambda},x_\lambda)_{\lambda     =1\cdots
P}$.  No   restriction on the  location  of $\M{x}_\lambda$  along the
\loss\, applies here. Under these assumptions, the above section 
translate as:
\begin{equation}
{\left\langle    {\delta^{\rm    ({\rm 3D})}}    \right\rangle}(\M{x}_\lambda)    =
  \sum_{\ell}\int   K^{\rm (3D)}_{\lambda   \ell}  (\M{x}_{\lambda},\M{x'}_\ell)
  \delta_{\ell} (x')  \d x' \,,  \,\, \mbox{ where }  \,\, K^{\rm (3D)}_{\lambda
  \ell}(\M{x}_\lambda,\M{x}_\ell')   \equiv   \frac{1}{(2\pi)^3}   \int   \,
  \exp\left({i   \M{k}   \cdot  (\M{x}_\lambda-\M{x}_\ell')}\right)   \left(
  \M{\tilde  \Xi_{\rm  3D}}  \cdot \M{\tilde  \Delta}^{-1}  \right)_{\lambda
  \ell} \, \d{}^3 \M{k} \EQN{defK}
\end{equation}
with $\M{\tilde \Delta}$ obeying \Eq{ptilde} and
\begin{equation}
 \M{\tilde  \Xi_{\rm 3D}}=\left[ {\matrix{{P_{_{\rm  3D }}^{11}(k_x)}&\cdots
&{P_{_{\rm  3D  }}^{L1}(k_x)}\cr \vdots  &\ddots  &\vdots  \cr {P_{_{\rm  3D
}}^{1P}(k_x)}&\cdots &{P_{_{ \rm 3D }}^{LP}(k_x)}\cr }} \right] \, , \quad %
\mbox{given}
\quad    P_{_{\rm    3D    }}^{\ell    \lambda}(k_x)=\int    {\exp    \left(
{i\,{\M{k}_\perp}\cdot\left\{      {\M{x}_{\perp,\ell}-\M{x}_{\perp,\lambda}}
\right\}}  \right)P_{{\rm 3D}}(\M{k})\,W_{{\rm  J},{\bar R}}(\M{k})}\,\d{}^2
\M{k}_\perp\,. \EQN{pijk3}
\end{equation}
We   check   that   when   we    consider   a   point   on   the   \loss   ,
$\M{x}=(\M{x}_{\perp,\ell},x)$,    $K^{\rm (3D)}_{\lambda    \ell}(\M{x},x')   =
\delta_{\rm D}(x-x') {\tilde\delta}_{\ell}^{\lambda}$ where
${\tilde\delta}_{\ell}^{\lambda}$ stands for the Kronecker symbol.

%%%%%%%%%%%%%%%%%%%%%%%%%
\section{Properties of the simulation}
%%%%%%%%%%%%%%%%%%%%%%%%%
\label{a:simul}

Note from Table~\ref{table:carasim} that the simulation boxes are rectangular.
This long box technique might be questionable. Indeed, 
the number of modes available in Fourier space is
different along each coordinate axis. This anisotropic mode sampling
contaminates the simulation, and the effect augments with the ratio
between the largest and the smallest side of the box. 

\begin{figure}%[hbt]
\centerline{\hbox{\psfig{figure=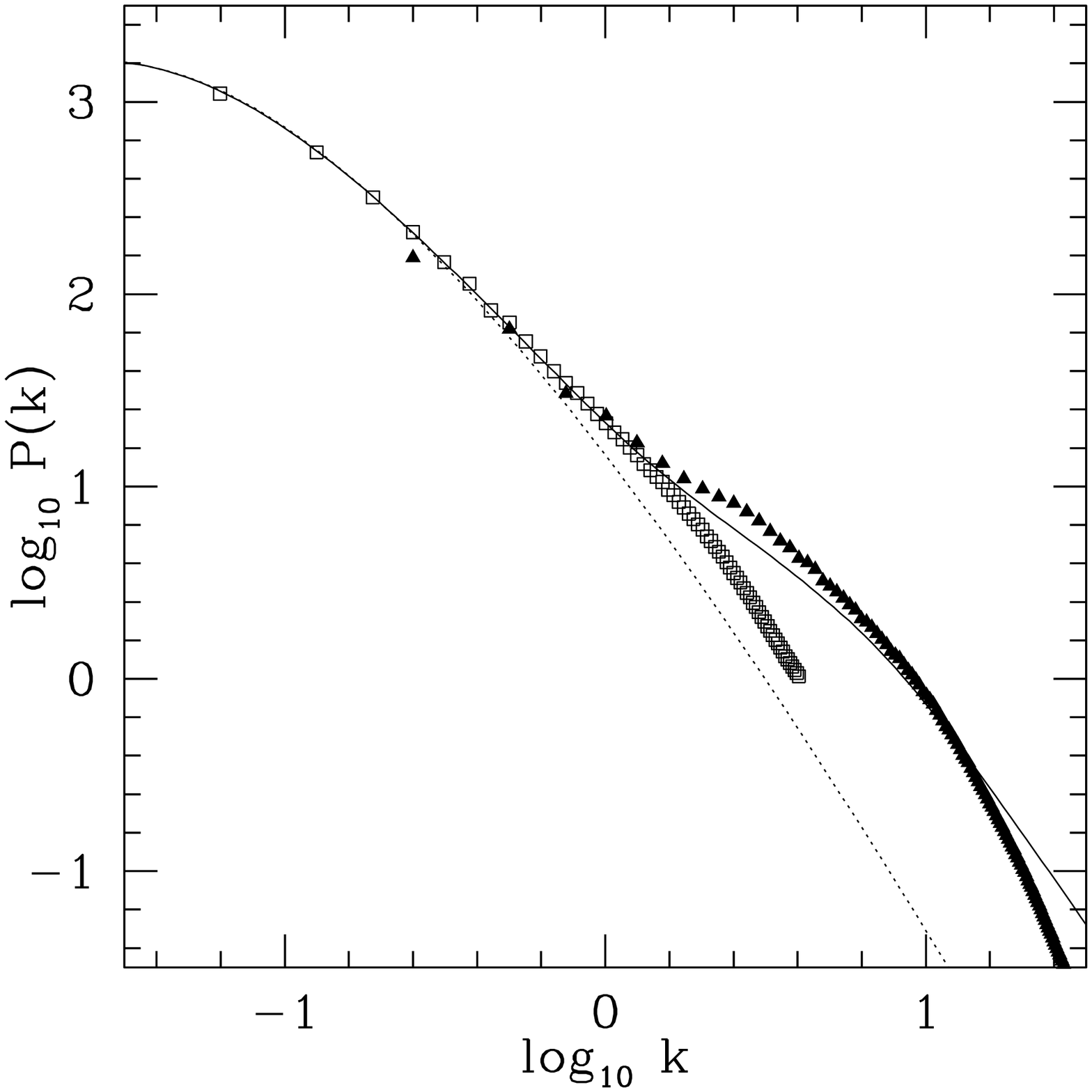,bbllx=18pt,bblly=145pt,bburx=572pt,bbury=700pt,width=5.5cm}\psfig{figure=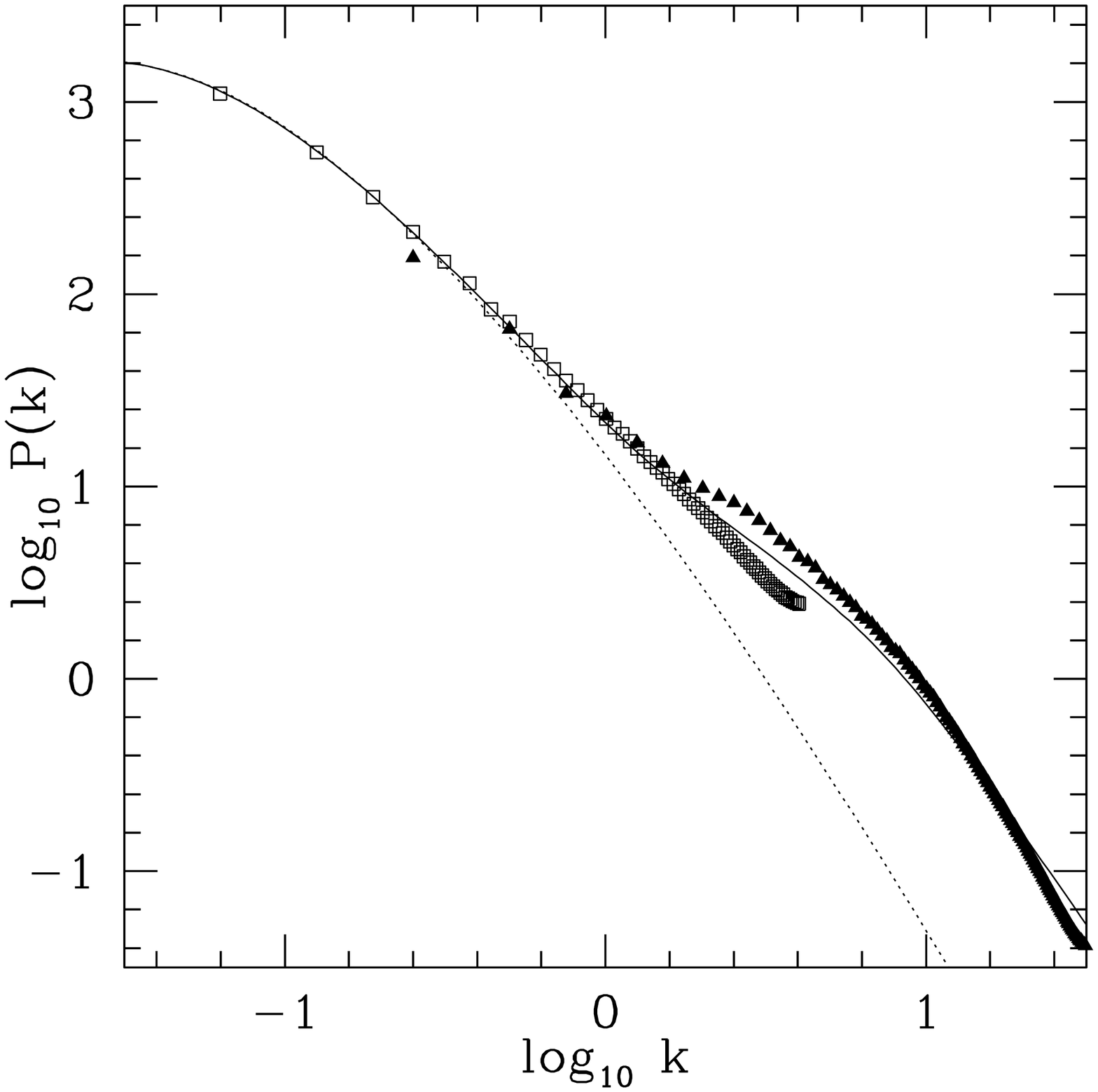,bbllx=18pt,bblly=145pt,bburx=572pt,bbury=700pt,width=5.5cm}\psfig{figure=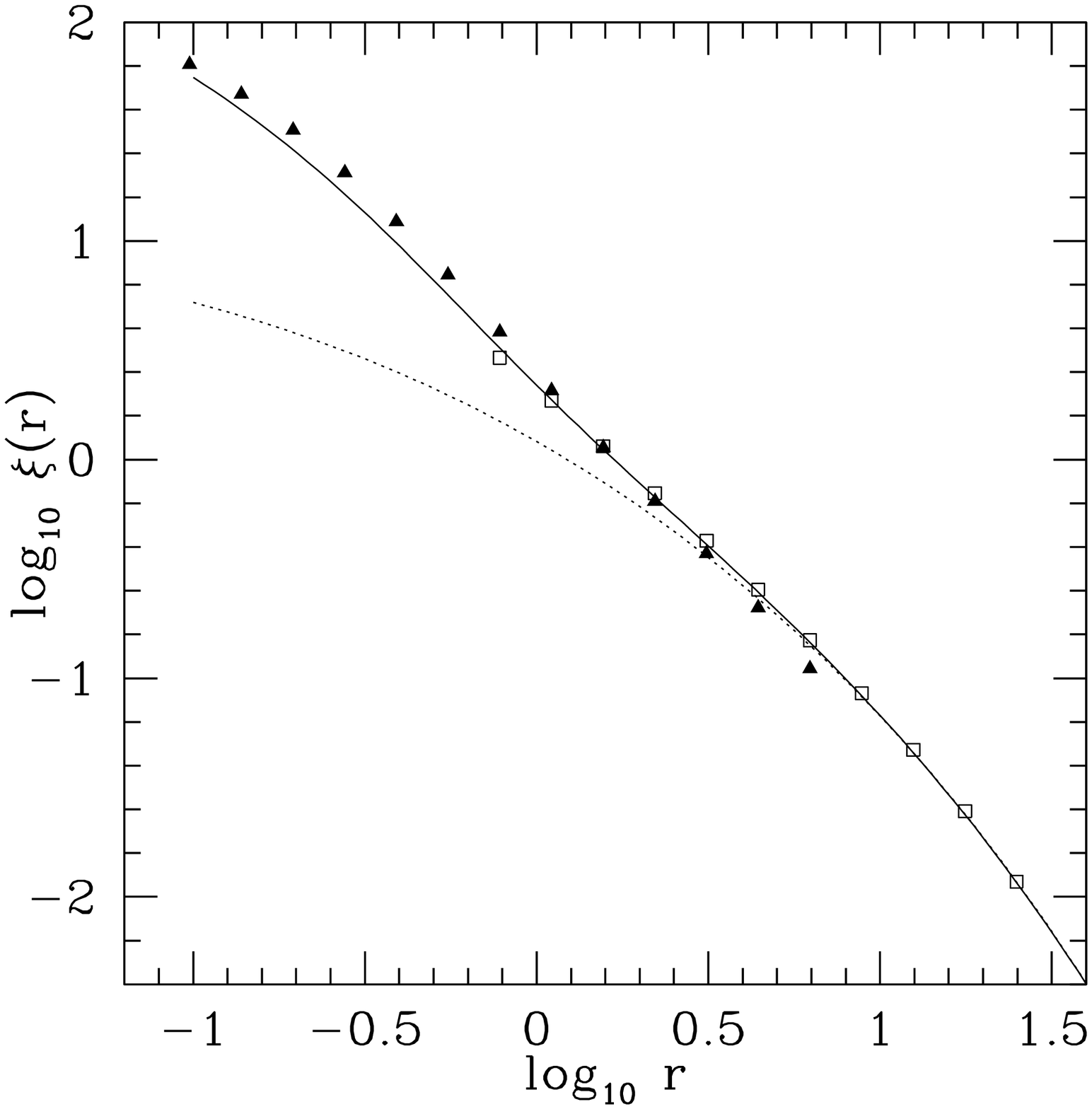,bbllx=18pt,bblly=145pt,bburx=572pt,bbury=700pt,width=5.5cm}}}
\caption[]{{\em Left panel:}
the power-spectrum measured at  $z=2$ in the S (filled triangles)
and B (open squares) simulations after adaptive smoothing, 
in logarithmic coordinates (wavenumber $k$
is expressed in Mpc$^{-1}$).  It is  compared to linear theory (dots) and to
non-linear ansatz of Peacock \& Dodds (1996, solid curve). {\em Middle panel:}
same as left panel, except that a correction for NGP
damping was applied to the data  prior to measurement of $P(k)$. {\em
Right panel:}
the variance of  the smoothed density field with a  spherical cell of radius
$r$ is shown  in logarithmic coordinates as a function  of $r$, as explained
in the text.}
\label{f:xiav}
\end{figure}

One way to test, at least partly, the quality  of our $N$-body experiments  is  
to compare second-order statistics measured in the simulations 
to  theoretical predictions, as illustrated by \Fig{xiav}.
Left panel shows the  measured power-spectrum, $P(k)=\langle |\delta_k|^2
\rangle$, in the density field  smoothed with the procedure described 
in \S~\ref{s:simu}.
Agreement with  linear theory is  appropriate at large scales,  as expected.
For comparison,  we plot  as well  the result obtained  from the  non-linear
Ansatz  of Hamilton  et  al.   (1991) optimized  for  the power-spectrum  by
Peacock  \& Dodds (1996).   The overall  agreement between  measurements and
non-linear  theory is  quite good,  except  at large  values of  $k$ in  both
simulations.  This is mainly the effect of the grid and to a lesser extent a
consequence  of the  adaptive Gaussian  smoothing.  Indeed,  any procedure
inferring  on a grid  a density  from a  particle distribution  implies some
smoothing with a  window of approximately the mesh  cell size.  This induces
large-$k$ damping  of the  power-spectrum. Here, the  smoothing is  not well
defined, but  most of the particles  are in dense regions,  due to non-linear
clustering,  and therefore  the corresponding  smoothing length,  $\ell$, is
likely to be much smaller than the grid size.  Thus, for most particles, all
the contribution  to the density is  given to the nearest  grid point (NGP).
As a  result, the Gaussian adaptive  smoothing has a  damping effect quite
close, although slightly  larger, to top hat smoothing with  a mesh cell (at
least for sufficiently evolved stages).   This is illustrated by middle panel
of \Fig{xiav} which shows the power-spectrum after correction for damping
due to NGP assignment.  Most of the missing power is recovered, as expected,
and the agreement with theory is much improved. Note that the triangles tend
to be  slightly above the solid  curve in the neighborhood  of $\log_{10} k
\simeq 0.4$.  This irregularity is  not surprising, given the small physical
size  of simulation  S.  It  is  probably associated  to a  rare event,  for
example an atypical cluster, although this does not show up significantly on
\Fig{figure_S}.

%-------------------------------------------------------
%-------------------------------------------------------

%LA FONCTION DE CORRELATION A 2 PTS
Right panel of \Fig{xiav}   shows    the   real   space  counterpart    of   the
power-spectrum. More precisely,  it displays  the  variance of the  smoothed
density field with a  sphere of radius $\ell$   as a function of $\ell$.  To
measure it,  we computed the  density from  the particle distribution  on a
grid twice thinner   than  the one used to    do the simulation,   using  the
cloud-in-cell method   (CIC,  e.g., Hockney   \&  Eastwood 1981).    Then we
corrected for CIC  damping  and smoothed  with the top  hat window  of  size
$\ell$ in  Fourier space.  Finally, back in  real space, the variance of the
density field was computed with the appropriate corrections for discreteness
(e.g., Peebles 1980), i.e.~$\sigma^2=\langle \delta^2 \rangle - 1/{\bar N}$,
where ${\bar N}$ is  the average particle count in  a cell of radius $\ell$.
The scale range considered was  $\lambda_{\rm g} \leq  \ell \leq L/4$, where
$L$ is the smallest dimension of  the box and  $\lambda_{\rm g}$ the spatial
resolution  of the simulation. As can  been seen in \Fig{xiav}, the
agreement with theoretical  predictions is quite  good, even  at $\ell\simeq
\lambda_{\rm g}$ although the effect of  softening of the forces is slightly
felt at this point.  Note as well that the triangles are somewhat shifted up
compared to the non-linear ansatz (except at  very large scales, where finite
volume effect contamination reduces the value  of $\sigma^2$, e.g., Colombi,
Bouchet \& Schaeffer 1994), as already noticed for the power-spectrum.

%%%%%%%%%%%%%%%%%%%%%%%%%%%%%%%%%%%%%%%%%%%
\section{Implementation of the inverse method.}
%%%%%%%%%%%%%%%%%%%%%%%%%%%%%%%%%%%%%%%%%%%

%%%%%%%%%%%%%%%%%%%%%%%%%%%%%%%%%%%%%%%%%%%
\subsection{Neglecting peculiar velocities.}
%%%%%%%%%%%%%%%%%%%%%%%%%%%%%%%%%%%%%%%%%%%

%%%%%%%%%%%%%%%%%%%%%%%%%%%%%%%%%%%%%%%%%%%
\subsubsection{High resolution spectra.}
%%%%%%%%%%%%%%%%%%%%%%%%%%%%%%%%%%%%%%%%%%%
\label{s:AT}

When the spectral  resolution is higher than $100$  km/s, thermal broadening
cannot be neglected and our model reads
\begin{equation}
  g_{i\ell}(\gamma)=A(\overline{z})c_1 \iint \left(\int_{-\infty}^{+\infty}
(D_{0}(x,\xp)\exp[\gamma(x,\xp)])^{\alpha-\beta}\exp\left(-           
c_2            \frac{(w_{i\ell}-
x)^2}{(D_{0}(x,\xp)\exp[\gamma(x,\xp)])^{2\beta}} \right) \d x 
\right) \delta_{D}(\xp-x_{\perp\ell})\d{}^{2} \xp \,, \EQN{defg1}
\end{equation}
where  $\alpha$,  $A({\bar z})$,  $c_1$,  $c_2$,  $\beta$, $D_0(x,\xp)$  and
$w_{i\ell}$  are  defined in  \Eqs{EOS}{defc1}  and  \Eq{defD0}.  Since  the
model,  $\M{M}\equiv  \gamma(x,\xp)$  is  a  continuous field,  we  need  to
interpret \Eq{sp1} in terms of convolutions, and functional derivatives.  In
particular  the  matrix   of  partial  functional  (Fr\'echet)  derivatives,
$\M{G}$, has the following kernel:
\begin{equation}
(\M{G})_{i\ell}(x,\xp)   \equiv   \left(\frac{\partial   g_{i\ell}}{\partial
\gamma}\right)(x,\xp)=    A(\overline{z})    c_1   D_0^{\alpha-\beta}(x,\xp)
\exp\left[(\alpha-\beta)\gamma(x,\xp)\right]                 B_{i\ell}(x,\xp)
\delta_D(\xp-\M{x}_{\perp,\ell}) \,, \EQN{defG}
\end{equation}
with $\delta_D(\xp-\M{x}_{\perp,\ell}) $ the Dirac delta function accounting
 for the singular distribution of \loss\, and:
\begin{equation}
B_{i\ell}(x,\xp)=  ( (\alpha-\beta)
+ c_2 2 \beta (w_{i\ell}-x)^2 D_0^{-2\beta}(x,\xp) 
\exp\left[-2\beta\gamma(x,\xp)\right] ){\cal B}_{i\ell}(x,\xp) \,,
\end{equation}
where :
\begin{equation}
{\cal B}_{i\ell}(x,\xp)= \exp\left(-c_2 \frac{(w_{i\ell}-x)^2}
{(D_0 (x,\xp)
\exp\left[\gamma(x,\xp)\right])^{2\beta}} \right) \, .
\end{equation}  The operator,  $\M{G}$,  defined  by
\Eq{defG} contracts over a given field, $\eta$, as:
\begin{equation}
%\begin{array}{rl}
(\M{G})_{il} \cdot \eta =
\int A(\overline{z}) c_1  D_0^{\alpha-\beta}(x,\xp) \exp\left[(\alpha-\beta)\gamma(x,\xp)\right] B_{i\ell}(x,\M{x}_{\perp,\ell})
\eta(x,\M{x}_{\perp,\ell}) \d x \, .
%\end{array}
\end{equation}

%%%%%%%%%%%%%%%%%%%%%%%%%%%%%%%%%%%%%%%%%%%
\subsubsection{Low resolution spectra.}
%%%%%%%%%%%%%%%%%%%%%%%%%%%%%%%%%%%%%%%%%%%
\label{s:Azero}

At low spectral resolution,  the model spells
\begin{equation}
  g_{i\ell}(\gamma)
  =A(\overline{z}) \iiint (D_{0}(x,\xp)\exp[\gamma(x,\xp)])^{\alpha} 
  \delta_D\left(x - w_{i\ell} \right)
\delta_D(\xp-\M{x}_{\perp,\ell}) \d x \d{}^{2} \xp \,, \EQN{}
\end{equation}
 which corresponds     to the limit  $c_{2}    \rightarrow  \infty$ in
 \Eq{defg1}.   The  kernel of  partial  functional derivatives $\M{G}$
 obeys:
\begin{equation}
(\M{G})_{i\ell}(x,\xp)=A(\overline{z})                          \alpha
D_0^{\alpha}(x,\M{x}_{\perp,\ell})                    \exp\left[\alpha
\gamma(x,\M{x}_{\perp,\ell})\right]    \delta_D\left(x   -  w  \right)
\delta_D(\xp-\M{x}_{\perp,\ell})\,.
\end{equation}
For instance $(\M{G} \cdot  \M{C}_0 \cdot  \M{G}^{\perp})_{i\ell,jm}$ in 
\Eq{eq1} reads
\begin{equation}
A(\overline{z})^2 \alpha^2  C_{\gamma\gamma}\left(w_{i\ell}
,w_{jm},\M{x}_{\perp,\ell}
,\M{x}_{\perp,m}\right)
D_0^{\alpha}(w_{i\ell},\M{x}_{\perp,\ell})
D_0^{\alpha}(w_{jm},\M{x}_{\perp,m}) 
\exp\left[\alpha \gamma({w_{i\ell}}
{},\M{x}_{\perp,\ell}) 
+ \alpha \gamma
(w_{jm},\M{x}_{\perp,m}) \right] \, .
\end{equation}

%%%%%%%%%%%%%%%%%%%%%%%%%%%%%%%%%%%
\subsection{Implementation of the inverse method  with peculiar velocities}
%%%%%%%%%%%%%%%%%%%%%%%%%%%%%%%%%%%
%%%%%%%%%%%%%%%%%%%%%%%%%%%%%%%%%%%

%%%%%%%%%%%%%%%%%%%%%%%%%%%%%%%%%%%
\subsubsection{Strong prior: peculiar velocity equals most likely velocity}
%%%%%%%%%%%%%%%%%%%%%%%%%%%%%%%%%%%
\label{s:AvelSP}

Restricting ourselves to a unique \los, our model reads
\begin{equation}
  g_{i\ell}(\gamma)=A(\overline{z})c_1 \int_{-\infty}^{+\infty}
(D_{0}(x)\exp[\gamma(x)])^{\alpha-\beta}\exp\left(-           
c_2            \frac{(w_{i\ell}-
x -v_{p}(x) )^2}{(D_{0}(x)\exp[\gamma(x)])^{2\beta}}  
\right) \, \d x  \,,  \EQN{defgv}
\end{equation}
where peculiar velocity, $v_{p}(x)$, equals the most likely velocity
\begin{equation}
\langle v_{p}(x) \rangle = \int K^{(v)}(x,y) \gamma{(y)} \d y \, . \EQN{defvel}
\end{equation}
The matrix of  partial functional derivatives, $\M{G}_{i}$ is  defined by its
contraction over a given field, $\eta$, as:
\begin{equation} 
(\M{G})_i\cdot \M{\eta  } \equiv \int  G_i(x) \eta(x)
\d  x   =\int  {\cal A}_i(x)  \eta(x)  \d  x   +   \int 
{\cal D}_i(x)
\left\{\int K^{(v)}(x,y) \eta(y) \d y \right\} \d x \,, \EQN{first}
\end{equation}
with :
\begin{equation}
{\cal A}_i(x)=A(\overline{z}) c_1 D_0^{\alpha-\beta}(x)
\exp\left((\alpha-\beta)\gamma(x)\right) 
\left\{\alpha-\beta+2\beta c_2 D_0^{-2\beta}\exp(-2\beta\gamma(x))
\left(w_i-x-v_{p}(x) \right) \right\}E_i(x) \,,
\end{equation}
\begin{equation}
{\cal D}_i(x)=A(\overline{z})c_1 D_0^{(\alpha-3\beta)}(x)
\exp\left((\alpha-3\beta)\gamma(x)\right) 2 c_2 
\left(w_i-x-v_{p}(x) \right) E_i(x)\,,
\end{equation}
\begin{equation}
 E_i(x)=
\exp\left( -c_2\frac{(w_i-x-v_{p}(x))^2}{D_0^{2\beta}(x)
\exp(2\beta\gamma(x))} \right) \,.
\EQN{fix}
\end{equation}
The double integration in the last term of \Eq{first}
 arises  because  $g$ is  effectively a double convolution.

%%%%%%%%%%%%%%%%%%%%%%%%%%%%%%%%%%%%%%%%%%%%%%%%%%%
\subsubsection{Weak prior: floating peculiar velocity }
%%%%%%%%%%%%%%%%%%%%%%%%%%%%%%%%%%%%%%%%%%%%%%%%%%%
\label{s:AvelWP}

We aim  to determine directly the  density and the  velocity, while assuming
the  correlations between  these  two  quantities are  known.  The model  is
identical  to \Eq{defgv},  but  the  peculiar  velocity  does  not  obey
\Eq{defvel}.
% \begin{equation}
%  g_{i\ell}(\gamma,v)=A(\overline{z})c_1 \int_{-\infty}^{+\infty}
% (D_{0}(x)\exp[\gamma(x)])^{\alpha-\beta}\exp\left(-           
% c_2            \frac{(w_{i\ell}-
% x -v_p(x))^2}{(D_{0}(x)\exp[\gamma(x)])^{2\beta}} 
% \right)\, \d x  \,.  \EQN{defgvw}
% \end{equation}
The matrix of  partial functional derivatives is :  $ \M{G}=\left( {\partial
g}/{\partial  \gamma}, {\partial  g}/{\partial  v_p} \right)  $.  The  first
component  of  $\M{G}$ is  given  by \Eq{defG}.  The  kernel  of the  second
component is computed as follow:
\begin{equation}
\frac{\partial g}{\partial v_p} =
A(\overline{z}) c_1 D_0^{\alpha-3\beta}(x)
\exp\left((\alpha-3\beta)\gamma(x)\right) 2c_2(w_i-x-v_{p}(x))
E_i(x) \equiv {\cal E}_i(x)\,
\end{equation}
where $E_i(x)$ is given by \Eq{fix}.
The matrix $ \M{G} \cdot \M{C}_0   \cdot   \T{\M{G}}$ [where
$M{C}_0$ is given by \Eq{defCgv}]
is computed as follow :
\begin{equation}
 \iint  \left[ {\cal A}_i(x)  {\cal A}_j  (y) C_{\gamma\gamma}(x,y)  + {\cal
A}_i(x) {\cal E}_j (y) C_{\gamma v}(x,y) + {\cal E}_i(x) {\cal A}_j (y) C_{v
\gamma}(x,y) + {\cal E}_i(x) {\cal E}_j  (y) C_{v v}(x,y) \right] \, \d x \d
y \,. \EQN{excont}
\end{equation}
Note that this  is a double
integral  to  be  compared  to  the  quadruple  integral  involved  in  the
computation of the equivalent term in the strong prior method (where contraction 
already involves a double convolution).

\bsp

\label{lastpage}

\end{document}